# The Pierre Auger Observatory III: Other Astrophysical Observations

THE PIERRE AUGER COLLABORATION

*Observatorio Pierre Auger, Av. San Martín Norte 304, 5613 Malargüe, Argentina*

# The Pierre Auger Collaboration

P. Abreu[74], M. Aglietta[57], E.J. Ahn[93], I.F.M. Albuquerque[19], D. Allard[33], I. Allekotte[1], J. Allen[96], P. Allison[98], J. Alvarez Castillo[67], J. Alvarez-Muñiz[84], M. Ambrosio[50], A. Aminaei[68], L. Anchordoqui[109], S. Andringa[74], T. Antičić[27], A. Anzalone[56], C. Aramo[50], E. Arganda[81], F. Arqueros[81], H. Asorey[1], P. Assis[74], J. Aublin[35], M. Ave[41], M. Avenier[36], G. Avila[12], T. Bäcker[45], M. Balzer[40], K.B. Barber[13], A.F. Barbosa[16], R. Bardenet[34], S.L.C. Barroso[22], B. Baughman[98], J. Bäuml[39], J.J. Beatty[98], B.R. Becker[106], K.H. Becker[38], A. Bellétoile[37], J.A. Bellido[13], S. BenZvi[108], C. Berat[36], X. Bertou[1], P.L. Biermann[42], P. Billoir[35], F. Blanco[81], M. Blanco[82], C. Bleve[38], H. Blümer[41, 39], M. Boháčová[29, 101], D. Boncioli[51], C. Bonifazi[25, 35], R. Bonino[57], N. Borodai[72], J. Brack[91], P. Brogueira[74], W.C. Brown[92], R. Bruijn[87], P. Buchholz[45], A. Bueno[83], R.E. Burton[89], K.S. Caballero-Mora[99], L. Caramete[42], R. Caruso[52], A. Castellina[57], O. Catalano[56], G. Cataldi[49], L. Cazon[74], R. Cester[53], J. Chauvin[36], S.H. Cheng[99], A. Chiavassa[57], J.A. Chinellato[20], A. Chou[93, 96], J. Chudoba[29], R.W. Clay[13], M.R. Coluccia[49], R. Conceição[74], F. Contreras[11], H. Cook[87], M.J. Cooper[13], J. Coppens[68, 70], A. Cordier[34], U. Cotti[66], S. Coutu[99], C.E. Covault[89], A. Creusot[33, 79], A. Criss[99], J. Cronin[101], A. Curutiu[42], S. Dagoret-Campagne[34], R. Dallier[37], S. Dasso[8, 4], K. Daumiller[39], B.R. Dawson[13], R.M. de Almeida[26], M. De Domenico[52], C. De Donato[67, 48], S.J. de Jong[68, 70], G. De La Vega[10], W.J.M. de Mello Junior[20], J.R.T. de Mello Neto[25], I. De Mitri[49], V. de Souza[18], K.D. de Vries[69], G. Decerprit[33], L. del Peral[82], O. Deligny[32], H. Dembinski[41], N. Dhital[95], C. Di Giulio[47, 51], J.C. Diaz[95], M.L. Díaz Castro[17], P.N. Diep[110], C. Dobrigkeit [20], W. Docters[69], J.C. D'Olivo[67], P.N. Dong[110, 32], A. Dorofeev[91], J.C. dos Anjos[16], M.T. Dova[7], D. D'Urso[50], I. Dutan[42], J. Ebr[29], R. Engel[39], M. Erdmann[43], C.O. Escobar[20], A. Etchegoyen[2], P. Facal San Luis[101], I. Fajardo Tapia[67], H. Falcke[68, 71], G. Farrar[96], A.C. Fauth[20], N. Fazzini[93], A.P. Ferguson[89], A. Ferrero[2], B. Fick[95], A. Filevich[2], A. Filipčič[78, 79], S. Fliescher[43], C.E. Fracchiolla[91], E.D. Fraenkel[69], U. Fröhlich[45], B. Fuchs[16], R. Gaior[35], R.F. Gamarra[2], S. Gambetta[46], B. García[10], D. García Gámez[83], D. Garcia-Pinto[81], A. Gascon[83], H. Gemmeke[40], K. Gesterling[106], P.L. Ghia[35, 57], U. Giaccari[49], M. Giller[73], H. Glass[93], M.S. Gold[106], G. Golup[1], F. Gomez Albarracin[7], M. Gómez Berisso[1], P. Gonçalves[74], D. Gonzalez[41], J.G. Gonzalez[41], B. Gookin[91], D. Góra[41, 72], A. Gorgi[57], P. Gouffon[19], S.R. Gozzini[87], E. Grashorn[98], S. Grebe[68, 70], N. Griffith[98], M. Grigat[43], A.F. Grillo[58], Y. Guardincerri[4], F. Guarino[50], G.P. Guedes[21], A. Guzman[67], J.D. Hague[106], P. Hansen[7], D. Harari[1], S. Harmsma[69, 70], J.L. Harton[91], A. Haungs[39], T. Hebbeker[43], D. Heck[39], A.E. Herve[13], C. Hojvat[93], N. Hollon[101], V.C. Holmes[13], P. Homola[72], J.R. Hörandel[68], A. Horneffer[68], M. Hrabovský[30, 29], T. Huege[39], A. Insolia[52], F. Ionita[101], A. Italiano[52], C. Jarne[7], S. Jiraskova[68], M. Josebachuili[2], K. Kadija[27], K.-H. Kampert[38], P. Karhan[28], P. Kasper[93], B. Kégl[34], B. Keilhauer[39], A. Keivani[94], J.L. Kelley[68], E. Kemp[20], R.M. Kieckhafer[95], H.O. Klages[39], M. Kleifges[40], J. Kleinfeller[39], J. Knapp[87], D.-H. Koang[36], K. Kotera[101], N. Krohm[38], O. Krömer[40], D. Kruppke-Hansen[38], F. Kuehn[93], D. Kuempel[38], J.K. Kulbartz[44], N. Kunka[40], G. La Rosa[56], C. Lachaud[33], P. Lautridou[37], M.S.A.B. Leão[24], D. Lebrun[36], P. Lebrun[93], M.A. Leigui de Oliveira[24], A. Lemiere[32], A. Letessier-Selvon[35], I. Lhenry-Yvon[32], K. Link[41], R. López[63], A. Lopez Agüera[84], K. Louedec[34], J. Lozano Bahilo[83], A. Lucero[2, 57], M. Ludwig[41], H. Lyberis[32], M.C. Maccarone[56], C. Macolino[35], S. Maldera[57], D. Mandat[29], P. Mantsch[93], A.G. Mariazzi[7], J. Marin[11, 57], V. Marin[37], I.C. Maris[35], H.R. Marquez Falcon[66], G. Marsella[54], D. Martello[49], L. Martin[37], H. Martinez[64], O. Martínez Bravo[63],

H.J. Mathes[39], J. Matthews[94, 100], J.A.J. Matthews[106], G. Matthiae[51], D. Maurizio[53], P.O. Mazur[93], G. Medina-Tanco[67], M. Melissas[41], D. Melo[2, 53], E. Menichetti[53], A. Menshikov[40], P. Mertsch[85], C. Meurer[43], S. Mićanović[27], M.I. Micheletti[9], W. Miller[106], L. Miramonti[48], S. Mollerach[1], M. Monasor[101], D. Monnier Ragaigne[34], F. Montanet[36], B. Morales[67], C. Morello[57], E. Moreno[63], J.C. Moreno[7], C. Morris[98], M. Mostafá[91], C.A. Moura[24, 50], S. Mueller[39], M.A. Muller[20], G. Müller[43], M. Münchmeyer[35], R. Mussa[53], G. Navarra[57 †], J.L. Navarro[83], S. Navas[83], P. Necesal[29], L. Nellen[67], A. Nelles[68, 70], J. Neuser[38], P.T. Nhung[110], L. Niemietz[38], N. Nierstenhoefer[38], D. Nitz[95], D. Nosek[28], L. Nožka[29], M. Nyklicek[29], J. Oehlschläger[39], A. Olinto[101], V.M. Olmos-Gilbaja[84], M. Ortiz[81], N. Pacheco[82], D. Pakk Selmi-Dei[20], M. Palatka[29], J. Pallotta[3], N. Palmieri[41], G. Parente[84], E. Parizot[33], A. Parra[84], R.D. Parsons[87], S. Pastor[80], T. Paul[97], M. Pech[29], J. Pękala[72], R. Pelayo[84], I.M. Pepe[23], L. Perrone[54], R. Pesce[46], E. Petermann[105], S. Petrera[47], P. Petrinca[51], A. Petrolini[46], Y. Petrov[91], J. Petrovic[70], C. Pfendner[108], N. Phan[106], R. Piegaia[4], T. Pierog[39], P. Pieroni[4], M. Pimenta[74], V. Pirronello[52], M. Platino[2], V.H. Ponce[1], M. Pontz[45], P. Privitera[101], M. Prouza[29], E.J. Quel[3], S. Querchfeld[38], J. Rautenberg[38], O. Ravel[37], D. Ravignani[2], B. Revenu[37], J. Ridky[29], S. Riggi[84, 52], M. Risse[45], P. Ristori[3], H. Rivera[48], V. Rizi[47], J. Roberts[96], C. Robledo[63], W. Rodrigues de Carvalho[84, 19], G. Rodriguez[84], J. Rodriguez Martino[11, 52], J. Rodriguez Rojo[11], I. Rodriguez-Cabo[84], M.D. Rodríguez-Frías[82], G. Ros[82], J. Rosado[81], T. Rossler[30], M. Roth[39], B. Rouillé-d'Orfeuil[101], E. Roulet[1], A.C. Rovero[8], C. Rühle[40], F. Salamida[47, 39], H. Salazar[63], G. Salina[51], F. Sánchez[2], M. Santander[11], C.E. Santo[74], E. Santos[74], E.M. Santos[25], F. Sarazin[90], B. Sarkar[38], S. Sarkar[85], R. Sato[11], N. Scharf[43], V. Scherini[48], H. Schieler[39], P. Schiffer[43], A. Schmidt[40], F. Schmidt[101], O. Scholten[69], H. Schoorlemmer[68, 70], J. Schovancova[29], P. Schovánek[29], F. Schröder[39], S. Schulte[43], D. Schuster[90], S.J. Sciutto[7], M. Scuderi[52], A. Segreto[56], M. Settimo[45], A. Shadkam[94], R.C. Shellard[16, 17], I. Sidelnik[2], G. Sigl[44], H.H. Silva Lopez[67], A. Śmiałkowski[73], R. Šmída[39, 29], G.R. Snow[105], P. Sommers[99], J. Sorokin[13], H. Spinka[88, 93], R. Squartini[11], S. Stanic[79], J. Stapleton[98], J. Stasielak[72], M. Stephan[43], E. Strazzeri[56], A. Stutz[36], F. Suarez[2], T. Suomijärvi[32], A.D. Supanitsky[8, 67], T. Šuša[27], M.S. Sutherland[94, 98], J. Swain[97], Z. Szadkowski[73], M. Szuba[39], A. Tamashiro[8], A. Tapia[2], M. Tartare[36], O. Taşcău[38], C.G. Tavera Ruiz[67], R. Tcaciuc[45], D. Tegolo[52, 61], N.T. Thao[110], D. Thomas[91], J. Tiffenberg[4], C. Timmermans[70, 68], D.K. Tiwari[66], W. Tkaczyk[73], C.J. Todero Peixoto[18, 24], B. Tomé[74], A. Tonachini[53], P. Travnicek[29], D.B. Tridapalli[19], G. Tristram[33], E. Trovato[52], M. Tueros[84, 4], R. Ulrich[99, 39], M. Unger[39], M. Urban[34], J.F. Valdés Galicia[67], I. Valiño[84, 39], L. Valore[50], A.M. van den Berg[69], E. Varela[63], B. Vargas Cárdenas[67], J.R. Vázquez[81], R.A. Vázquez[84], D. Veberič[79, 78], V. Verzi[51], J. Vicha[29], M. Videla[10], L. Villaseñor[66], H. Wahlberg[7], P. Wahrlich[13], O. Wainberg[2], D. Walz[43], D. Warner[91], A.A. Watson[87], M. Weber[40], K. Weidenhaupt[43], A. Weindl[39], S. Westerhoff[108], B.J. Whelan[13], G. Wieczorek[73], L. Wiencke[90], B. Wilczyńska[72], H. Wilczyński[72], M. Will[39], C. Williams[101], T. Winchen[43], L. Winders[109], M.G. Winnick[13], M. Wommer[39], B. Wundheiler[2], T. Yamamoto[101 a], T. Yapici[95], P. Younk[45], G. Yuan[94], A. Yushkov[84, 50], B. Zamorano[83], E. Zas[84], D. Zavrtanik[79, 78], M. Zavrtanik[78, 79], I. Zaw[96], A. Zepeda[64], M. Zimbres-Silva[20, 38] M. Ziolkowski[45]

[1] *Centro Atómico Bariloche and Instituto Balseiro (CNEA- UNCuyo-CONICET), San Carlos de Bariloche, Argentina*
[2] *Centro Atómico Constituyentes (Comisión Nacional de Energía Atómica/CONICET/UTN-FRBA), Buenos Aires, Argentina*
[3] *Centro de Investigaciones en Láseres y Aplicaciones, CITEFA and CONICET, Argentina*
[4] *Departamento de Física, FCEyN, Universidad de Buenos Aires y CONICET, Argentina*
[7] *IFLP, Universidad Nacional de La Plata and CONICET, La Plata, Argentina*
[8] *Instituto de Astronomía y Física del Espacio (CONICET- UBA), Buenos Aires, Argentina*
[9] *Instituto de Física de Rosario (IFIR) - CONICET/U.N.R. and Facultad de Ciencias Bioquímicas y Farmacéuticas U.N.R., Rosario, Argentina*
[10] *National Technological University, Faculty Mendoza (CONICET/CNEA), Mendoza, Argentina*
[11] *Observatorio Pierre Auger, Malargüe, Argentina*
[12] *Observatorio Pierre Auger and Comisión Nacional de Energía Atómica, Malargüe, Argentina*
[13] *University of Adelaide, Adelaide, S.A., Australia*
[16] *Centro Brasileiro de Pesquisas Fisicas, Rio de Janeiro, RJ, Brazil*
[17] *Pontifícia Universidade Católica, Rio de Janeiro, RJ, Brazil*

[18] *Universidade de São Paulo, Instituto de Física, São Carlos, SP, Brazil*
[19] *Universidade de São Paulo, Instituto de Física, São Paulo, SP, Brazil*
[20] *Universidade Estadual de Campinas, IFGW, Campinas, SP, Brazil*
[21] *Universidade Estadual de Feira de Santana, Brazil*
[22] *Universidade Estadual do Sudoeste da Bahia, Vitoria da Conquista, BA, Brazil*
[23] *Universidade Federal da Bahia, Salvador, BA, Brazil*
[24] *Universidade Federal do ABC, Santo André, SP, Brazil*
[25] *Universidade Federal do Rio de Janeiro, Instituto de Física, Rio de Janeiro, RJ, Brazil*
[26] *Universidade Federal Fluminense, EEIMVR, Volta Redonda, RJ, Brazil*
[27] *Rudjer Bošković Institute, 10000 Zagreb, Croatia*
[28] *Charles University, Faculty of Mathematics and Physics, Institute of Particle and Nuclear Physics, Prague, Czech Republic*
[29] *Institute of Physics of the Academy of Sciences of the Czech Republic, Prague, Czech Republic*
[30] *Palacky University, RCATM, Olomouc, Czech Republic*
[32] *Institut de Physique Nucléaire d'Orsay (IPNO), Université Paris 11, CNRS-IN2P3, Orsay, France*
[33] *Laboratoire AstroParticule et Cosmologie (APC), Université Paris 7, CNRS-IN2P3, Paris, France*
[34] *Laboratoire de l'Accélérateur Linéaire (LAL), Université Paris 11, CNRS-IN2P3, Orsay, France*
[35] *Laboratoire de Physique Nucléaire et de Hautes Energies (LPNHE), Universités Paris 6 et Paris 7, CNRS-IN2P3, Paris, France*
[36] *Laboratoire de Physique Subatomique et de Cosmologie (LPSC), Université Joseph Fourier, INPG, CNRS-IN2P3, Grenoble, France*
[37] *SUBATECH, École des Mines de Nantes, CNRS-IN2P3, Université de Nantes, Nantes, France*
[38] *Bergische Universität Wuppertal, Wuppertal, Germany*
[39] *Karlsruhe Institute of Technology - Campus North - Institut für Kernphysik, Karlsruhe, Germany*
[40] *Karlsruhe Institute of Technology - Campus North - Institut für Prozessdatenverarbeitung und Elektronik, Karlsruhe, Germany*
[41] *Karlsruhe Institute of Technology - Campus South - Institut für Experimentelle Kernphysik (IEKP), Karlsruhe, Germany*
[42] *Max-Planck-Institut für Radioastronomie, Bonn, Germany*
[43] *RWTH Aachen University, III. Physikalisches Institut A, Aachen, Germany*
[44] *Universität Hamburg, Hamburg, Germany*
[45] *Universität Siegen, Siegen, Germany*
[46] *Dipartimento di Fisica dell'Università and INFN, Genova, Italy*
[47] *Università dell'Aquila and INFN, L'Aquila, Italy*
[48] *Università di Milano and Sezione INFN, Milan, Italy*
[49] *Dipartimento di Fisica dell'Università del Salento and Sezione INFN, Lecce, Italy*
[50] *Università di Napoli "Federico II" and Sezione INFN, Napoli, Italy*
[51] *Università di Roma II "Tor Vergata" and Sezione INFN, Roma, Italy*
[52] *Università di Catania and Sezione INFN, Catania, Italy*
[53] *Università di Torino and Sezione INFN, Torino, Italy*
[54] *Dipartimento di Ingegneria dell'Innovazione dell'Università del Salento and Sezione INFN, Lecce, Italy*
[56] *Istituto di Astrofisica Spaziale e Fisica Cosmica di Palermo (INAF), Palermo, Italy*
[57] *Istituto di Fisica dello Spazio Interplanetario (INAF), Università di Torino and Sezione INFN, Torino, Italy*
[58] *INFN, Laboratori Nazionali del Gran Sasso, Assergi (L'Aquila), Italy*
[61] *Università di Palermo and Sezione INFN, Catania, Italy*
[63] *Benemérita Universidad Autónoma de Puebla, Puebla, Mexico*
[64] *Centro de Investigación y de Estudios Avanzados del IPN (CINVESTAV), México, D.F., Mexico*
[66] *Universidad Michoacana de San Nicolas de Hidalgo, Morelia, Michoacan, Mexico*
[67] *Universidad Nacional Autonoma de Mexico, Mexico, D.F., Mexico*
[68] *IMAPP, Radboud University Nijmegen, Netherlands*
[69] *Kernfysisch Versneller Instituut, University of Groningen, Groningen, Netherlands*
[70] *Nikhef, Science Park, Amsterdam, Netherlands*
[71] *ASTRON, Dwingeloo, Netherlands*
[72] *Institute of Nuclear Physics PAN, Krakow, Poland*

[73] University of Łódź, Łódź, Poland
[74] LIP and Instituto Superior Técnico, Lisboa, Portugal
[78] J. Stefan Institute, Ljubljana, Slovenia
[79] Laboratory for Astroparticle Physics, University of Nova Gorica, Slovenia
[80] Instituto de Física Corpuscular, CSIC-Universitat de València, Valencia, Spain
[81] Universidad Complutense de Madrid, Madrid, Spain
[82] Universidad de Alcalá, Alcalá de Henares (Madrid), Spain
[83] Universidad de Granada & C.A.F.P.E., Granada, Spain
[84] Universidad de Santiago de Compostela, Spain
[85] Rudolf Peierls Centre for Theoretical Physics, University of Oxford, Oxford, United Kingdom
[87] School of Physics and Astronomy, University of Leeds, United Kingdom
[88] Argonne National Laboratory, Argonne, IL, USA
[89] Case Western Reserve University, Cleveland, OH, USA
[90] Colorado School of Mines, Golden, CO, USA
[91] Colorado State University, Fort Collins, CO, USA
[92] Colorado State University, Pueblo, CO, USA
[93] Fermilab, Batavia, IL, USA
[94] Louisiana State University, Baton Rouge, LA, USA
[95] Michigan Technological University, Houghton, MI, USA
[96] New York University, New York, NY, USA
[97] Northeastern University, Boston, MA, USA
[98] Ohio State University, Columbus, OH, USA
[99] Pennsylvania State University, University Park, PA, USA
[100] Southern University, Baton Rouge, LA, USA
[101] University of Chicago, Enrico Fermi Institute, Chicago, IL, USA
[105] University of Nebraska, Lincoln, NE, USA
[106] University of New Mexico, Albuquerque, NM, USA
[108] University of Wisconsin, Madison, WI, USA
[109] University of Wisconsin, Milwaukee, WI, USA
[110] Institute for Nuclear Science and Technology (INST), Hanoi, Vietnam
† Deceased
[a] at Konan University, Kobe, Japan

# Anisotropies and Chemical Composition of Ultra-High Energy Cosmic Rays Using Arrival Directions Measured by the Pierre Auger Observatory

EDIVALDO M. SANTOS[1], FOR THE PIERRE AUGER COLLABORATION[2]
[1]*Instituto de Física, Universidade Federal do Rio de Janeiro, 21941-972, Rio de Janeiro, Brazil*
[2]*Observatorio Pierre Auger, Av. San Martin Norte 304, 5613 Malargüe, Argentina*
*(Full Author list: http://www.auger.org/archive/authors_2011_05.html)*
*auger_spokespersons@fnal.gov*

**Abstract:** The Pierre Auger Collaboration has reported evidence for anisotropies in the arrival directions of cosmic rays with energies larger than $E_{th} = 55$ EeV. There is a correlation above the isotropic expectation with nearby active galaxies and the largest excess is in a celestial region around the position of the radio galaxy Cen A. If these anisotropies are due to nuclei of charge $Z$, the protons accelerated in those sources are expected, under reasonable assumptions, to lead to excesses in the same regions of the sky at energies above $E_{th}/Z$. We here report the lack of anisotropies at these lower energies for illustrative values of $Z = 6$, 13 and 26. These observations set stringent constraints on the allowed proton fraction at the sources.

## 1 Introduction

Measurements of the anisotropies in the distribution of arrival directions of Ultra-High Energy Cosmic Rays (UHECR), when combined with information on their chemical composition and spectral features can provide valuable information on the sources and acceleration mechanisms capable of producing subatomic particles with macroscopic energies.

The Pierre Auger Observatory, the largest cosmic ray detector ever built, has observed [1] a flux suppression above 40 EeV (where 1 EeV = $10^{18}$ eV) consistent with that expected from the interaction of protons or heavy nuclei with the cosmic microwave background [2, 3]. In addition, it has reported evidence for anisotropy in the distribution of arrival directions of the highest energy events [4, 5, 6]. The arrival directions of the events with energies above 55 EeV show a degree of correlation within an angular scale of $\sim 3°$ with the positions of nearby (within $\sim 75$ Mpc) Active Galactic Nuclei (AGN) from the VCV catalog [7], which is above that expected from chance coincidences in an isotropic sky. However, one cannot identify AGN as the actual sources of UHECR since these trace the distribution of matter in the local Universe where other potential acceleration sites (such as Gamma Ray Bursts) are also present. Another interesting feature observed in the data sample is an excess of arrival directions towards the celestial position of Cen A, which is most significant in an angular window of radius 18°. This is the nearest radio loud AGN at $\sim 4$ Mpc from Earth, and is located at equatorial coordinates $(\alpha,\delta) = (201.4°, -43.0°)$.

The determination of the composition of primary CRs at the energies for which their flux is measured to be strongly suppressed is an active area of study. This stems from both the low observed flux and the reliance on Monte Carlo models that require large extrapolations from currently measured physics. A method was recently proposed linking the anisotropy measurements to the cosmic ray composition by exploiting that particles with the same rigidity follow the same path through a magnetic field [8]. Given generic assumptions about the acceleration process at the source, neglecting interactions with the photon background and assuming that the anisotropies at energies $E$ are caused by heavy primaries with charge $Z$, it relates the strength of an anisotropy at energy $E/Z$ to the fraction of protons at that energy in the same source. We here describe observations related to a search for this kind of effect using data collected by the Pierre Auger Observatory [9].

## 2 The Detector and the Data Sample

Located in the city of Malargüe, Mendoza, Argentina, the Pierre Auger Observatory is a hybrid detector consisting of a Surface Detector (SD) with 1660 stations covering an area of $\sim 3000$ km$^2$ and a Fluorescence Detector (FD) comprised of 27 fluorescence telescopes in four locations around the border and overlooking the array. As the shower develops in the atmosphere, the nitrogen scintillation light is detected by the telescopes which are able to record the ultraviolet radiation emitted during the de-excitation of molecular nitrogen. When shower particles reach ground level they are detected through water-Cherenkov light produced within the SD stations [10].

The reconstruction of the event direction is done by fitting a certain shower front model propagating at the speed of light to the measured arrival times and particle densities in the stations triggered by the air shower. By profiting from the unique hybrid nature of the Auger Observatory, events which are detected simultaneously by the SD and the FD are used to inter-calibrate these two detectors, providing an energy estimate almost independent of Monte Carlo simulations. Firstly, the estimated signal at 1000 m from the reconstructed shower core, $S(1000)$, is corrected for atmospheric attenuation, and gives rise to a signal value at a reference zenith angle ($S_{38}$). Finally, this signal can then be correlated to the calorimetric energy measurement performed by the FD. Such a calibration curve has been determined for the hybrid events and can be used for the whole high statistics sample measured by the SD [11].

The data used in this analysis were collected by the SD from 1 January 2004 to 31 December 2009 and contain showers with reconstructed zenith angle $\theta < 60$ degrees. Only events for which the station with the highest signal was surrounded by an entire hexagon of active detectors at the time of detection have been included. Recording the number of active detector configurations able to trigger such showers allows one to obtain the array exposure as a function of time. Also, by monitoring the communications between individual stations and the Central Data Acquisition System, we are able to identify dead times in the detectors. After accounting for these and removing periods of large fluctuations in the array aperture we are left with a livetime for the SD array of about 87%.

## 3 Low Energy Anisotropy Searches

In ref. [8] Lemoine and Waxman explored the consequences of the assumption that the anisotropies observed at the highest energies (above a threshold $E_{th}$) were caused by a predominantly heavy component. Assuming the presence of protons in the same source, and considering the fact that the Larmor radius in a given magnetic field depends only on rigidity, $E/Z$ for relativistic particles, if the high energy anisotropy is due to particles with charge $Z$ there should be a corresponding low energy anisotropy of protons at energies above $E_{th}/Z$.

In ref. [6] the most significant excess for a top-hat window centered on Cen A was found for a radius of 18°, and we will hence focus on this region. There are a total of 60 events in this data set, and 10 are at a distance smaller than 18° of the position of Cen A[1]. The number of events expected by random correlations inside this angular window for the case of a completely isotropic sky, taking into account also the detector exposure, is estimated as $N_{bkg} = (N_{tot}-N_{obs})x/(1-x) = 2.44$, where $x \simeq 0.0466$ is the fraction of the sky, weighted by the observatory's exposure, covered by the 18° circular window around Cen A. The *a posteriori* nature of the observed excess around Cen A (the location of the excess, the energy threshold and angular size were chosen so as to maximize the excess) implies that new independent data will be required to determine the actual strength of the source and establish its significance in an *a priori* way.

Taking as representative values for the atomic number of heavy primaries $Z = 6, 13$ and 26, we have searched for anisotropies in an 18° window around Cen A above threshold energies of $E_{th}/Z = 9.2$ EeV, 4.2 EeV and 2.1 EeV, respectively. Table 1 presents the total, observed, and expected number of events adopting different values of $Z$. No significant excesses have been found.

| $Z$ | $E_{min}$ [EeV] | $N_{tot}$ | $N_{obs}$ | $N_{bkg}$ |
|---|---|---|---|---|
| 6 | 9.2 | 4455 | 219 | $207 \pm 14$ |
| 13 | 4.2 | 16640 | 797 | $774 \pm 28$ |
| 26 | 2.1 | 63600 | 2887 | $2920 \pm 54$ |

Table 1: Total number of events, $N_{tot}$, and those observed in an angular window of 18° around Cen A, $N_{obs}$, as well as the expected background $N_{bkg}$. Results are given for different energy thresholds, corresponding to $E_{min} = E_{th}/Z$ for the indicated values of $Z$ and $E_{th} = 55$ EeV.

Above an energy of $E_{th} = 55$ EeV, the arrival directions measured by Auger have a degree of correlation above isotropic expectations with the positions of nearby AGN in the VCV catalog at less than 75 Mpc ($z_{max} = 0.018$) in an angular window of 3.1° [6]. Therefore, we have also looked for anisotropies above the same low energy thresholds for events in $\psi = 3.1°$ windows around the same VCV AGN. Once again, no statistically significant excesses have been identified and a summary of the searches is shown in table 2. It is worth mentioning that the data collected during the exploratory scan, i.e., the period during which the collected data were used to tune the correlation parameters ($E_{th}, z_{max}, \psi$) in order to maximize the correlation signal (see [6] for details), were not used to produce this table.

For the heaviest primaries considered here ($Z = 26$), the low energy threshold (2.1 EeV) falls below the region of full SD efficiency ($E > 3$ EeV). In this case, we have performed a fit to the observed zenith angle distribution of the events in order to account for the zenith angle dependent detection efficiency in the estimate of the isotropic expectations in the windows considered.

1. In ref. [6], 13 out of 69 arrival directions are reported within 18°. of Cen A. The difference with the numbers reported here is due to a stricter event selection necessary for an accurate estimate of the exposure at low energies.

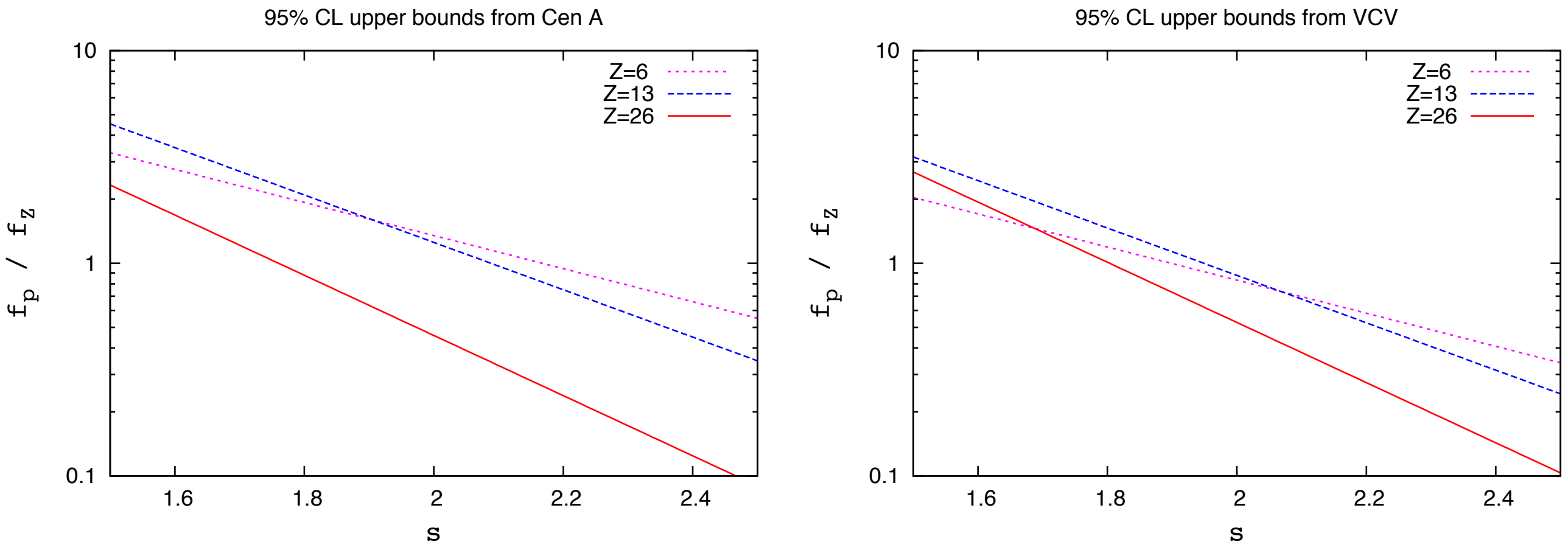

Figure 1: Upper bounds at 95%CL on the allowed proton to heavy fractions in the source as a function of the assumed low energy spectral index $s$. The different lines are for charges Z = 6, 13 and 26, as indicated. Left: bounds from the Cen A analysis. Right: bounds from the VCV analysis.

| $Z$ | $E_{min}$ [EeV] | $N_{tot}$ | $N_{obs}$ | $N_{bkg}$ |
|---|---|---|---|---|
| 6 | 9.2 | 3626 | 763 | $770 \pm 28$ |
| 13 | 4.2 | 13482 | 2852 | $2860 \pm 54$ |
| 26 | 2.1 | 51641 | 10881 | $10966 \pm 105$ |

Table 2: Total number of events, $N_{tot}$, and those observed within 3.1° from objects with $z \leq 0.018$ in the VCV catalog, $N_{obs}$, as well as the expected isotropic background $N_{bkg}$. Results are given for different energy thresholds, corresponding to $E_{min} = E_{th}/Z$ for the indicated values of $Z$ and $E_{th} = 55$ EeV.

## 4 Chemical Composition Constraints

In astrophysical environments for which the acceleration processes are essentially dependent on the magnetic rigidity, one can write the differential cosmic ray energy spectrum for primaries of atomic number $Z$ as:

$$\frac{dn_Z}{dE} = k_Z \Phi \left( \frac{E}{Z} \right),$$

where $k_Z$ is a normalization constant for the spectrum. Under this assumption, the expected number of protons $N_p(E > E_{th}/Z)$ above $E_{th}/Z$ can be shown to be related to the number of heavy primaries $N_Z(E > E_{th})$ above $E_{th}$ through $N_p(E > E_{th}/Z) = \frac{k_p}{Zk_Z} N_Z(E > E_{th})$. Experimentally one can estimate the ratio of source event numbers above the low and high energy thresholds as

$$R_Z \equiv \frac{N(E > E_{th}/Z)}{N(E > E_{th})},$$

where $N = N_{obs} - N_{bkg}$. The numerator of this ratio is equal to the sum of protons ($N_p(E > E_{th}/Z)$) and heavy nuclei ($N_Z(E > E_{th}/Z)$), whereas the denominator is considered to be dominated essentially by heavy primaries, i.e., $N_Z(E > E_{th})$. Therefore, a conservative upper bound on the ratio is $R_Z > \frac{k_p}{Zk_Z} + 1$, where no extra assumption was made on the spectral shapes of both chemical species. This inequality can be interpreted as a lower bound on the spectral normalizations $k_p/k_Z < (R_Z - 1)Z$.

We use the profile likelihood method [12] to derive upper bounds on the ratio $R_Z$ both for events around Cen A and those around the positions of the VCV AGN. This method takes into account Poisson fluctuations in the signal and expected background at both high and low energies simultaneously. We find the following 95% CL bounds:

$$R_{26}^{\rm CenA} < 12.9, \quad R_{13}^{\rm CenA} < 17.3, \quad R_6^{\rm CenA} < 9.1$$

$$R_{26}^{\rm VCV} < 14.7, \quad R_{13}^{\rm VCV} < 12.4, \quad R_6^{\rm VCV} < 6.0$$

If we now assume that below a certain cutoff $E_1$ the energy spectra are proportional to power laws of rigidity, one can write

$$\Phi \left( \frac{E}{Z} \right) \propto \left( \frac{E}{Z} \right)^{-s}$$

and the ratio of spectral normalizations can be written in terms of the relative abundances of protons to species of charge $Z$ at the sources $\frac{f_p}{f_Z} = \frac{k_p}{k_Z} Z^{-s}$.

Figure 1 shows the corresponding upper limits on the proton to heavy primary abundances at the sources as a function of the spectral index $s$ for different $Z$. The bounds obtained from the analysis of Cen A are similar to those obtained from VCV AGN, becoming more stringent as the spectral index hardens. Even though we have not included energy losses in this analysis, these will eventually degrade the energy of the high energy nuclei, increasing the size of the predicted low energy anisotropy [8]. Therefore, the bounds discussed here are conservative.

Since the size of the angular window around Cen A was chosen *a posteriori*, an unbiased estimate of the significance will only be found with new independent data. However, it is worth mentioning that varying the energy threshold to 50 or 60 EeV leads to similar results. Also, vary-ing the angular window to 10° has no large impact on the bounds, with the main effects coming from the change in the expected background, the limits being relaxed by a factor $\sim 2$ in this case.

## 5 Conclusions

We have presented observations of the distribution of events at energies above $E_{th}/Z$ in the directions where anisotropies have been previously observed above $E_{th} = 55$ EeV. We pursued the idea that the anisotropies at high energies could be caused by heavy primaries, either for the excess of events around Cen A at an angular scale of 18° or for an angular scale of 3.1° around the positions of VCV AGN. We have taken as representative values for the atomic numbers present in the sources $Z = 6, 13$, and 26. The low energy ($E_{th}/Z$) anisotropy caused by the protons in the same sources are not observed, allowing us to derive upper bounds on the light to heavy composition abundances at the sources. The bounds from both the VCV and the Cen A analyses are similar, and their dependence with the source spectral index is such that softer spectra produce less stringent upper limits. Low energy abundance measurements have been performed by the ATIC-2 experiment [13] and they point to $f_p/f_Z$ values, as measured on Earth, above the upper limits presented here (for example, $f_p \simeq f_{He} \simeq 2f_{CNO} \simeq 2f_{Ne-Si} \simeq 2f_{Z>17} \simeq 4f_{Fe}$). At these low energies (100 TeV), cosmic rays are believed to be of galactic origin, and the larger diffusion coefficient of protons in our galaxy's magnetic field as compared to heavier nuclei imply that the corresponding $f_p/f_Z$ at the sources are even larger. However, the probable extragalactic origin of UHECR, as well as their much higher energies, implies that the ATIC measured abundances do not necessarily apply to the sources contributing to the Auger data and should be taken only as indicative values of the expected ratios.

Therefore, scenarios in which a rigidity dependent acceleration mechanism leads to a heavy primary domination at the highest energies and in which there is an abundant proton component at low energies are not favored (see Fig. 1). How these conclusions are modified in the presence of strong structured magnetic fields and taking into account the relevant energy losses remains to be seen. Finally, we mention that this joint composition-anisotropy study is independent of measurements of the average depth of the maximum of shower development [14, 15]. Instead, it depends on assumptions related to propagation and acceleration mechanisms at the sources.

# Bounds on the density of sources of ultra high energy cosmic rays from the Pierre Auger Observatory data

MANLIO DE DOMENICO[1,2], FOR THE PIERRE AUGER COLLABORATION[3]
[1] *Laboratorio sui Sistemi Complessi, Scuola Superiore di Catania, Via Valdisavoia 9, 95123 Catania, Italy*
[2] *Istituto Nazionale di Fisica Nucleare, Sez. di Catania, Via S. Sofia 64, 95123 Catania, Italy*
[3] *Observatorio Pierre Auger, Av. San Martín Norte 304, 5613 Malargüe, Argentina*
*(Full author list: http://www.auger.org/archive/authors_2011_05.html)*
*auger_spokespersons@fnal.gov*

**Abstract:** We present constraints on the density of sources obtained by analyzing the clustering (or absence of clustering) of the arrival directions of ultra-high energy cosmic rays detected at the Pierre Auger Observatory. We consider bounds for isotropically distributed sources and for sources distributed according to the 2MRS catalog.

## 1 Introduction

The identification of the sources of ultra-high energy cosmic rays (UHECRs) is a major challenge in astroparticle physics. Only few astrophysical objects in the universe are expected to be able to accelerate particles up to 100 EeV (1 EeV is $10^{18}$ eV) [1]. It is likely that those sources are extragalactic, and only sources closer than about 200 Mpc from Earth can contribute appreciably to the observed flux above 60 EeV. Interactions with the cosmic microwave background by cosmic ray protons, or nuclei, with larger energies lead to strong attenuation of their flux from more distant sources (the Greisen-Zatsepin-Kuz'min (GZK) effect [2, 3]). Observing in the southern hemisphere, the Pierre Auger Collaboration has reported the measurement of a correlation above the isotropic expectation between the arrival directions of cosmic rays with energies exceeding $\sim 60$ EeV and the positions of active galactic nuclei (AGN) within 75 Mpc [4, 5, 6], at angular scales of $\sim 3°$. This observation, along with the measurement of a suppression of the flux at the highest energies [7, 8] is consistent with an extragalactic origin of the UHECRs and with the expectation from the GZK effect. Note however that the HiRes Collaboration has reported an absence of a comparable correlation in observations in the northern hemisphere [9].

If the deflections in the trajectories of UHECRs caused by intervening magnetic fields are small, the distribution of their arrival directions in the energy range above the GZK threshold is expected to reflect the clustering properties of those local sources. A large number of multiplets of arrival directions is expected if the local density of sources is sufficiently small, whereas fewer multiplets are expected for larger values of the density. Indeed, the lower the density of sources is, the larger is the probability that more than one of the observed cosmic rays come from the same source. Hence, a statistical analysis of clustering in the observed UHECR arrival directions should shed light on the density of their sources, further reducing the list of candidate astrophysical sources. Conversely, if the deviations in the trajectories of UHECRs are large, as expected if heavy nuclei are the dominant composition or if intervening magnetic fields have a strong effect, this approach may not be suitable for establishing constraints on the density of sources, since the clustering signal could be similar to that expected for smaller deflections and a larger density.

Estimates of the density of sources in our cosmic neighborhood have been obtained in the range $10^{-6} - 10^{-4}$ Mpc$^{-3}$ (with large uncertainties), using data from previous experiments, under various assumptions on the sources and their distribution [10, 11, 12, 13, 14]. More recently, approaches involving the two-point autocorrelation function or its variants have been used to constrain the source density. Representative studies can be found in [15], in which source models that trace the distribution of matter in the nearby universe as well as a model with a continuous, uniform distribution of sources were analysed in an autocorrelation study of the first 27 arrival directions of UHECRs with energies larger than 56 EeV measured by the Pierre Auger Observatory [5]. Results from such analyses suggest a source density ranging from $0.2 \times 10^{-4}$ Mpc$^{-3}$ to $5 \times 10^{-4}$ Mpc$^{-3}$ with an upper bound $\approx 10^{-2}$ Mpc$^{-3}$ at 95% CL.

In the present study, we derive bounds on the density of sources through an autocorrelation analysis of the set of 67 arrival directions of UHECRs with energies larger than 60 EeV measured by the Pierre Auger Observatory through 31 December 2010. We compare the autocorrelation properties in the data with the expectation from simulation sets of arrival directions drawn from randomly located sources with varying density. We consider two astrophysical scenarios: one with sources distributed uniformly in the nearby universe, and another in which the source distribution follows the large scale structure of nearby matter according to the 2MASS Redshift Survey (2MRS) catalog of galaxies. The bounds apply if the deflections of CR trajectories by intervening magnetic fields do not erase the clustering properties expected from the models at the angular scales considered.

## 2 Data set

The surface detector of the Auger Observatory consists of 1660 water-Cherenkov stations that detect photons and charged particles in air showers at ground level. A triangular grid of detectors with 1.5 km spacing spans over 3000 km$^2$, and operates with a duty cycle of almost 100%. The energy resolution is 15%, with a systematic uncertainty of 22% [16]. The angular resolution, defined as the angular radius that would contain 68% of the reconstructed events, is better than $0.9°$ above 10 EeV. The data set consists of 67 events recorded by the Auger Observatory from 1 January 2004 to 31 December 2010, with reconstructed energies above 60 EeV and zenith angles smaller than $60°$. The event selection implemented in the present analysis requires that at least five active nearest-neighbors surround the station with the highest signal when the event was recorded, and that the reconstructed shower core be inside an active equilateral triangle of detectors. The integrated exposure for this event selection amounts to $2.58 \times 10^4$ km$^2$ sr yr.

## 3 Statistical method and astrophysical models

As an estimator of the clustering, in this study we make use of the two-point autocorrelation function (ACF), i.e. the cumulative number of pairs within the angular distance $\theta$, defined by

$$n_p(\theta) = \sum_{i=2}^{n}\sum_{j=1}^{i-1} \Theta\left(\theta - \theta_{ij}\right) \quad (1)$$

where $n$ is the number of UHECRs being considered, $\Theta$ is the step function and $\theta_{ij}$ is the angular distance between events $i$ and $j$. In figure 1 (left panel) we show the ACF of the arrival directions of CRs with energy larger than 60 EeV measured by the Auger Observatory and the 90% confidence region for the isotropic expectation. In the right panel of figure 1, the autocorrelation of the same set of arrival directions, but restricted to galactic latitudes $|b| > 10°$, is shown. This cut in galactic latitude is needed for the comparison with the scenario based in the 2MRS catalog of galaxies, due to its incompleteness near the galactic plane.

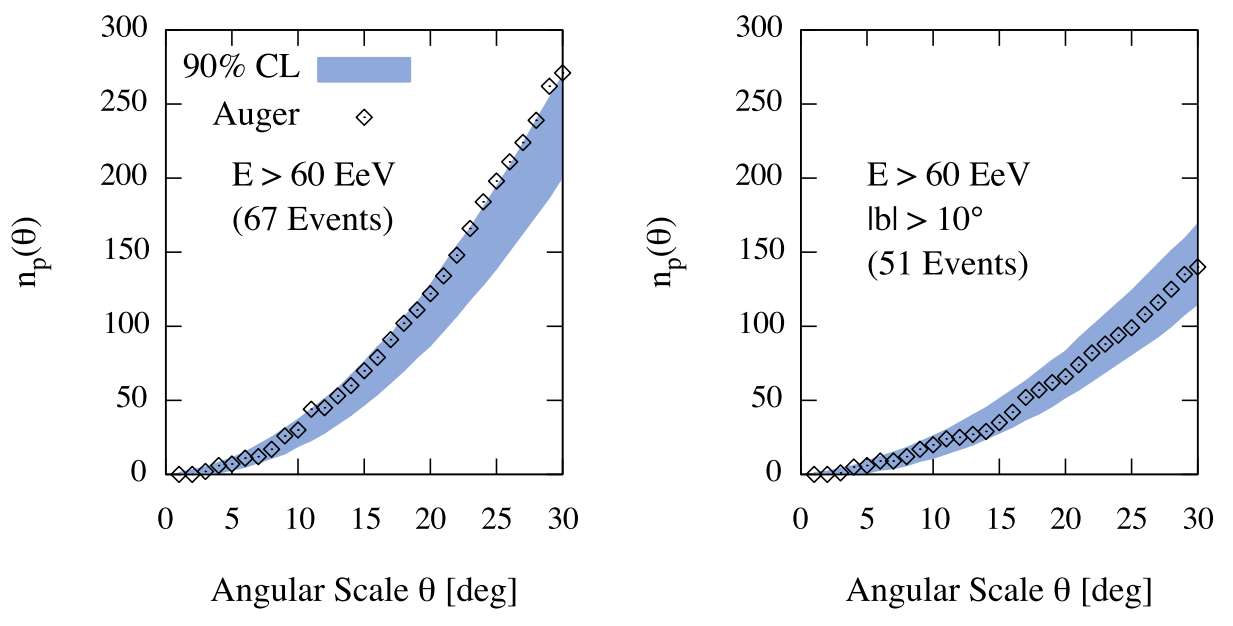

Figure 1: Number of pairs $n_p$ as a function of the angular scale $\theta$ for the data (diamonds) and 90% confidence region for the isotropic expectation (shaded area). *Left:* 67 events with energy above 60 EeV. *Right:* 51 events with energy above 60 EeV and galactic latitude $|b| > 10°$.

In our analysis, we only consider angular scales larger than $5°$ to constrain the source density from the ACF. Deflections of this size are likely to affect the trajectories of protons, and they may be larger for heavier nuclei. The effect of magnetic fields, which are not known in enough detail to be taken into account in this analysis, could smooth away the clustering pattern expected from a particular source scenario at scales smaller than the typical deflections. For angular scales ranging from $5°$ to $30°$, we measure the number of pairs $n_p(\theta)$ in the data and we compare it to that in simulated sets of arrival directions with distributions expected in a given astrophysical model, as a function of the source density $\rho$. This allows us to obtain the range of densities compatible with the observations at a given confidence level. We chose the scenario based on 2RMS galaxies to illustrate the expectations from sources that trace the distribution of matter in the nearby universe, and we investigated the clustering differences with a scenario based on a finite number of random uniformly distributed sources.

The particular choices of the uniform and the 2MRS models is justified by the fact that, for a fixed value of the source density $\rho$, we are interested in investigating the clustering differences between sets of events following the distribution of matter in the nearby universe and sets of events generated by a finite number of random uniformly distributed sources. In both cases, we assume a power-law injection spectrum at the source with spectral index $s = 2.7$ and an equal intrinsic luminosity of cosmic rays. The simulated particles are successively propagated in a $\Lambda-$Cold Dark Matter universe (Hubble constant at present time $H_0 = 70.0$ km/s/Mpc, density of matter $\Omega_m = 0.27$ and density of energy $\Omega_\Lambda = 0.73$) [17], taking into account non-negligible energy-loss processes in the cosmic microwave background photon field. For a given energy threshold $E_{\text{thr}}$ of the events, the probability for a source to generate an event is proportional to the inverse square of its distance $D$ and to a factor accounting for the expected flux attenuation of UHECRs due to the GZK effect. Such a probability is defined by

$$\omega(D, E_{\text{thr}}) \propto \frac{1}{D^2} \frac{s-1}{E_{\text{thr}}^{-s+1}} \int_{E_i(D,E_{\text{thr}})}^{\infty} E^{-s} dE, \qquad (2)$$

where $E_i\,(D, E_{\text{thr}})$ is the initial energy, estimated as in [18], required by the particle to reach the Earth with final energy $E_{\text{thr}}$. Moreover, events are generated by taking into account the non-uniform exposure of the Auger Observatory. The GZK horizon $R_{\text{GZK}}$ is defined as the distance within which 90% of the observed flux above the energy threshold is expected to be produced, i.e. $\omega(R_{\text{GZK}}, E_{\text{thr}}) = 0.1$. It is similar for both UHE protons and iron nuclei, but typically much shorter for nuclei with intermediate mass. In what follows we evaluate the predictions from the astrophysical scenarios using the GZK attenuation expected for protons. We tested the density of sources from $10^{-6}$ Mpc$^{-3}$ to $10^{-3}$ Mpc$^{-3}$ and present the results for three different values of the energy threshold: 60 EeV, 70 EeV and 80 EeV. For higher values of the energy threshold, the number of events becomes too small to perform a reliable clustering analysis. Conversely, lower energy thresholds imply larger GZK horizons, and the incompleteness of galaxy catalogs limits the discrimination power of the method, as will be discussed at the end of this section. For each value of the density $\rho$, $N = \frac{4}{3}\pi\rho R_{\text{GZK}}^3$ sources are generated in a sphere with radius $R_{\text{GZK}}(E_{\text{thr}})$ for each energy threshold considered. We make use of the 2MRS catalog because it is the most densely sampled all-sky redshift survey to date. It is a compilation [19] of the redshifts of the $K_{\text{mag}} < 11.25$ brightest galaxies from the 2MASS catalog [20]. It contains approximately 22,000 galaxies within 200 Mpc, providing an unbiased measure of the distribution of galaxies in the local universe, out to a mean redshift of z = 0.02, and to within $10°$ of the Galactic plane. To avoid biases due to its incompleteness in the galactic plane region, we exclude galaxies (as well as event arrival directions) with galactic latitudes $|b| < 10°$ from all analyses. We use galaxies with magnitude $M < -23.1$, which makes the sample complete up to 80 Mpc with density $\approx 10^{-3}$ Mpc$^{-3}$, the largest values we test. At larger distances, the density of a complete sample is smaller, for instance $\approx 10^{-4}$ Mpc$^{-3}$ for $D = 200$ Mpc. In order to test higher values, we extend the original catalog between 80 Mpc and 200 Mpc with sources isotropically distributed in the sky in number such that the density is also $\approx 10^{-3}$ Mpc$^{-3}$. Our approach is rather conservative, reducing the clustering signal in the skies obtained in the 2MRS case and providing, as a consequence, smaller values of the lower bounds of the density of sources. The incompleteness of the catalog represents the main impediment for performing our analysis with a lower energy threshold for the events. The GZK horizon increases for decreasing energy thresholds and, as a consequence, a greater isotropic contamination is required to complete the catalog, further reducing the clustering signal due to large scale structure. On the other hand, the number of events decreases by increasing the energy threshold, reducing the discrimination power of clustering detection.

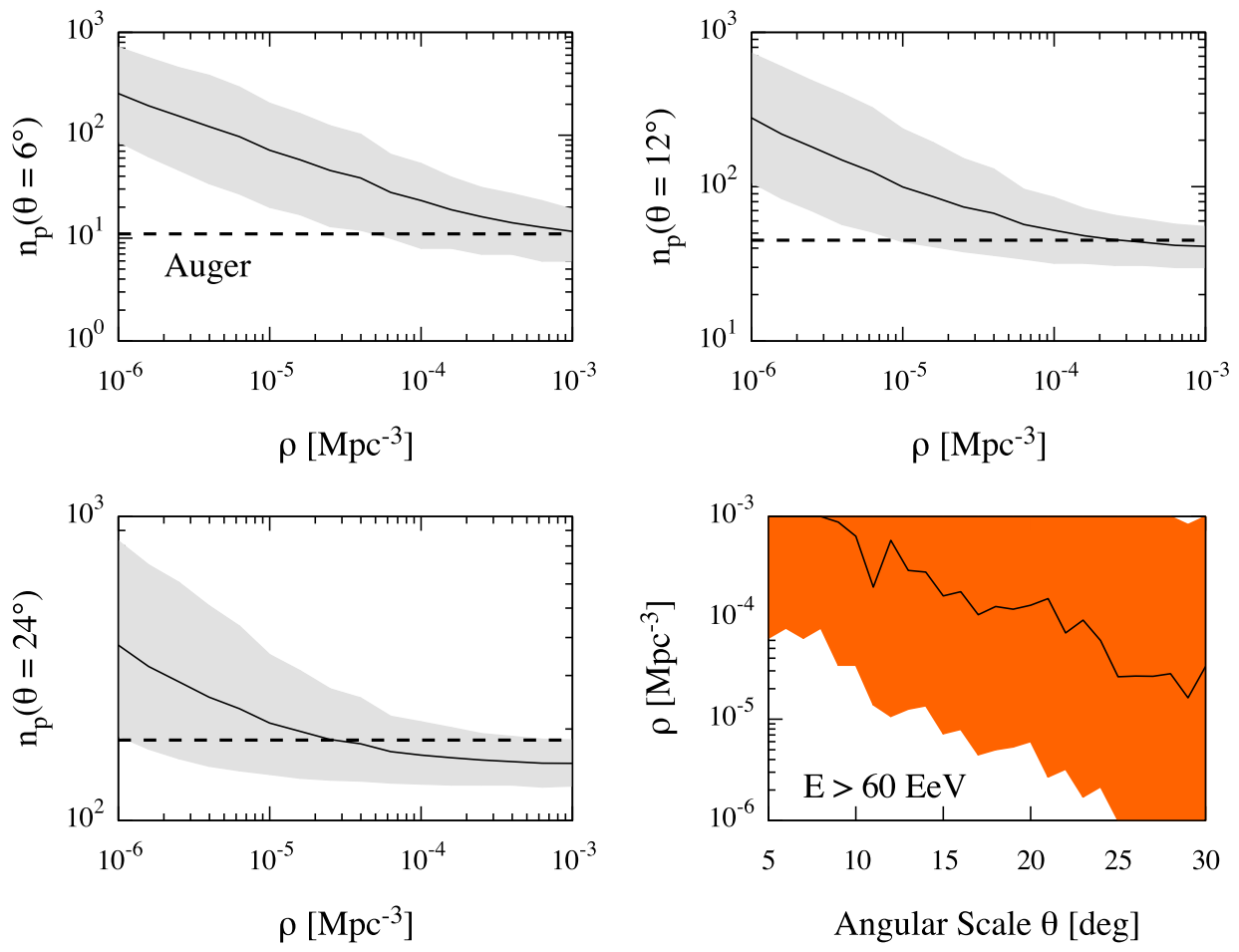

Figure 2: Events with $E > 60$ EeV and a uniform distribution of sources. *Top and bottom-left:* Number of pairs as a function of the source density, for three different values of the angular scale ($\theta = 6°, 12°$ and $24°$). Solid lines indicate the average number of pairs in the case of Monte-Carlo simulations, the shaded area denotes the 90% confidence region and the dashed line indicates the value obtained from the data. *Bottom-right:* source density obtained from the average number of pairs (solid line) and the allowed region for source density with 90% CL (shaded area).

## 4 Application to the data

The procedure for constraining the source density from the clustering properties of the UHECRs measured with the Auger Observatory is as follows. We evaluate the ACF function of a large number of simulated sets of arrival directions drawn (in number equal to the events in the dataset) from the two astrophysical scenarios under consideration and for different values of the source density. The 95% CL upper (lower) bounds on the source density are the values for which only 5% of the simulated sets show more (less) clustering than the data, at a given angular scale.

We illustrate the procedure in figure 2 (top and bottom-left) for the particular case of the scenario with a uniform distribution of sources, for an energy threshold $E_{\text{thr}} = 60$ EeV, and for three different angular scales, namely $\theta = 6°, 12°$ and $24°$. The solid line is the average number of pairs predicted in this scenario as a function of the source density and the shaded area represents the dispersion in the number of pairs within 90% of the simulations. The dashed line corresponds to the number of pairs in the data. The 95% CL lower and upper limits are the ends of the range in source density for which $n_p$ in the data is within the shaded area. In figure 2 (bottom-right) we show the result of this procedure as a function of the angular scale. The solid line is the value of the source density for which the average number of pairs coincides with that in the data at the angular scale considered. The shaded area incorporates the 95% CL limits on the source density. The bounds are (typically) more restrictive at smaller angular scales and their validity depends on the uncertain strength of magnetic deflections. Moreover, such bounds apply if typical magnetic deflections do not significantly modify the clustering properties above the angular scale considered. In practice, the clustering observed in the current data set is insufficient to establish upper bounds on the density of sources at 95%CL for the astrophysical scenarios considered here, and only lower bounds can be derived. In figure 3 we show the lower bound $\rho_{\rm LB}$ (95% CL) for the three energy thresholds considered, for both the uniform (left panel) and the 2MRS (right panel) models. The bounds decrease with increasing angular scales and can also differ by up to one order of magnitude for the same angular scale and different energy thresholds. At relatively small angular scales, the bounds derived from lower energy thresholds are more stringent, being of order of $10^{-4}$ Mpc$^{-3}$, regardless of the astrophysical scenario.

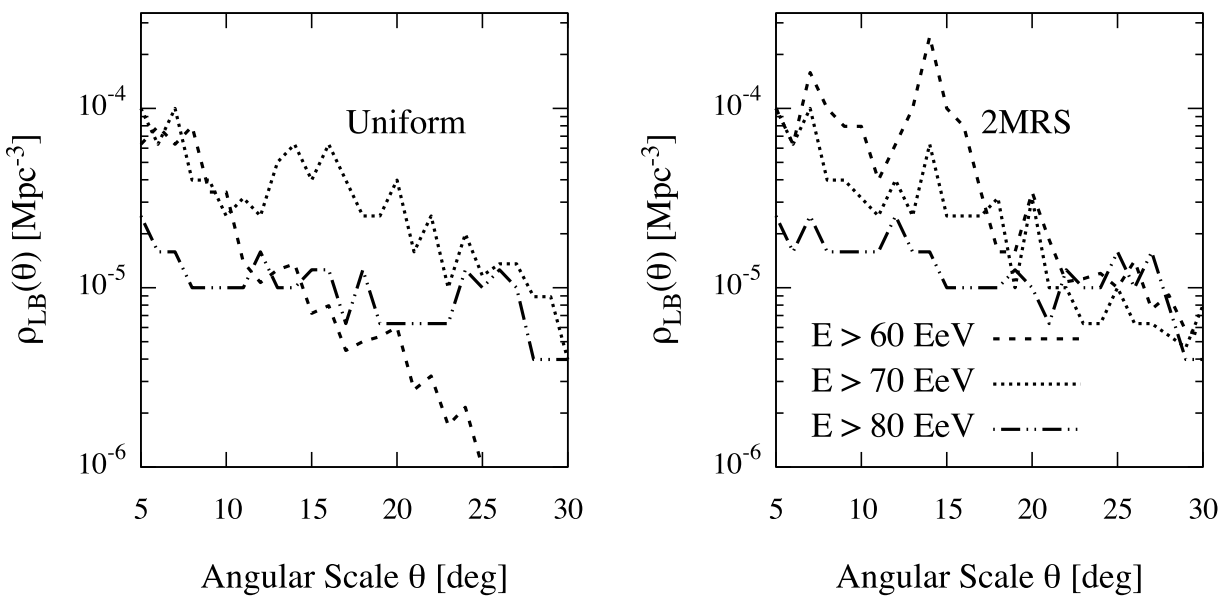

Figure 3: Lower bound (95% CL) on the source density of UHECRs, as a function of the angular scale and for different values of the energy threshold ($E_{\rm thr} = 60, 70$ and 80 EeV). The number of events corresponding to each energy threshold is 67, 33 and 17, respectively (if the cut $|b| > 10°$ is not applied, otherwise it is 51, 26 and 15, respectively). *Left:* uniform case. *Right:* 2MRS case.

## 5 Conclusions

In this study we have shown that the number of pairs of arrival directions of UHECRs detected with the Pierre Auger Observatory, with energy larger than 60 EeV, can be used to constrain the local density of their sources in particular astrophysical models. We have investigated two scenarios, one with sources uniformly distributed in the nearby universe, and another one with sources distributed following the large scale structure of nearby matter. In both cases, equal intrinsic luminosity of the sources has been assumed. If the effects of intervening magnetic fields do not smooth out the clustering properties of UHECRs on scales of about $5°$ (as can be expected in the case of a proton composition), the measurements imply a 95%CL lower limit on the source density of order $10^{-4}$ Mpc$^{-3}$. Conversely, if magnetic deflections are larger, and such that the clustering properties observed reflect the expectation from the source scenario only at larger angular scales, then less stringent lower bounds apply. They are about one order of magnitude smaller for angular scales around $25°$. The bounds apply to specific scenarios, since they depend on the overall distribution of sources.

# Search for energy-position correlated multiplets in Pierre Auger Observatory data

GERALDINA GOLUP[1] FOR THE PIERRE AUGER COLLABORATION[2]
[1]*Centro Atómico Bariloche, Instituto Balseiro (CNEA-UNCuyo-CONICET), S. C. de Bariloche, Argentina*
[2] *Observatorio Pierre Auger, Av. San Martín Norte 304, 5613 Malargüe, Argentina*
*(Full author list: http://www.auger.org/archive/authors_2011_05.html)*
*auger_spokespersons@fnal.gov*

**Abstract:** We present the results of an analysis of data recorded at the Pierre Auger Observatory in which we search for groups of directionally-aligned events (or 'multiplets') which exhibit a correlation between arrival direction and the inverse of the energy. These signatures are expected from sets of events coming from the same source after having been deflected by intervening coherent magnetic fields. We here report the largest multiplets found in the data and compute the probability that they arise by chance from an isotropic distribution of events. There is no statistically significant evidence for the presence of multiplets arising from magnetic deflections in the present data.

## 1 Introduction

The identification of the sources of cosmic rays is greatly complicated by the fact that cosmic rays traverse magnetic fields as they propagate from their sources to Earth. However, the deflections caused by magnetic fields are expected to be inversely proportional to the energy of the cosmic rays. Therefore, it may be possible to identify several cosmic ray events from the same source by looking for spatial alignments in their arrival directions and large correlations between the directions and the inverse of the energy[1]. The identification of these kind of multiplets would not only allow for the accurate location of the direction of the source, but would also provide a measurement of the integral of the component of the magnetic field orthogonal to the trajectory of the cosmic rays.

Cosmic rays are deflected by galactic and extragalactic magnetic fields. The strength of extragalactic fields is not well known, and the importance of their effect is a matter of debate [1, 2, 3]. In this study, we focus on the effect of the galactic field. The galactic field is also poorly constrained, although there are considerable efforts underway to provide measurements of its amplitude and orientation [4, 5, 6]. This field is usually described as the superposition of a large-scale regular component and a turbulent one. The regular component has a few $\mu$G strength and is coherent on scales of a few kpc with a structure related to the spiral arms of the galactic disk. The deflection of cosmic rays with energy $E$ and charge $Z$ by the regular component of the magnetic field $\bar{B}$ after traversing a distance $L$ is given by

$$\delta \simeq 16^\circ \frac{20\ \mathrm{EeV}}{E/Z} \left| \int_0^L \frac{\mathrm{d}\bar{\mathrm{l}}}{3\ \mathrm{kpc}} \times \frac{\bar{\mathrm{B}}}{2\ \mu\mathrm{G}} \right| . \quad (1)$$

This is the predominant deflection because, although the turbulent component has a root mean square amplitude of $B_{\mathrm{rms}} \simeq (1-2) B_{\mathrm{reg}}$, it has a much smaller coherence length (typically $L_c \simeq$ 50-100 pc) [7, 8], leading to a smaller deflection,

$$\delta_{\mathrm{rms}} \simeq 1.5^\circ \frac{20\ \mathrm{EeV}}{E/Z} \frac{\mathrm{B_{rms}}}{3\ \mu\mathrm{G}} \sqrt{\frac{L}{1\ \mathrm{kpc}}} \sqrt{\frac{L_c}{50\ \mathrm{pc}}} . \quad (2)$$

In this study, we perform a search for correlated multiplets in the data set of events with energy above 20 EeV recorded at the Pierre Auger Observatory. This analysis relies on the acceleration at the source of at least one abundant light component. Due to the magnitude of the magnetic fields involved, heavy nuclei at these energies would appear spread over a very large region of the sky, probing regions with different amplitudes and directions of the magnetic field, and hence losing their alignment and correlation with the inverse of energy.

1. To detect several events from the same source, the sources of cosmic rays should be steady, in the sense that the lifetime of the source is larger than the difference in the time delays due to the propagation in the intervening magnetic fields for the energies considered. Moreover, magnetic fields should also be steady in the same sense so that cosmic rays traverse approximately the same fields.

## 2 The Pierre Auger Observatory and the data set

The Pierre Auger Observatory, located in Malargüe, Argentina, at 1400 m a.s.l., is the largest air shower array in the world and its main purpose is to measure ultra-high energy cosmic rays (energy $E > 10^{18}$ eV $\equiv$ 1 EeV). It consists of a surface array of 1660 water Cherenkov stations. The surface array is arranged in an equilateral triangular grid with 1500 m spacing, covering an area of approximately 3000 km$^2$ [9]. The array is overlooked by 27 telescopes at four sites [10] which constitute the fluorescence detector. The surface and air fluorescence detectors are designed to perform complementary measurements of air showers created by cosmic rays. The surface array is used to observe the lateral distribution of the air shower particles at ground level, while the fluorescence telescopes are used to record the longitudinal development of the shower as it moves through the atmosphere.

The data used for this analysis are 1509 events with $E > 20$ EeV and zenith angles smaller than 60° recorded by the surface detector array from 1st January 2004 to 31st December 2010. The events are required to have at least five active detectors surrounding the station with the highest signal, and the reconstructed core must be inside an active equilateral triangle of stations [11]. The angular resolution, defined as the 68$^{\text{th}}$ percentile of the distribution of opening angles between the true and reconstructed directions of simulated events, is better than 0.9° for events that trigger at least six surface detectors ($E > 10$ EeV) [12]. The absolute energy scale, given by the fluorescence calibration, has a systematic uncertainty of 22% and the energy resolution is about 15% [13].

## 3 Method for searching multiplets

If the magnetic deflections are small, it is a good approximation to consider a linear relation between the cosmic ray observed arrival directions $\bar{\theta}$ and the inverse of the energy $E$,

$$\bar{\theta} = \bar{\theta}_s + \frac{Ze}{E}\int_0^L \mathrm{d}\bar{l} \times \bar{B} \simeq \bar{\theta}_s + \frac{\bar{D}(\bar{\theta}_s)}{E}, \qquad (3)$$

where $\bar{\theta}_s$ denotes the actual source direction, and $\bar{D}(\bar{\theta}_s)$ will be called the deflection power[2]. In the case of proton sources, departures from the linear approximation are relevant for energies below 20 EeV for typical galactic magnetic field models [14].

In order to identify sets of events coming from the same source, the main requirement will be that they appear aligned in the sky and have a high value of the correlation coefficient between $\theta$ and $1/E$. We will further require that the multiplets contain at least one event[3] with energy above 45 EeV and that the multiplets do not extend more than 20° in the sky.

To compute the correlation coefficient for a given subset of $N$ nearby events, we first identify the axis along which the correlation is maximal. For this we initially use an arbitrary coordinate system $(x, y)$ in the tangent plane to the celestial sphere (centered in the average direction to the events) and compute the covariances $\mathrm{Cov}(x, 1/E) = \frac{1}{N}\sum_{i=1}^{N}(x_i - \langle x\rangle)(1/E_i - \langle 1/E\rangle)$ and $\mathrm{Cov}(y, 1/E)$. We then rotate the coordinates to a system $(u, w)$ in which $\mathrm{Cov}(w, 1/E) = 0$, and hence $\mathrm{Cov}(u, 1/E)$ is maximal. This corresponds to a rotation angle between the $u$ and $x$ axes given by

$$\alpha = \arctan\left(\frac{\mathrm{Cov}(y, 1/E)}{\mathrm{Cov}(x, 1/E)}\right). \qquad (4)$$

The correlation between $u$ and $1/E$ is measured through the correlation coefficient

$$C(u, 1/E) = \frac{\mathrm{Cov}(u, 1/E)}{\sqrt{\mathrm{Var}(u)\mathrm{Var}(1/E)}}, \qquad (5)$$

where the variances are given by $\mathrm{Var}(x) = \left\langle (x - \langle x\rangle)^2 \right\rangle$.

A given set of events will be identified as a correlated multiplet when $C(u, 1/E) > C_{\min}$ and when the spread in the transverse direction $w$ is small, $W = \max(|w_i - \langle w\rangle|) < W_{\max}$. The values for $C_{\min}$ and $W_{\max}$ were chosen as a compromise between maximizing the signal from a true source and minimizing the background arising from chance alignments. We performed numerical simulations of sets of events from randomly-located extragalactic sources. In these simulations, protons were propagated through a bisymmetric magnetic field with even symmetry (BSS-S) [15, 16] and the effect of the turbulent magnetic field included by simply adding a random deflection with root mean square amplitude $\delta_{\mathrm{rms}} = 1.5°(20\ \mathrm{EeV}/E)$. We considered one hundred extragalactic sources located at random directions and simulated sets of $N$ events with energies following an $E^{-2}$ spectrum at the source and adding random gaussian uncertainties in the angular directions and energies to account for the experimental resolution. As an example we show in Figure 1 (left panel) the resulting distribution of $W$ for multiplets of 14 events. The significance of a given multiplet can be quantified by computing the fraction of isotropic simulations in which a multiplet with the same or larger multiplicity and passing the same cuts appears by chance. We note that when reducing $W_{\max}$, some of the events of the multiplets will be missed and their multiplicity will be reduced. However, the significance of a smaller multiplet passing a tighter bound on $W_{\max}$ can

2. The deflection power will be given in units of 1° 100 EeV, which is $\approx 1.9\ e\ \mu$G kpc.

3. Note that the energy of the most energetic event of a set of 10 events with $E > 20$ EeV from a source with spectral index $s = 2.5$ is larger than 45 EeV with a probability of 97% (for a spectral index $s = 3$ this probability is $\sim 90\%$).

be larger than the significance of the complete multiplet with a looser $W_{\max}$ cut. It turns out that the largest mean significance for the simulated sources appears when a cut $W_{\max} \simeq 1.5°$ is applied. In the case of 14-plets, in $50\%$ of the simulations all the events pass this cut and the multiplet will be reconstructed as a 14-plet, while in $38\%$ of the cases one event is lost and in $11\%$ of the cases two events are lost. The angular scale of $1.5°$ provides in fact a reasonable cut which accounts for the angular resolution and the mean value of the turbulent field deflections.

A similar analysis can be performed to fix the cut on the correlation coefficient $C_{\min}$. The distribution of $C(u, 1/E)$ for the simulated 14-plets is shown in Figure 1 (right panel). The largest mean significance is attained now for values of $C_{\min}$ in the range from 0.85 to 0.9, depending on the multiplicity considered. For a cut $C_{\min} = 0.9$, in $57\%$ of simulations with 14 events all events pass the cuts, in $12\%$ of the simulations one event is lost and in $11\%$ of them two events are lost. We will then fix in the following $W_{\max} = 1.5°$ and $C_{\min} = 0.9$. We note that the choice of the optimal cut slightly depends on the galactic magnetic field model considered in the simulations and on the modeling of the turbulent field deflections.

When a correlated multiplet is identified it is possible to reconstruct the position of its potential source $(u_s, 0)$ (in the $u$-$w$ coordinate system) and estimate the deflection power $D$ by performing a linear fit to the relation

$$u = u_s + \frac{D}{E}. \tag{6}$$

## 4 Results

A search for correlated multiplets was performed in the Pierre Auger Observatory data with events with energies above 20 EeV. The largest multiplet found is one 12-plet and there are also two independent decaplets. We show the arrival directions in galactic coordinates of these multiplets in Figure 2. In Table 1, we list their deflection power, position of the potential source location and correlation coefficient[4]. The uncertainties in the reconstruction of the position of the potential sources have been calculated by propagating the uncertainties in energy and arrival direction to an uncertainty in the rotation angle (Eq. 4) and in the linear fit performed to the deflection vs. $1/E$ (Eq. 6).

We performed the same analysis applied to simulations of events with random arrival directions, weighted by the geometric exposure of the experiment [17], and with the energies of the observed events. From these realizations we computed the probability that the observed number (or more) of correlated multiplets appears by chance. The fraction of simulations with at least one multiplet with 12 or more events is $6\%$, and the fraction having at least three multiplets with 10 or more events is $20\%$. From these chance probabilities we conclude that, in the present data set, there is no statistically significant evidence for the presence of multiplets from actual sources. We note that with the present statistics, a multiplet passing the required selection cuts should have at least 14 correlated events in order that its chance probability be $10^{-3}$.

## 5 Conclusions

We performed a search for energy-position correlated multiplets in the data collected by the Pierre Auger Observatory with energy above 20 EeV. The largest multiplet found was one 12-plet. The probability that it appears by chance from an isotropic distribution of events is $6\%$. Therefore, there is no significant evidence for the presence of correlated multiplets arising from magnetic deflections in the present data set. We will continue analyzing future data and check if some of the observed multiplets grow significantly or if some new large multiplets appear.

4. Decaplet II in Table 1 consists of three dependent sets of ten events (a-c) that are formed by the combination of a set of twelve events. These three decaplets are not independent of each other since they have most events in common.

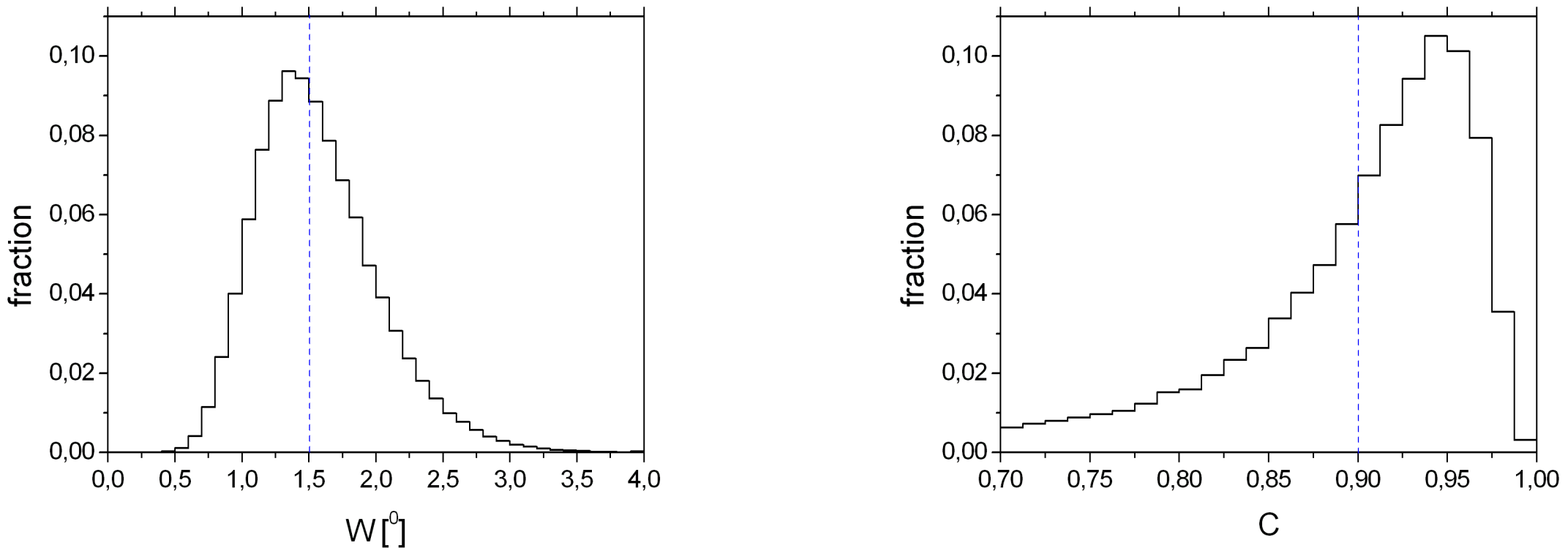

Figure 1: Distribution for 100 simulated 14-plets of $W$ (left panel) and $C(u, 1/E)$ (right panel). The vertical dashed lines indicate the cuts on $W$ and $C$ optimized for multiplicity and significance (Section 3).

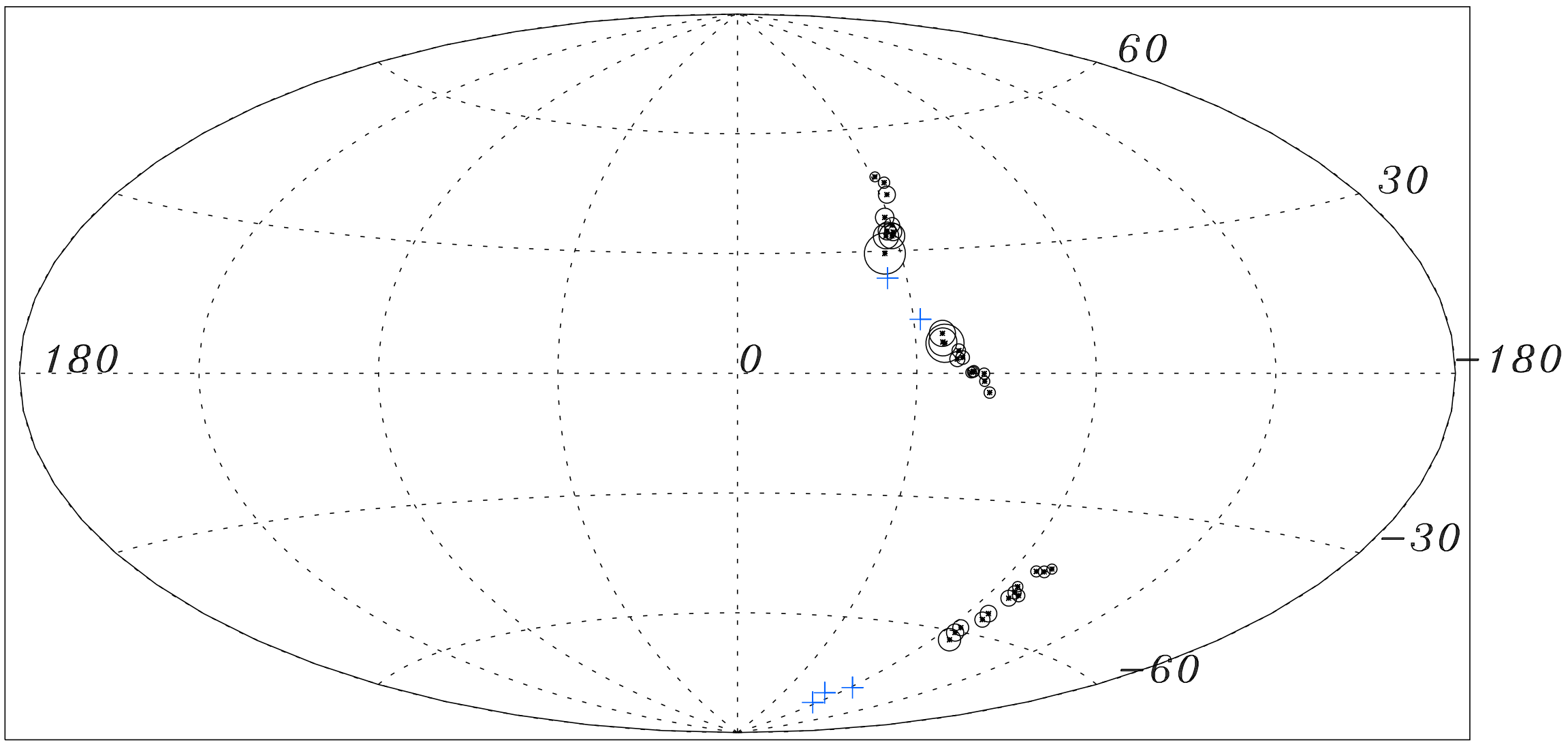

Figure 2: Observed multiplets with 10 or more events in galactic coordinates. The size of the circle is proportional to the energy of the event. Plus signs indicate the positions of the potential sources for each multiplet. One decaplet is in fact three dependent decaplets that are formed by the combination of twelve events and the three corresponding reconstructions of the potential sources are shown.

| multiplet | $D[°100\ \text{EeV}]$ | $(l, b)_S[°]$ | $\Delta u_S[°]$ | $\Delta w_S[°]$ | $C$ |
|---|---|---|---|---|---|
| 12 − plet | $4.3 \pm 0.7$ | $(-46.7, 13.2)$ | 2.4 | 0.9 | 0.903 |
| 10 − plet I | $5.1 \pm 0.9$ | $(-39.9, 23.4)$ | 2.7 | 0.9 | 0.901 |
| 10 − plet IIa | $8.2 \pm 1.3$ | $(-85.6, -80.4)$ | 4.3 | 1.9 | 0.920 |
| 10 − plet IIb | $7.6 \pm 1.2$ | $(-79.6, -77.9)$ | 4.0 | 1.6 | 0.919 |
| 10 − plet IIc | $6.5 \pm 1.1$ | $(-91.5, -75.7)$ | 3.9 | 1.6 | 0.908 |

Table 1: Deflection power, $D$; reconstructed position of the potential source in galactic coordinates, $(l, b)_S$; uncertainty in the reconstructed position of the potential source along the direction of deflection, $\Delta u_S$, and orthogonal to it, $\Delta w_S$; and linear correlation coefficient, $C$, for the largest correlated multiplets found. The data correspond to events with energy above 20 EeV from 1st January 2004 to 31st December 2010.

.

# Search for Galactic point-sources of EeV neutrons

BENJAMIN ROUILLÉ D'ORFEUIL[1] FOR THE PIERRE AUGER COLLABORATION[2]
[1]*Kavli Institute for Cosmological Physics and Enrico Fermi Institute, The University of Chicago, Chicago, IL, USA*
[2]*Observatorio Pierre Auger, Av. San Martin Norte 304, (5613) Malargüe, Mendoza, Argentina*
*Full author list: http://www.auger.org/archive/authors_2011_05.html*
*auger_spokespersons@fnal.gov*

**Abstract:** The Pierre Auger Observatory has sensitivity to neutron fluxes produced at cosmic ray acceleration sites in the Galaxy. Because of relativistic time dilation, the neutron mean decay length is $(9.2 \times E)$ kpc, where $E$ is the neutron energy in EeV. A blind search over the field of view of the Auger Observatory for a point-like excess yields no statistically significant candidates. The neutron flux upper limit is reported as a celestial function for three different energy thresholds. Also a search for excesses of cosmic rays in the direction of selected populations of candidate Galactic sources is performed. The bounds obtained constrain models for persistent discrete sources of EeV cosmic rays in the Galaxy.

## 1 Motivations for EeV neutron astronomy

At EeV (1 EeV = $10^{18}$ eV) energies, the Galactic magnetic field isotropizes the charged particle fluxes, making it impossible to pick out possible Galactic proton sources. On the other hand, neutron astronomy inside our Galaxy is possible. Neutrons travel indeed undeflected by magnetic fields, and their mean decay length $\lambda_{\rm n} = (9.2 \times E)$ kpc, where $E$ is the neutron energy in EeV, is comparable to the Earth distance from the Galactic center. Hence, neutron induced extensive air showers (EAS) could produce a directional excess of cosmic rays (CRs) in the sky, clustered within the observatory's angular resolution.

High energy neutrons can be produced by the interaction of accelerated protons or heavier nuclei with the radiation and baryonic backgrounds inside the sources or in their surroundings. They can take over most of the initial CR energy per nucleon and would not be magnetically bound to the accelerating region. Gamma-rays can also be generated via these interactions, but they acquire a lesser fraction of the primary CR energy.

If one assumes that CRs are produced with a continuous power-law spectrum that extends from GeV to EeV with an injection spectral index of $-2$, the energy deposited in each decade should be comparable. Accordingly, the observed GeV-TeV gamma-ray fluxes, provided that they have a significant component of hadronic origin, would motivate the search for neutron fluxes in the EeV range.

In terms of high energy CR astrophysics, it is crucial to look for Galactic sources that could accelerate particles up to EeV energies. A time-honored picture is that the transition between particles produced in Galactic and extragalactic sources happens at the 'ankle', a hardening of the slope in the CR energy spectrum appearing in the middle of the EeV energy decade [1], that could naturally be explained by the emergence of a dominant extragalactic component (see [2] for a review). This model requires that particles be accelerated above $\sim 1$ EeV by sources inside our Galaxy.

## 2 Methodology

The array of surface detectors (SD) of the Pierre Auger Observatory is used to search for point-like excesses at EeV energies that would be indicative of a flux of neutral particles from a discrete source. The sensitivity of the SD in this energy range and its large aperture ensures that constraining limits can be set over a large fraction of the sky. These upper limits will be interpreted as upper limits on neutron fluxes since (i) above any fixed energy, the emission rate of neutrons from a CR source in our Galaxy is expected to be well above the emission rate of gamma-rays of hadronic origin and (ii) in the search for an excess of arrival directions in a small solid angle, the SD is far more sensitive to neutrons than to gamma-rays. Indeed, roughly half of the signal in hadronic EAS is due to muons traversing the water Cherenkov-stations. Gamma-ray EAS, being muon poor, should, for a given energy, produce a smaller signal, they hence have a reduced trigger efficiency and are also harder to identify in the larger background of lower energy CRs.

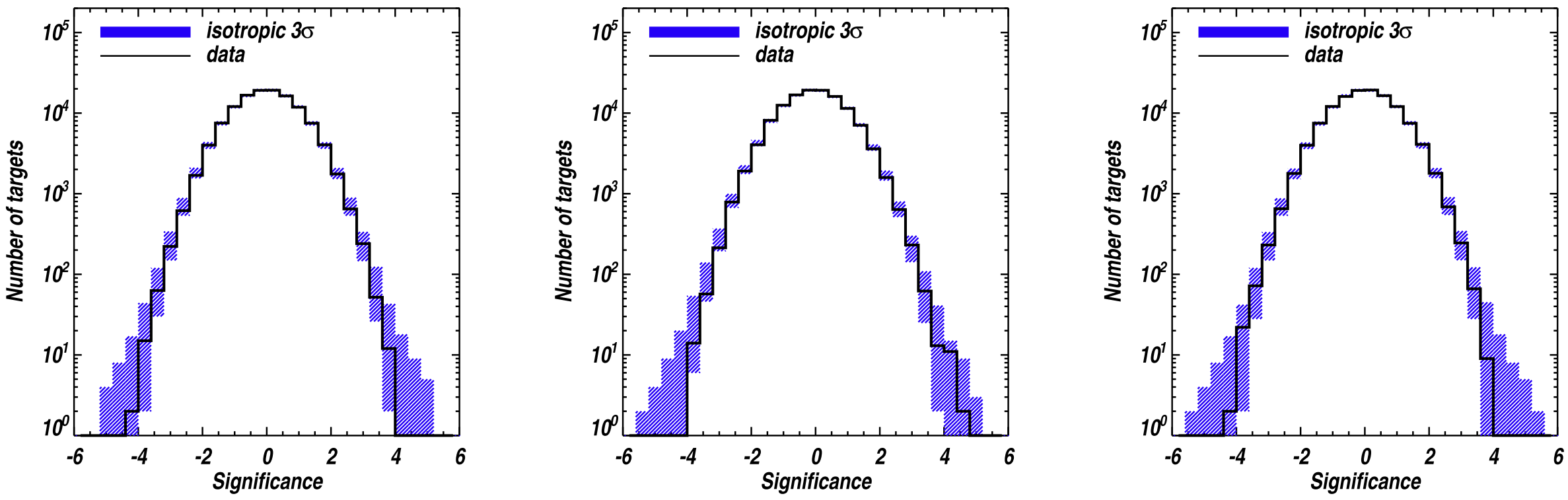

Figure 1: Distribution of the Li-Ma significances of the blind searches together with the $3\sigma$ containment of 5000 Monte-Carlo samples of an isotropic sky. From left to right: $[1-2]$ EeV, $[2-3]$ EeV, and $E \geq 1$ EeV.

We perform here two analyses to constrain the neutron flux from Galactic sources in three energy bands: $[1-2]$ EeV, $[2-3]$ EeV, and $E \geq 1$ EeV. First, a blind search for localized excesses in the CR flux over the exposed sky was carried out. The search compared the number of observed events with that expected from an isotropic background, in top-hat counting regions matching the angular resolution of the instrument. Flux upper limits were derived and plotted on celestial maps. Second, a stacking analysis was performed in the direction of bright Galactic gamma-ray sources detected by the Fermi LAT (100 MeV – 100 GeV) and the H.E.S.S. (100 GeV – 100 TeV) telescopes.

These analyses use high quality events with zenith angles $\theta < 60°$ recorded by the SD between 1 January 2004 and 30 October 2010. Periods of unstable acquisition were removed. More than 340000 SD events have been reconstructed with energies above 1 EeV.

## 3 Blind search over the covered sky

To study the possible presence of overdensities, one needs first to obtain the background expectations for the different sky directions under the assumption of an isotropic CR distribution. This is achieved by parametrizing the zenith angle distribution of the observed events in the energy range under study to smooth out statistical fluctuations [3].

Sensitivity to point sources is optimized by matching the target region size to the angular resolution of the instrument. The angular resolution of the SD, $\psi$, corresponding to the 68% containment radius, is better than $1.8°$ and $1.5°$ above 1 EeV and 2 EeV, respectively [5]. For a gaussian point spread function characterized by $\sigma$, the signal-to-noise ratio is optimized for a top-hat radius given by $1.59\sigma = 1.05\psi$.

We use an HEALPix [4] grid with resolution parameter $N_{\rm side} = 128$ to define the center point of each target region. The size of a pixel being small ($27.5'$) compared to a target region, there is a significant overlap between neighboring targets. The number of arrival directions (observed or expected) in any target is taken as the sum of the counts in the pixels (using a higher resolution: $N_{\rm side} = 1024$) whose center is contained in the target region.

We evaluate the Li-Ma significance[1] [6] in each target. The distribution of the significances of the blind searches are shown in Figure 1. The blind search over the field of view (FOV) of the SD reveals no candidate point on the sky that clearly stands up above the expected distribution of significances in isotropic simulations (shaded region). It is therefore sensible to derive a flux upper limit in each target.

We adopt the definition of [7] to compute the upper limit $\bar{s}_{\rm UL}$ of confidence level CL $= 1-\alpha$ on the expected signal $\bar{s}$, when an observation results in a count $n$ in the presence of a Poisson background distribution with mean value $\bar{b}$:

$$P(\leq n|\bar{b}+\bar{s}_{\rm UL}) = \alpha \times P(\leq n|\bar{b}) \qquad (1)$$

The CL is set to 95%. For each target we derive the bounds on the neutron flux by dividing $\bar{s}_{\rm UL}$ by the exposure (in km$^2$ yr). The latter is obtained by dividing, for each region, the expected number of background events per target solid angle by the intensity of CRs in the energy bin under study, which is obtained from the measured CR energy spectrum [1]. As the target circle encompasses 71.75% of the total gaussian-distributed signal, the final upper limit to the flux is obtained by scaling the above bound by 1/0.7175. Figure 2 presents sky maps of the flux upper limits for the three energy bins considered for the analysis. The upper limits become less stringent near the border of the FOV because of the reduced statistics. We hence only present the results for $\delta < 15°$ to avoid the lowest exposure regions.

Note that if the background were due to a heavier composition, since the efficiency for detection of heavy nuclei is expected to be slightly larger at EeV energies than for the potential neutron signal, the bounds could be slightly relaxed. We note however that the measurements of the depth of shower maximum are consistent with a predominantly light composition at EeV energies [8].

The galactic center is a particularly interesting target because of the presence of a massive black hole. The results

1. For the $\alpha$ parameter in the expression of the Li-Ma significance, we use $\alpha_{\rm LM} = n_{\rm exp}/n_{\rm tot}$ with $n_{\rm exp}$ the background expected in the target and $n_{\rm tot}$ the total number of events.

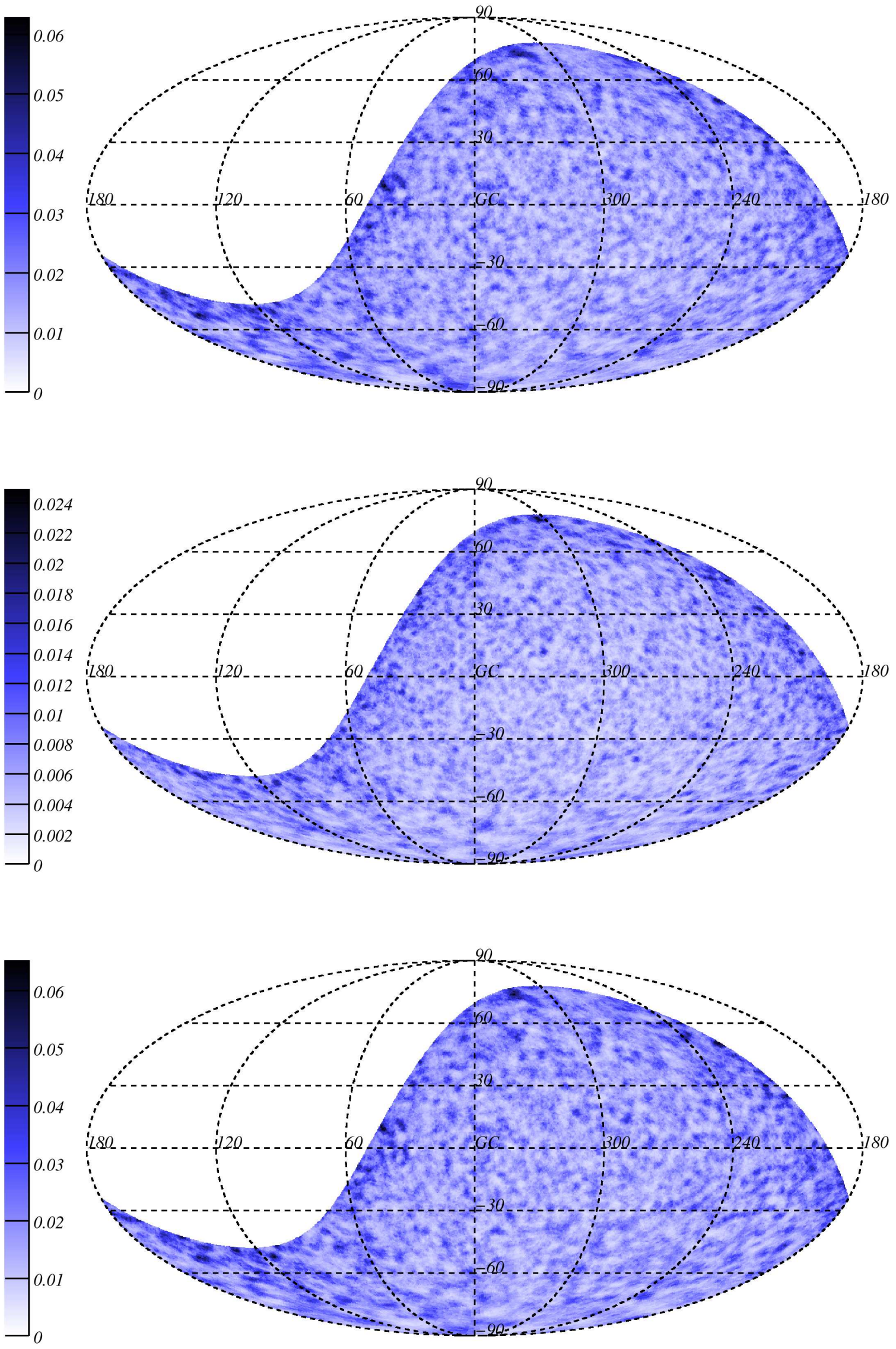

Figure 2: Flux upper limits celestial maps (in unit of km$^{-2}$ yr$^{-1}$) in Galactic coordinates. From top to bottom: $[1-2]$ EeV, $[2-3]$ EeV, and $E \geq 1$ EeV.

for the window centered on it and for $E \geq 1$ EeV shows no excess ($S = -1.43$) and hence we obtain a 95% CL upper limit on the flux from a point source in this direction of 0.01 km$^{-2}$ yr$^{-1}$, which updates the bounds obtained previously in [9]. We note that for directions along the Galactic plane the upper limits are below 0.024 km$^{-2}$ yr$^{-1}$, 0.014 km$^{-2}$ yr$^{-1}$ and 0.026 km$^{-2}$ yr$^{-1}$ for the energy bins $[1-2]$ EeV, $[2-3]$ EeV and $E \geq 1$ EeV, respectively.

## 4 Targeted search

The targeted search involves the selection of bright gamma-ray sources and the search for an excess in their directions. We define the excess signal in a solid angle $\Omega$ around one source as: $S = N_{\rm s}/\sqrt{N_{\rm iso}}$, where $N_{\rm s}$ is the difference between the observed and expected ($N_{\rm iso}$) number of events in the target region around each source.

In order to improve the signal over background, we perform a stacking analysis on sets of $\mathcal{N}_{\rm s}$ sources. The stacked excess signal reads $S_{\rm stacked} = \sum N_{\rm s}/\sqrt{\sum N_{\rm iso}}$ and scales as $S\sqrt{\mathcal{N}_{\rm s}}$ for the ideal case in which sources producing equal neutron flux on Earth are detected with uniform coverage.

The acceleration of particles above 1 EeV by sources inside our Galaxy is theoretically challenging. The most powerful Galactic objects either do not possess the required luminosity to accelerate particles to such high energy, or present acceleration environments that are too dense for particles to escape without losing energy. Pulsars and Pulsar Wind Nebulæ (PWN) however are considered to be good potential accelerators (see e.g., [10, 11]), and recent work shows that the maximum energy of accelerated iron nuclei may reach $5$ EeV in certain supernova remnants (SNR) [12]. Models predicting the production of neutrons at EeV energies from powerful Galactic sources have also been discussed (e.g., [13, 14]). The candidate sources are expected to be strong gamma-ray emitters at GeV and TeV energies.

For this reason, we apply this analysis to Galactic gamma-ray sources extracted from the Fermi LAT Point Source Catalog [15] and the H.E.S.S. Source Catalog[2], focussing on pulsars, PWN and SNR. Targets were selected among the sources located in the portion of the Galactic plane, defined as $|b| < 10°$, covered by the FOV of the SD, and located at a distance shorter than 9 kpc ($\lambda_{\rm n}$ at 1 EeV). As an example, we built two sets by selecting from each catalog the ten brightest sources (in flux observed on Earth) fulfilling these criteria. Our targets are listed in Tables 1 and 2.

| Name 1FGL | $l$ [deg] | $b$ [deg] | distance [kpc] |
|---|---|---|---|
| J0835.3-4510 | 263.55 | -2.79 | $0.29 \pm 0.02$ |
| J1709.7-4429 | 343.10 | -2.69 | $1.4 - 3.6$ |
| J1856.1+0122 | 34.70 | -0.42 | 2.8 |
| J1809.8-2332 | 7.39 | -1.99 | $1.7 \pm 1.0$ |
| J1801.3-2322c | 6.57 | -0.21 | 1.9 |
| J1420.1-6048 | 313.54 | 0.23 | $5.6 \pm 1.7$ |
| J1018.6-5856 | 284.32 | -1.70 | 2.2 |
| J1028.4-5819 | 285.06 | -0.49 | $2.3 \pm 0.7$ |
| J1057.9-5226 | 285.98 | 6.65 | $0.7 \pm 0.2$ |
| J1418.7-6057 | 313.33 | 0.14 | $2 - 5$ |

Table 1: Set of bright sources selected from the Fermi LAT Point Source Catalog. Distances are from [16] and the references cited in [17].

| Name HESS | $l$ [deg] | $b$ [deg] | distance [kpc] |
|---|---|---|---|
| J0852-463 | 266.28 | -1.24 | 0.2 |
| J0835-455 | 263.85 | -3.09 | 0.29 |
| J1713-397 | 347.28 | -0.38 | 1 |
| J1616-508 | 332.39 | -0.14 | 6.5 |
| J1825-137 | 17.82 | -0.74 | 3.9 |
| J1708-443 | 343.04 | -2.38 | 2.3 |
| J1514-591 | 320.33 | -1.19 | 5.2 |
| J1809-193 | 10.92 | 0.08 | 3.7 |
| J1442-624 | 315.41 | -2.30 | 2.5 |
| J1640-465 | 338.32 | -0.02 | 8.6 |

Table 2: Set of bright sources selected from the H.E.S.S. Source Catalog. Distances are from the TeVCat catalog (http://tevcat.uchicago.edu/).

The stacked signal computed from the SD data at the positions of the two sets of sources are presented in Table 3 for the three energy bins under study. No excess is found.

| Set of sources | Energy bin [EeV] | $S_{\rm stacked}$ |
|---|---|---|
| Table 1 | $[1-2]$ | 2.07 |
| Table 1 | $[2-3]$ | 0.51 |
| Table 1 | $\geq 1$ | 2.35 |
| Table 2 | $[1-2]$ | -0.75 |
| Table 2 | $[2-3]$ | -0.40 |
| Table 2 | $\geq 1$ | -0.89 |

Table 3: Application to sets of sources (Tables 1 and 2). Stacked excess signals, $S_{\rm stacked}$ derived for the SD data.

## 5 Conclusion

The data recorded by the Auger Observatory in the EeV energy range has been used to search for point like excesses that would be indicative of a flux of neutrons from Galactic sources. Two analyses were performed, (i) a blind search of the exposed sky and (ii) a stacking analysis in the direction of bright gamma-ray sources detected by the Fermi LAT and H.E.S.S. telescopes. Both analyses reveal no statistically significant excess. Upper limits were calculated for all parts of the sky. Above 1 EeV, the flux upper limit is less than $0.065$ km$^{-2}$ yr$^{-1}$ corresponding to an energy flux of $0.13$ EeV km$^{-2}$ yr$^{-1}$ $\simeq 0.4$ eV cm$^{-2}$ s$^{-1}$ in the EeV decade assuming an $1/E^2$ differential energy spectrum.

2. http://www.mpi-hd.mpg.de/hfm/HESS/pages/home/sources/

# An update on a search for ultra-high energy photons using the Pierre Auger Observatory

MARIANGELA SETTIMO[1] FOR THE PIERRE AUGER COLLABORATION[2]
[1]*University of Siegen, Department of Physics, 57068 Siegen, Germany*
[2]*Observatorio Pierre Auger, Av. San Martín Norte 304, 5613 Malargüe, Argentina*
*(Full author list: http://www.auger.org/archive/authors_2011_05.html)*
*auger_spokespersons@fnal.gov*

**Abstract:** The large collection area of the Pierre Auger Observatory and the availability of a variety of composition-sensitive parameters provide an excellent opportunity to search for photons in the cosmic ray flux above $10^{18}$ eV. Upper limits published previously, using data from the Observatory, placed severe constraints on top-down models. Now, the increase in exposure by more than a factor 2 since 2008 together with the combination of different observables means that the detection of GZK-photons predicted using bottom-up models is almost within reach. Current results will be presented and their implications will be discussed.

## 1 Introduction

Photons are one of the theoretical candidates for ultra-high energy cosmic rays (UHECR) with energies larger than $10^{18}$ eV. A large fraction ($\sim$ 50%) of photons in the cosmic-ray spectrum at the highest energies is indeed predicted within several "top-down" models to explain the origin of cosmic rays. Severe constraints to these models were imposed by previous photon searches above $10^{19}$ eV [1]. A smaller contribution of typically (0.01 - 1)% above $10^{19}$ eV [2] is additionally expected as the product of the photoproduction of pions with the microwave background (GZK effect [3, 4]). The Pierre Auger Observatory [5] has reported a suppression of the cosmic ray energy spectrum beyond $10^{19.6}$ eV [6] which is consistent with the predicted GZK cut-off for protons but could also be due to the photon disintegration of heavy nuclei or due to a limit in the maximum particle energy reached at the sources. The observation of a photon flux compatible with this theoretical prediction could provide an independent proof of the GZK process. The upper limits on the photon fraction were extended to 2 EeV in [7] using the hybrid detection mode provided by the Pierre Auger Observatory. The analysis was based on the measurement of the depth of the shower maximum, $X_{\max}$, since photon induced showers are expected to develop deeper in the atmosphere compared to hadrons. In addition, they are also characterized by a smaller number of secondary muons and a more compact "footprint" at the ground.

In this work we improve the search for EeV photons with hybrid events by: (i) combining observables of the fluorescence detector and the surface array for a better photon-hadron discrimination; (ii) extending the energy range by a factor 2, down to 1 EeV; and (iii) determining bounds on the flux of photons.

## 2 Photon search

The Pierre Auger Observatory, located in Malargüe, Argentina, consists of a surface array (SD) [8] of 1660 water Cherenkov stations spread over an area of 3000 km$^2$ and overlooked by 27 air fluorescence telescopes [9]. The SD samples the density of the secondary particles of the air shower at the ground while the fluorescence detector (FD) observes the longitudinal development of the shower. The analysis presented in this work uses *hybrid* data (detected by at least one FD telescope and one SD station) collected between January 2005 and September 2010. Due to the FD duty cycle ($\sim$ 13%) the event statistics is reduced compared to the SD-only detection mode. However, the hybrid detection technique provides a precise geometry and energy determination with the additional benefit of a smaller energy threshold for detection (around the EeV range).

To improve the photon-hadron discrimination power we complement the previous analysis, based on the $X_{\max}$ measurement, with an SD observable, $S_b$, defined in [10] as

$$S_b = \sum_i S_i \left(\frac{R_i}{R_{\rm ref}}\right)^b \qquad (1)$$

where the sum runs over the triggered stations, $S_i$ is the recorded signal in the station at distance $R_i$ from the hybrid reconstructed axis and $R_{\mathrm{ref}}$ is a reference distance equal to 1000 m for this analysis. The exponent $b$ is chosen equal to 4 for maximizing the separation power between photons and hadrons. The $S_b$ parameter combines the different amplitude of the signal in the surface detector and the sharper lateral distribution function (i.e. the signals recorded in the SD stations as a function of distance from the axis) expected for photon induced showers. Events with zenith angle smaller than 60° and with a good geometry reconstruction are selected for the analysis. To ensure a reliable profile reconstruction we require: a reduced $\chi^2$ of the longitudinal profile fit to the Gaisser-Hillas function smaller than 2.5, a $\chi^2$ of a linear fit to the longitudinal profile exceeding the Gaisser-Hillas fit $\chi^2$ by at least a factor of 1.1, the $X_{\mathrm{max}}$ observed within the field of view of the telescopes, the Cherenkov light contamination smaller than 50% and the uncertainty of the reconstructed energy less than 20%. To reject misreconstructed profiles, only time periods with the sky not obscured by clouds and with a reliable measurement of the vertical optical depth of aerosols [11, 12], are selected. On the SD side we require at least 4 active stations within 2 km from the hybrid reconstructed axis. This prevents an underestimation of $S_b$ (which would mimic the behavior of a photon event) due to missing or temporarily inefficient detectors. For the classification of photon candidates we perform a Fisher analysis [13] trained with a sample of a total of ~30000 photon and proton CORSIKA [14] showers generated according to a power law spectrum between $10^{17}$ and $10^{20}$ eV. QGSJET-II [15] and FLUKA [16] are used as hadronic interaction models. To carefully reproduce the operating conditions of the DAQ, time dependent simulations are performed according to the hybrid detector on-time [17]. The actual configurations of FD and SD and realistic atmospheric conditions are also taken into account. The correlation between $X_{\mathrm{max}}$ and $S_b$ is shown in Figure 1 for well reconstructed photon (empty blue circles) and proton (red crosses) showers, in the energy interval between $10^{18}$ and $10^{18.5}$ eV. Photon-like events are expected to lie in the top-left part of the plot because of the deeper $X_{\mathrm{max}}$ and of the smaller $S_b$. A Fisher analysis is performed in bins of 0.5 in the logarithm of energy and, for the moment, using only proton showers since they are expected to be the main source of background for the photon search. The impact of a mixed composition assumption will be discussed later. The Fisher response is shown in Figure 2, for the same conditions of Figure 1. The best performance of this combination of observables, compared to FD-only or SD-only, is reached at the lowest energies. Particularly at higher energies, the main contribution to the Fisher observable comes from $X_{\mathrm{max}}$. Photon-like events are selected by applying an "a priori" cut at 50% of the photon detection efficiency. This provides a conservative result in the upper limit calculation by reducing the dependence on the hadronic interaction models and on the mass composition assumption. With this choice the expected hadron contamination is about 1% in the lowest energy interval (between $10^{18}$ and $10^{18.5}$ eV) and it becomes smaller for increasing energies.

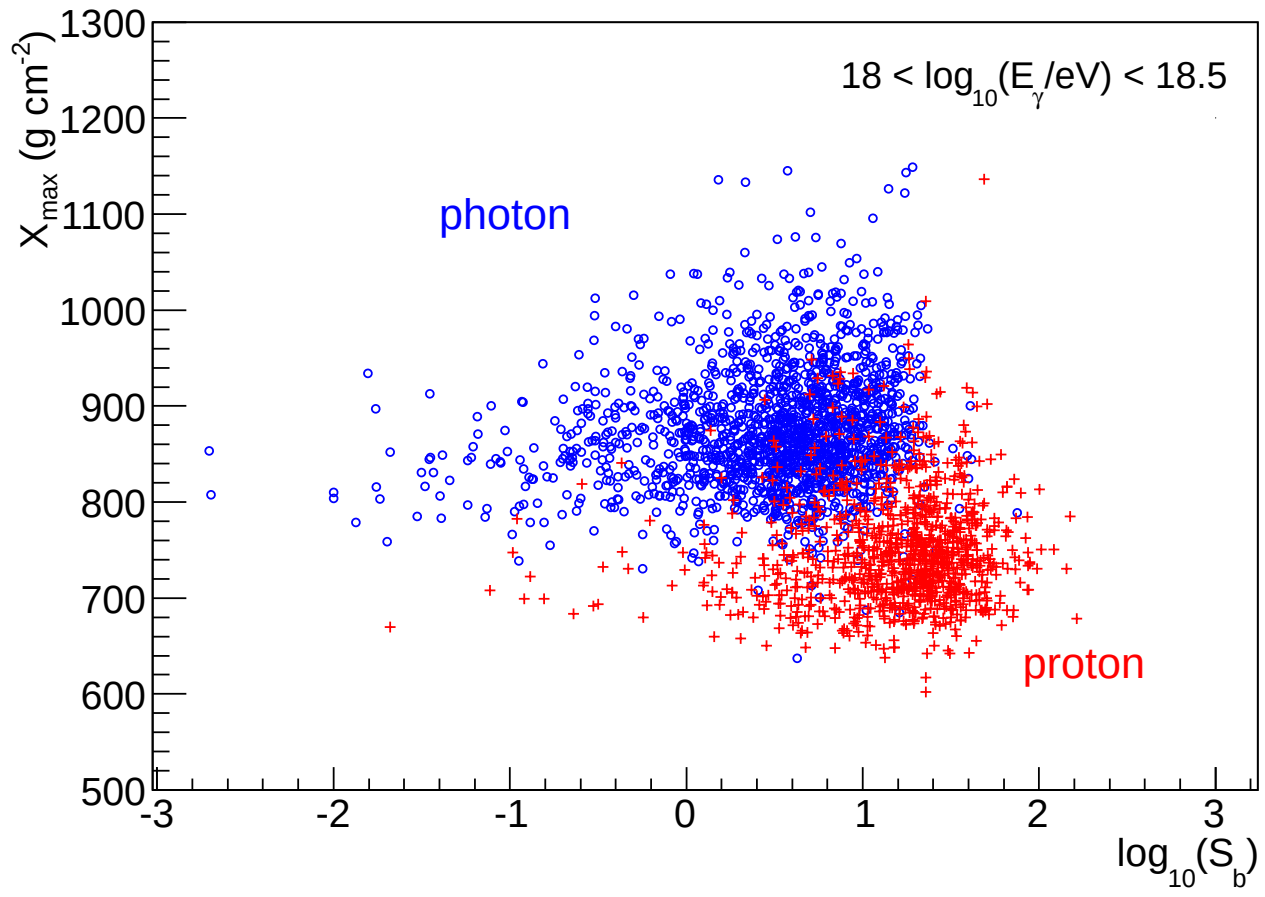

Figure 1: Scatter plot of $X_{\mathrm{max}}$ vs $\log_{10}(S_b)$ for proton (red crosses) and photon (empty blue circles) simulated showers with energy between $10^{18}$ and $10^{18.5}$ eV.

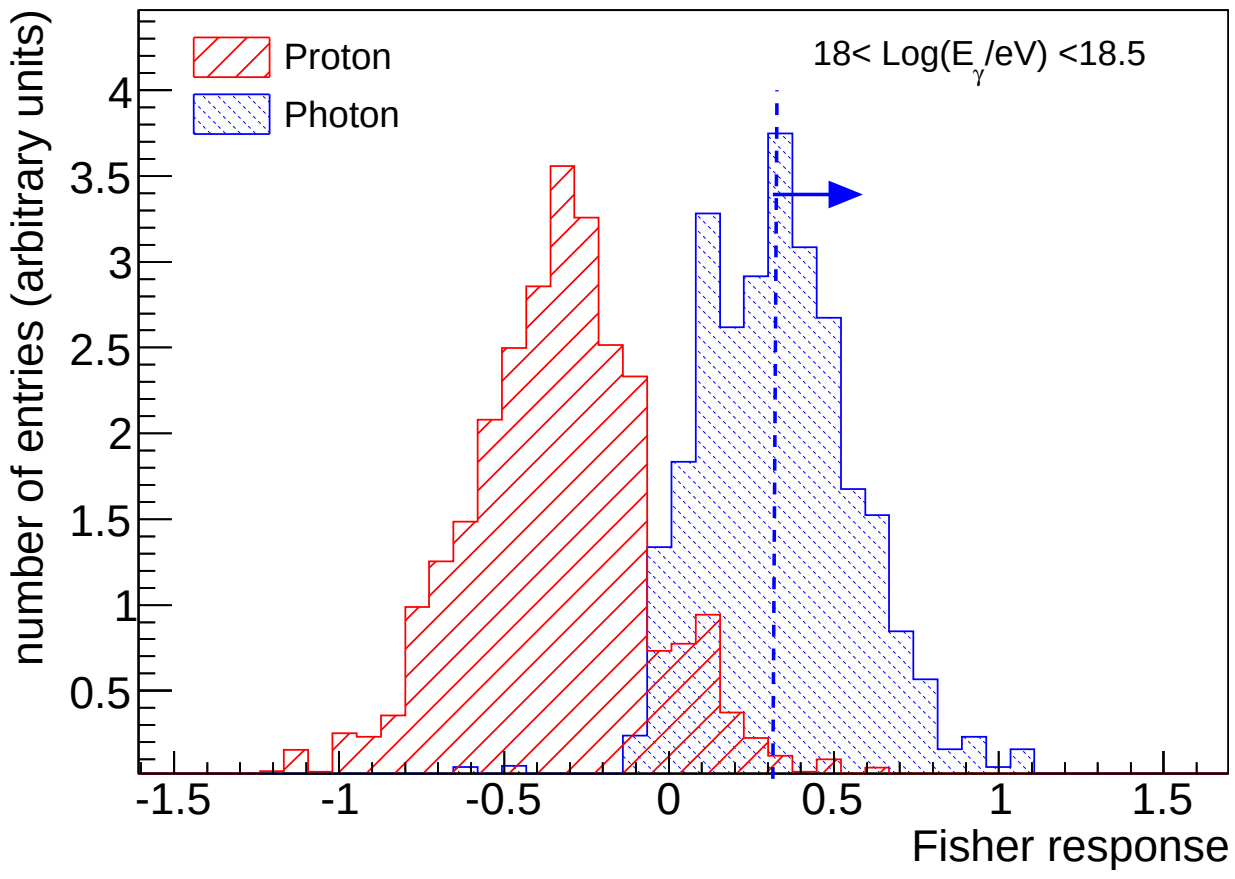

Figure 2: Distribution of the Fisher response for proton (red) and photon (blue) for simulations with energy between $10^{18}$ and $10^{18.5}$ eV. Photon-like events are selected requiring a Fisher value larger than $X_{cut}$ (dashed line) as indicated by the arrow.

Applying the method to data, 6, 0, 0, 0 and 0 photon candidates are found for energies above 1, 2, 3, 5 and 10 EeV. We checked with simulations that the observed number of photon candidates is consistent with the expectation for nuclear primaries, under the assumption of a mixed composition. For the two events with the deepest $X_{\mathrm{max}}$ (both larger than 1000 g cm$^{-2}$) the hadronic background has been individually checked by simulating 1000 dedicated proton CORSIKA showers with the same energy, arrival direction and core position as reconstructed for the real events. The actual SD and FD configurations at the detection time are considered. The profile of one candidate is shown in Fig-ure 3 (top). The values of $X_{\max}$ and $S_b$ are compared to the expectation for protons (bottom). A fraction of about 2% of such background events passes the cut on the Fisher observable (empty circles).

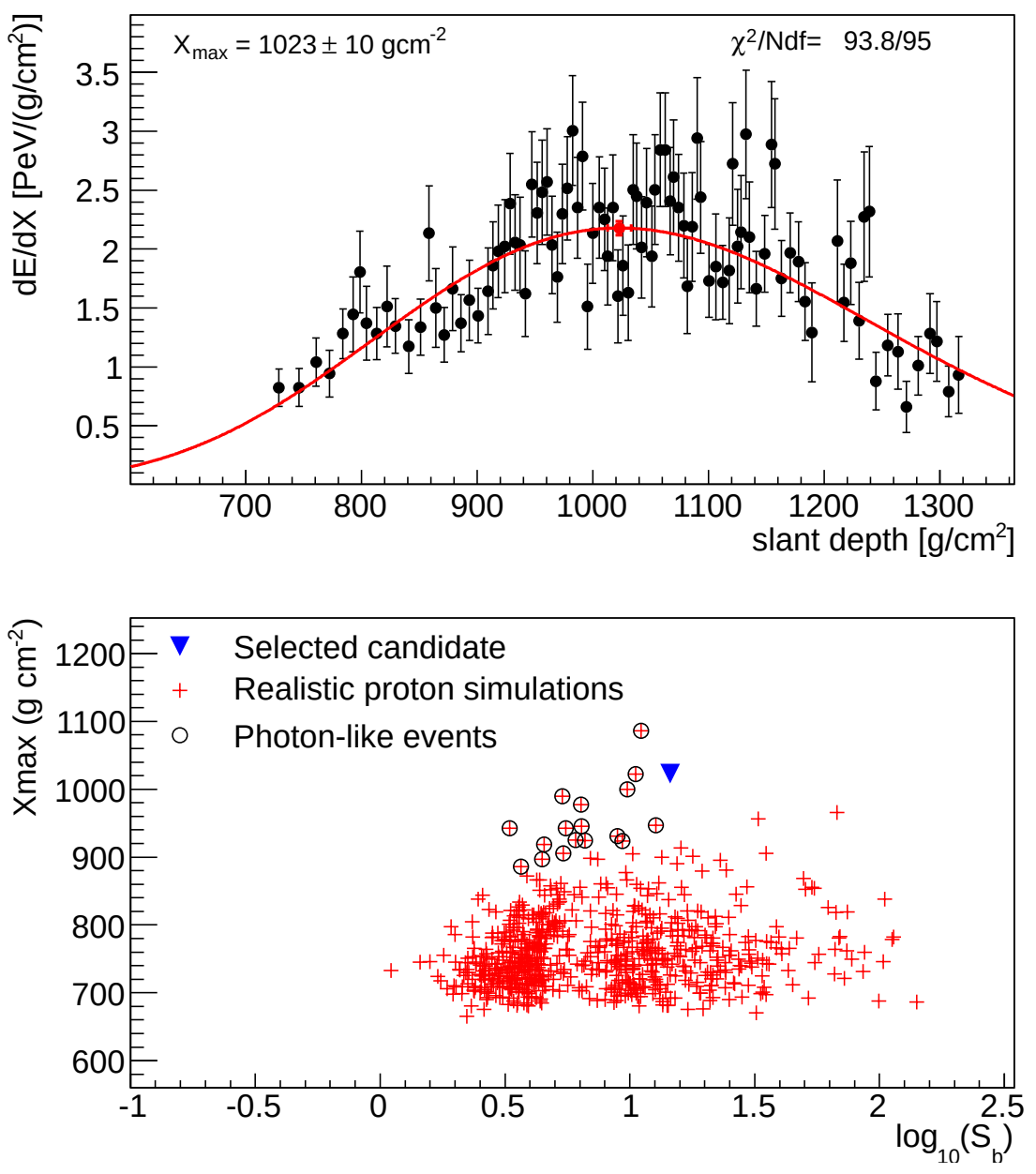

Figure 3: Example of one photon candidate. Top: longitudinal profile and a Gaisser-Hillas fit. Bottom: $X_{\max}$ vs $\log_{10}(S_b)$ for the candidate (triangle) compared to dedicated proton simulations (crosses). Photon-like events selected after the Fisher analysis are marked as empty circles.

## 3 Photon upper limits

The 95% CL upper limits on the photon flux $\Phi_\gamma^{95CL}$ integrated above an energy threshold $E_0$ is given by:

$$\Phi_\gamma^{95CL} = \frac{N_\gamma^{95CL}(E_\gamma > E_0)}{\mathcal{E}_{\gamma,\min}}. \quad (2)$$

where $E_\gamma$ is the reconstructed energy assuming that the primary particle is a photon (i.e., the calorimetric energy measured by FD plus a correction of about 1% due to the invisible energy [18]), $N_\gamma^{95CL}$ is the number of photon candidates above $E_0$ at 95% of confidence level and $\mathcal{E}_{\gamma,\min}$ is the exposure of the hybrid detector. To be conservative, in equation (2) we use the minimum value of the exposure above $E_0$ and a possible nuclear background is not subtracted for the calculation of $N_\gamma^{95CL}$. An additional independent sample of 20000 photon showers is used for determining the exposure of the hybrid detector using a procedure as the one discussed in [17]. Events are selected with the same criteria applied to data, and the final exposure is shown in Figure 4 for photon primaries after the Fisher analysis and the "a priori" cut discussed before. To reduce the impact of statistical fluctuation, a fit of a Gamma function to the exposure values has been performed and is shown as a dashed line. The arrow indicates the energy region of interest for the analysis presented in this work. Upper limits on the integral photon flux of $8.2 \cdot 10^{-2}$ km$^{-2}$ sr$^{-1}$ y$^{-1}$ above 1 EeV and $2.0 \cdot 10^{-2}$ km$^{-2}$ sr$^{-1}$ y$^{-1}$ above 2, 3, 5 and 10 EeV are derived. They are shown in Figure 5 compared to previous experimental results (SD [1], Hybrid 2009 [7], AGASA [19]) and Yakutsk [20] and to model predictions [2, 21]. The Hybrid 2009 limits on the photon fraction are converted to flux limits using the integrated Auger spectrum [6]. The bounds corroborate previous results disfavoring exotic models also in the lowest energy region. Comparing the flux limits on the measured Auger spectrum [6], upper bounds to the fraction of photons of about 0.4%, 0.5%, 1.0%, 2.6% and 8.9% are obtained for energies above 1, 2, 3, 5 and 10 EeV.

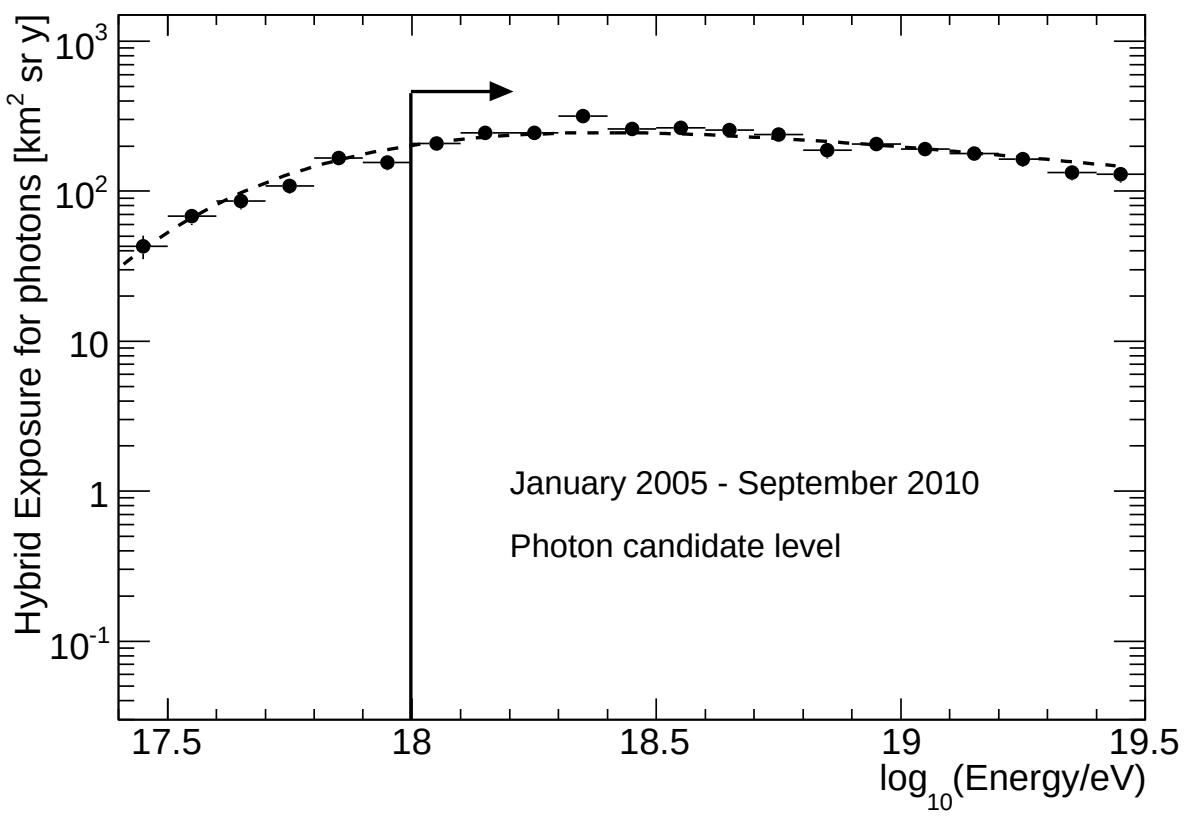

Figure 4: Exposure of the hybrid detector for photon primaries as a function of energy after all cuts.

We studied the robustness of the results against different sources of uncertainty. Increasing (reducing) all $X_{\max}$ values by the uncertainty $\Delta X_{\max}$ = 13 g cm$^{-2}$ [22] changes the number of photon candidates above 1 EeV by +1 (-2) not affecting the higher energies. As a consequence, this leads to an increase of ~10% (decrease of ~ 25%) of the first point of the upper limits. The uncertainty on the shower geometry determination corresponds to $\Delta S_b$ ~5%, changing the number of photon candidates by $\pm 0$ $(+1)$ above 1 EeV. The overall uncertainty on the hybrid exposure calculation for photons is about 5%. It includes the uncertainty due to on-time calculation (~4%), input spectra for Monte Carlo simulations and dependence of the trigger efficiency on the fluorescence yield model (~2%). Another source of systematic uncertainties is the energy scale which has been estimated to be about 22% [23]. An increase (reduction) of the energy scale, keeping the energy thresholds $E_0$ fixed, would change the upper limits by +14% (-54%) above 1 EeV and by +6% (-7%) above 2, 3, 5 and 10 EeV. This is a consequence of a different number of photon candidates ($^{+1}_{-4}$ in the first bin, unchanged in the others) and of the exposure ($^{-6\%}_{+7\%}$).

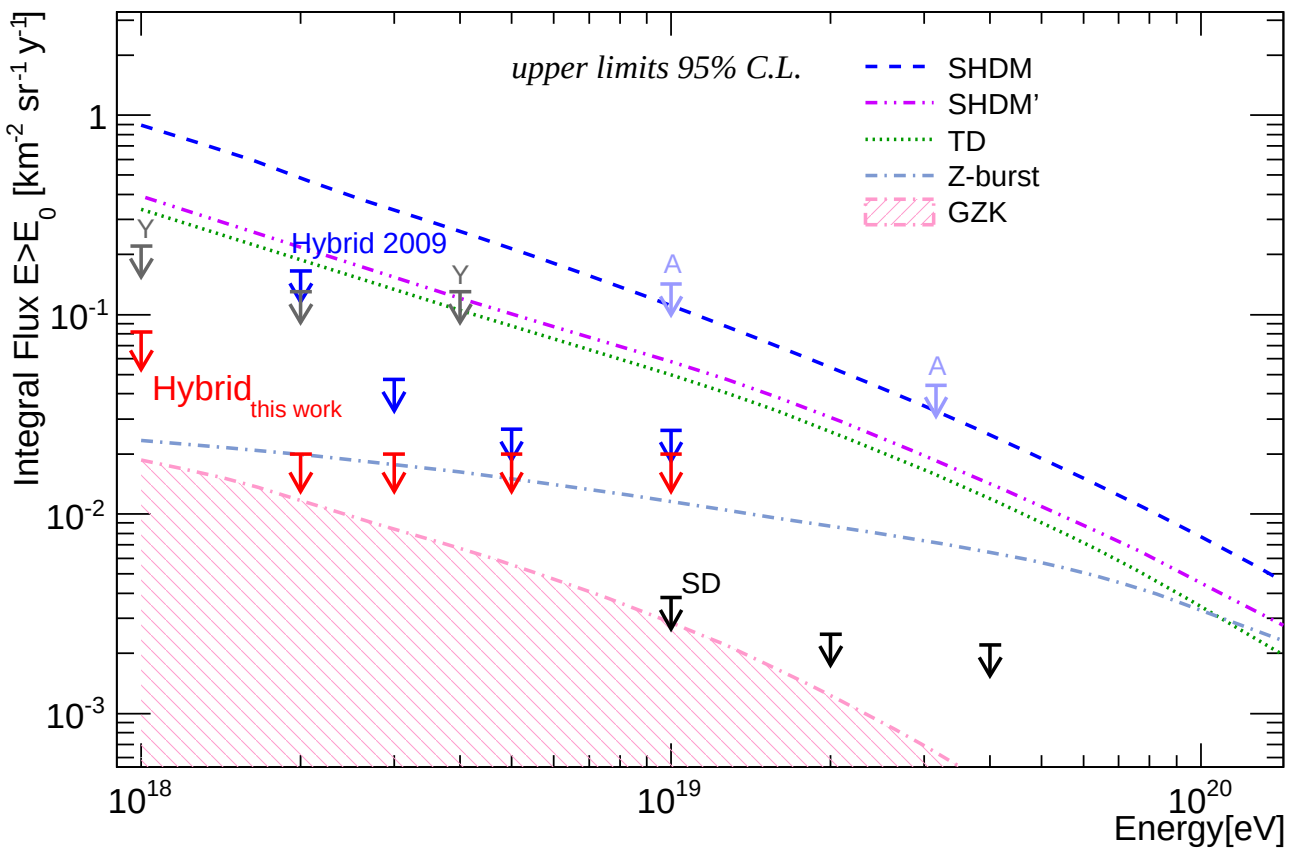

Figure 5: Upper limits on the photon flux above 1, 2, 3, 5 and 10 EeV derived in this work (red arrows) compared to previous limits from Auger (SD [1] and Hybrid 2009 [7]), from AGASA (A) [19] and Yakutsk (Y) [20]. The shaded region and the lines give the predictions for the GZK photon flux [2] and for top-down models (TD, Z-Burst, SHDM from [2] and SHDM' from [21]). The Hybrid 2009 limits on the photon fractions are converted to flux limits using the integrated Auger spectrum.

As the photon induced showers have an almost pure electromagnetic nature, no significant impact is expected when using another hadronic interaction model. However, since the Fisher analysis is also driven by the hadronic showers, we performed the same analysis using a sample of proton CORSIKA showers with QGSJET 01 [24]. In this case the separation capability improves by about 20% because this model predicts shallower $X_{\rm max}$ and a larger number of muons for proton showers. The number of photon candidates is then reduced by 1 above 1 EeV. The same effect is obtained when a 50% proton - 50% iron mixed composition assumption is used in the classification phase. The impact on the exposure is about a few percent.

## 4 Conclusions and Outlook

Using more than 5 years of hybrid data collected by the Pierre Auger Observatory we obtain an improved set of upper limits on the photon flux, in an energy region not covered by the SD-alone, and we extend the range of these limits down to $10^{18}$ eV. The derived limits on the photon fraction are 0.4%, 0.5%, 1.0%, 2.6% and 8.9% above 1, 2, 3, 5 and 10 EeV, significantly improving previous results at the lower energies, where limits well below the 1% level are reached now. These bounds also help reduce the systematic uncertainties on primary mass composition, energy spectrum and proton-air cross section measurements in the EeV range. The photon search conducted in this work benefits from the combination of complementary information provided by the fluorescence and surface detectors. While the focus of the current analysis was the low EeV range, future work will be performed to improve the photon-hadron separation also at higher energies using further information provided by the SD.

# The Pierre Auger Observatory and ultra-high energy neutrinos: upper limits to the diffuse and point source fluxes

YANN GUARDINCERRI[1] FOR THE PIERRE AUGER COLLABORATION[2]
[1]*Facultad de Ciencias Exactas y Naturales, Universidad de Buenos Aires, Buenos Aires, Argentina*
[2]*Observatorio Pierre Auger, Av. San Martín Norte 304, 5613 Malargüe, Argentina*
*(Full author list: http://www.auger.org/archive/authors_2011_05.html)*
*auger_spokespersons@fnal.gov*

**Abstract:** With the Surface Detector of the Pierre Auger Observatory, we can detect ultra-high energy neutrinos in the sub-EeV energy range and above. Neutrinos of all flavours can interact in the atmosphere and induce inclined showers close to the ground (down-going). The sensitivity of the Surface Detector to tau neutrinos is further enhanced through the "Earth-skimming" mechanism (up-going). Both types of neutrino interactions can be identified through the broad time structure of the signals induced in the Surface Detector stations. Two independent sets of identification criteria were designed to search for down and up-going neutrinos in the data collected from 1 January 2004 to 31 May 2010, with no candidates found. Assuming a differential flux $f(E_\nu) = kE_\nu^{-2}$, we place a 90% CL upper limit on the single flavour neutrino flux of $k < 2.8 \times 10^{-8}$ GeV cm$^{-2}$ s$^{-1}$ sr$^{-1}$ in the energy interval $1.6 \times 10^{17}$ eV $- 2.0 \times 10^{19}$ eV based on Earth-skimming neutrinos and $k < 1.7 \times 10^{-7}$ GeV cm$^{-2}$ s$^{-1}$ sr$^{-1}$ in the energy interval $1 \times 10^{17}$ eV $- 1 \times 10^{20}$ eV based on down-going neutrinos. We also show that the Auger Observatory is sensitive to ultra-high energy neutrinos from a large fraction of the sky, and we place limits on the neutrino flux from point-like sources as a function of declination, and in particular from the active galaxy Centaurus A.

## 1 Introduction

Essentially all models of Ultra High Energy Cosmic Ray (UHECR) production predict neutrinos as the result of the decay of charged pions, produced in interactions of the cosmic rays within the sources themselves or in their propagation through background radiation fields [1, 2]. Neutrinos are also copiously produced in top-down models proposed as alternatives to explain the production of UHECRs [1].

With the surface detector (SD) of the Pierre Auger Observatory [3] we can detect and identify UHE neutrinos (UHE$\nu$s) in the 0.1 EeV range and above. "Earth-skimming" tau neutrinos [4] are expected to be observed through the detection of showers induced by the decay products of an emerging $\tau$ lepton, after the propagation and interaction of a $\nu_\tau$ inside the Earth. "Down-going" neutrinos of all flavours can interact in the atmosphere and induce a shower close to the ground [5].

This contribution updates both, Earth-skimming [6, 7, 8] and down-going [8] analyses with data until the 31 May 2010 and shows, for the first time, the sensitivity of the Pierre Auger surface detector to neutrinos from point-like sources.

## 2 Identifying neutrinos in data

Identifying neutrino-induced showers in the much larger background of the ones initiated by nucleonic cosmic rays is based on a simple idea: neutrinos can penetrate large amounts of matter and generate "young" inclined showers developing close to the SD, exhibiting shower fronts extended in time. In contrast, UHE particles such as protons or heavier nuclei interact within a few tens of g cm$^{-2}$ after entering the atmosphere, producing "old" showers with shower fronts narrower in time. In Fig. 1 we show a sketch of these two kinds of showers together with an Earth-skimming shower and a $\nu_\tau$ interacting in the Andes, which can also be identified.

Although the SD is not directly sensitive to the nature of the arriving particles, the 25 ns time resolution of the FADC traces, with which the signal is digitised in the SD stations, allows us to distinguish the narrow signals in time expected from a shower initiated high in the atmosphere from the broad signals expected from a young shower. Several observables can be used to characterise the time structure and shape of the FADC traces. They are described in [9] where their discrimination power is also studied.

In this work we use two different sets of identification criteria to select neutrinos. One is used to define Earth-skimming tau neutrinos and the other for down-going neutrinos. They are given in Table 1 and described in the following.

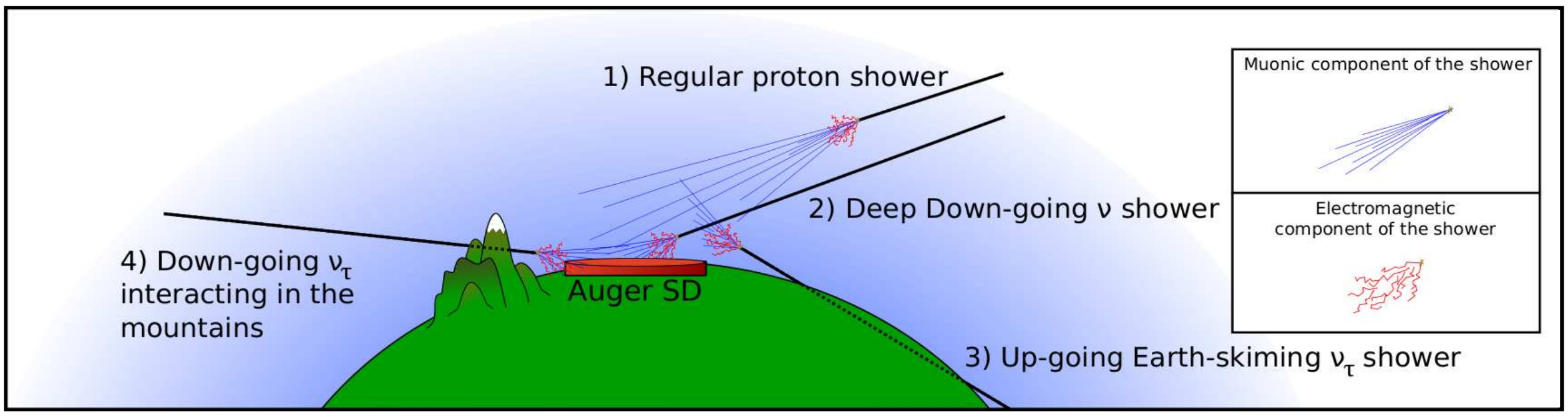

Figure 1: Sketch of different inclined showers which can be detected by the Pierre Auger Observatory. (1) An inclined shower induced by a proton interacting high in the atmosphere whose electromagnetic component is absorbed and only the muons reach the detector. Inclined showers presenting significant electromagnetic component at the detector level: (2) a deep down-going $\nu$ shower; (3) an Earth-skimming $\nu_\tau$ shower; (4) and a $\nu_\tau$ interacting in the mountains.

Table 1: Criteria to select Earth-skimming $\nu_\tau$ and down-going $\nu$. See text for details.

| | Earth-skimming | Down-going |
|---|---|---|
| Inclined Showers | N° of Stations $\geq 3$<br>$L/W > 5$<br>$0.29\frac{\text{m}}{\text{ns}} < V < 0.31\frac{\text{m}}{\text{ns}}$<br>RMS($V$)$< 0.08\frac{\text{m}}{\text{ns}}$<br>- | N° of Stations $\geq 4$<br>$L/W > 3$<br>$V < 0.313\frac{\text{m}}{\text{ns}}$<br>$\frac{\text{RMS}(V)}{V} < 0.08$<br>$\theta_{rec} > 75°$ |
| Young Showers | ToT fraction>0.6 | Fisher discriminator based on AoP |

The analyses start with the inclined shower selection (down-going:$\theta > 75°$ and Earth-skimming $\theta < 96°$). These showers usually have elongated patterns on the ground along the azimuthal arrival direction. A length $L$ and a width $W$ are assigned to the pattern and a cut on their ratio $L/W$ is applied. We also calculate the apparent speed $V$ of an event using the times of signals at ground and the distances between stations projected onto $L$. Finally, for down-going events, we reconstruct the zenith angle $\theta_{rec}$.

Once we have selected inclined showers we look for young showers. A station having signals extended in time usually has a Time over Threshold (ToT) local trigger while narrow signals have other local triggers [3, 10]. The Earth-skimming analysis identifies young showers placing a cut on the fraction of ToT stations (ToT fraction). For down-going events, to optimize the discrimination power, we use the Fisher discriminant method using AoP (area of the FADC trace over its peak value, which gives an estimate of the spread in time of the signal) as input variables. The advantage of the Fisher discriminant is that it allows us to place an optimized cut to reject backgrounds from regular hadronic showers, and that it provides an a priori measure of how neutrino-like a possible candidate is.

## 3 Exposure and limit on the diffuse flux

The Earth skimming and down going criteria are applied to data collected from 1 Jan 04 to 31 May 10, and from 1 Nov 07 to 31 May 10, respectively. The down-going sample is smaller than the Earth-skimming one because data from 1 Jan 04 to 31 Oct 07 was used as a training sample for the Fisher discriminator [1]. Due to the fact that the Observatory was continuously growing during the construction phase (2004 - 2008) and that the SD is a dynamic array (some stations can occasionally be not operative), the previous periods correspond to 3.5 yr (Earth-skimming) and 2 yr (down-going) of data of a full SD array. No neutrino candidates were found and an upper limit on the diffuse flux of ultra-high energy neutrinos can be placed.

For this purpose the exposure of the SD array to UHE neutrinos is calculated. For down-going neutrinos, this involves folding the SD array aperture with the interaction probability and the identification efficiency, and integrating in time, taking into account changes in the array configuration due to the installation of new stations and other changes. The identification efficiency $\varepsilon$ for the set of cuts defined above depends on the neutrino energy $E_\nu$, the slant depth $D$ from ground to the neutrino interaction point, the zenith angle $\theta$, the core position $\vec{r} = (x, y)$ of the shower in the surface $S$ covered by the array, and the time $t$ through the instantaneous configuration of the array. Moreover it depends on the neutrino flavour ($\nu_e$, $\nu_\mu$, or $\nu_\tau$), and the type of interaction – charged (CC) or neutral current (NC) – since the different combinations of flavour and interaction induce different types of showers. The efficiencies $\varepsilon$ were obtained through MC simulations of the first interaction between the $\nu$ and a nucleon with HERWIG [11], of the development of the shower in the atmosphere with AIRES [12], and of the response of the surface detector array, see [9] for more details. Assuming a 1:1:1 flavour

1. In the case of Earth-skimming analysis, data from 1 Nov to 31 Dec 04 was used as a test sample and excluded from the search sample.

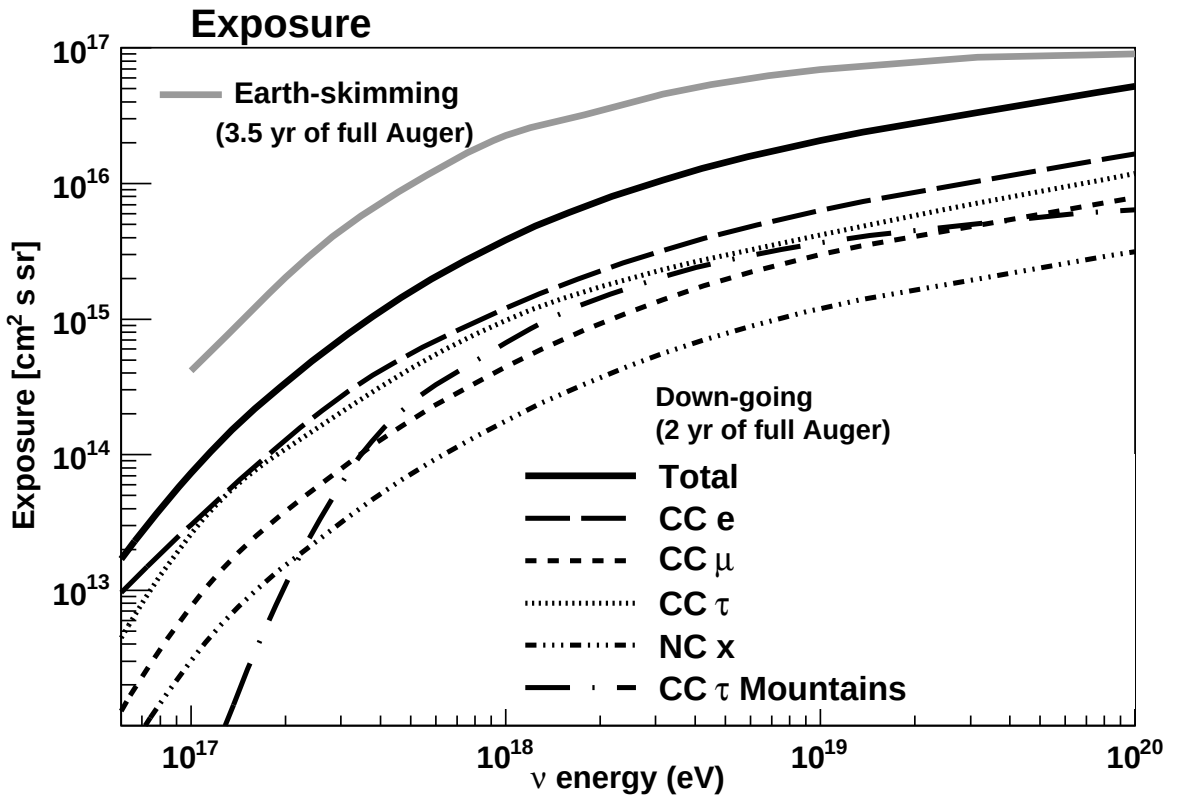

Figure 2: Exposure of the surface detector of the Pierre Auger Observatory for Earth-skimming neutrino initiated showers (3.5 yr of full Auger) and for down-going neutrino initiated showers for all the considered channels as a function of neutrino energy (2 yr of full Auger).

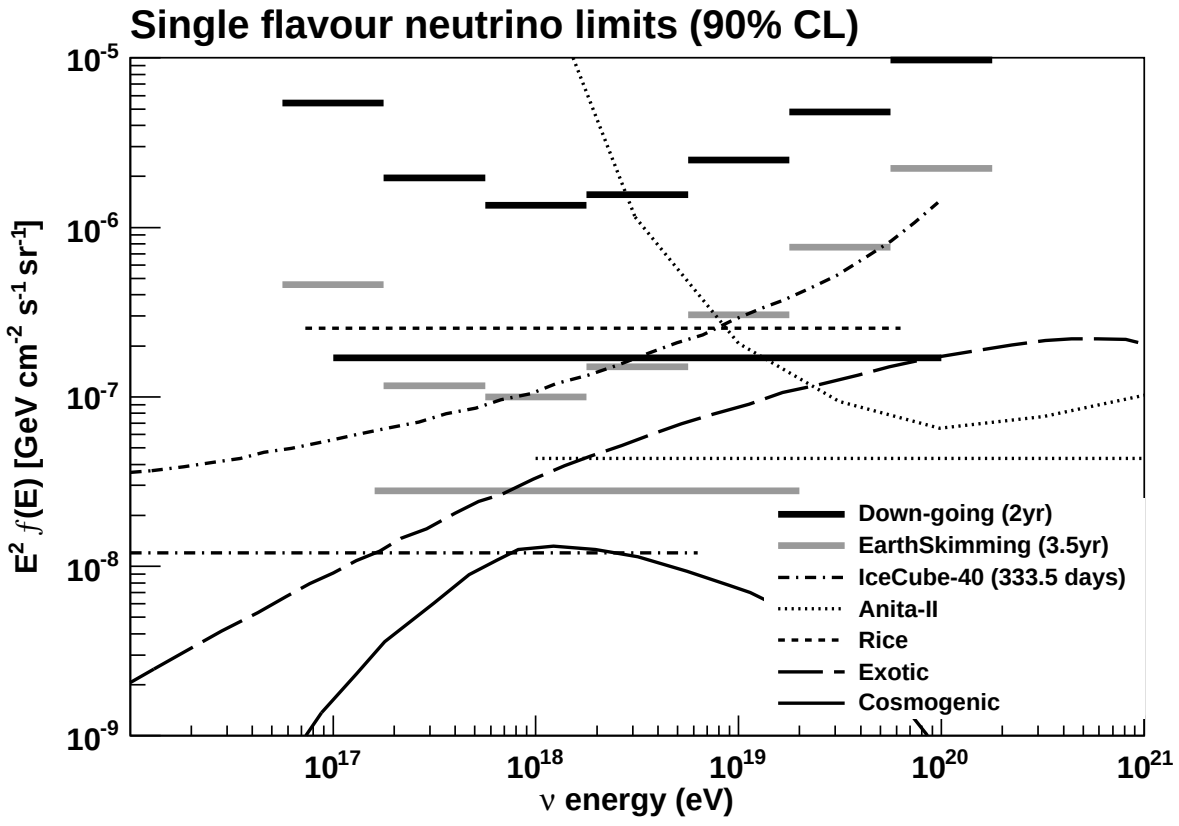

Figure 3: Differential and integrated upper limits (90% C.L.) from the Pierre Auger Observatory for a diffuse flux of down-going $\nu$ (2 yr of full Auger) and Earth-skimming $\nu_\tau$ (3.5 yr of full Auger). Limits from other experiments are also plotted [16]. Expected fluxes are shown for cosmogenic neutrinos [17] and for a theoretical exotic model [18].

ratio, the total exposure can be written as:

$$\mathcal{E}^{\rm DG}(E_\nu) = \frac{2\pi}{m} \sum_i \left[ \sigma^i(E_\nu) \int \, dt\, d\theta\, dD\, dS \right.$$
$$\left. \sin\theta\, \cos\theta\ \varepsilon^i(\vec{r},\theta,D,E_\nu,t) \right] \quad (1)$$

where the sum runs over the 3 neutrino flavours and the CC and NC interactions, $m$ is the mass of a nucleon, and $\sigma^i$ is the $\nu$ cross section with a nucleon. For $\nu_\tau$ we have taken into account the possibility that it produces a double shower in the atmosphere triggering the array – one in the $\nu_\tau$ CC interaction itself and another in the decay of the $\tau$ lepton. Furthermore, we consider the possibility of a $\nu_\tau$ interacting in the Andes inducing a shower through the decay products of the $\tau$ lepton.

For the Earth-skimming neutrinos the procedure is described in Ref [7].

In Fig. 2 we show both the Earth-skimming and down-going exposures for the respective search periods.

Several sources of systematic uncertainties have been taken into account and their effect on the exposure evaluated. For down-going neutrinos there is $[-30\%, 10\%]$ systematic uncertainty in the exposure due to the neutrino-induced shower simulations and the hadronic models. Another source of uncertainty comes from the neutrino cross section which is $\sim 10\%$ [13]. For the Earth-skimming showers the systematic uncertainties are dominated by the tau energy losses, the topography and the shower simulations [7].

Using the computed exposures and assuming a typical $f(E_\nu) = k \cdot E_\nu^{-2}$ differential neutrino flux and a 1:1:1 flavour ratio, an upper limit on the value of $k$ can be obtained. We use a semi-Bayesian extension [14] of the Feldman-Cousins approach [15] to include the uncertainties in the exposure. The updated single-flavour 90% C.L. limit based on Earth-skimming neutrinos is: $k <$ $2.8 \times 10^{-8}$ GeV cm$^{-2}$ s$^{-1}$ sr$^{-1}$ in the energy interval $1.6 \times 10^{17}$ eV $-$ $2.0 \times 10^{19}$ eV and the updated single-flavour 90% C.L. limit based on down-going neutrinos is: $k < 1.7\times10^{-7}$ GeV cm$^{-2}$ s$^{-1}$ sr$^{-1}$ in the energy interval $1\times10^{17}$ eV$-1\times10^{20}$ eV. These results are shown in Fig. 3 including the limit in different bins of width 0.5 in $\log_{10} E_\nu$ (differential limit) to show at which energies the sensitivity of the Pierre Auger Observatory peaks. The expected number of events from a cosmogenic [17] (neutrinos produced by the interaction of cosmic rays with background radiation fields) and an exotic model [18] (neutrinos produced due to the decay of heavy particles) are given in Table 2.

## 4 Limits to point-like sources

As we found no candidate events in the search period, we can place a limit on the UHE neutrino flux from a source at declination $\delta$.

A point source moves through the sky so that it is visible from the SD of the Pierre Auger Observatory with zenith angle $\theta(t)$ which depends on the sidereal time $t$. For an observatory located at a latitude $\lambda$ the relation between the zenith angle and the declination of the source $\delta$ is given by:

$$\cos\theta(t) = \sin\lambda\, \sin\delta + \cos\lambda\, \cos\delta\, \sin(\omega t - \alpha_0) \quad (2)$$

with $\omega = 2\pi/T$, where $T$ is the duration of one sidereal day and $\alpha_0$ depends on the right ascension.

The sensitivity to UHE$\nu$s is limited to large zenith angles so the rate of events from a point source in the sky depends strongly on its declination. The point-source exposure $\mathcal{E}^{\rm PS}(E_\nu,\delta)$ can be obtained in a similar way as the diffuse exposure but avoiding the integration in solid angle and taking into account that the probability of neutrino identification $\varepsilon$ depends on $\theta$, while the $\theta$ of the source depends on sidereal time through Eq. (2). Also $\varepsilon$ itself depends explicitly on time because the configuration of the SD array changes with time.

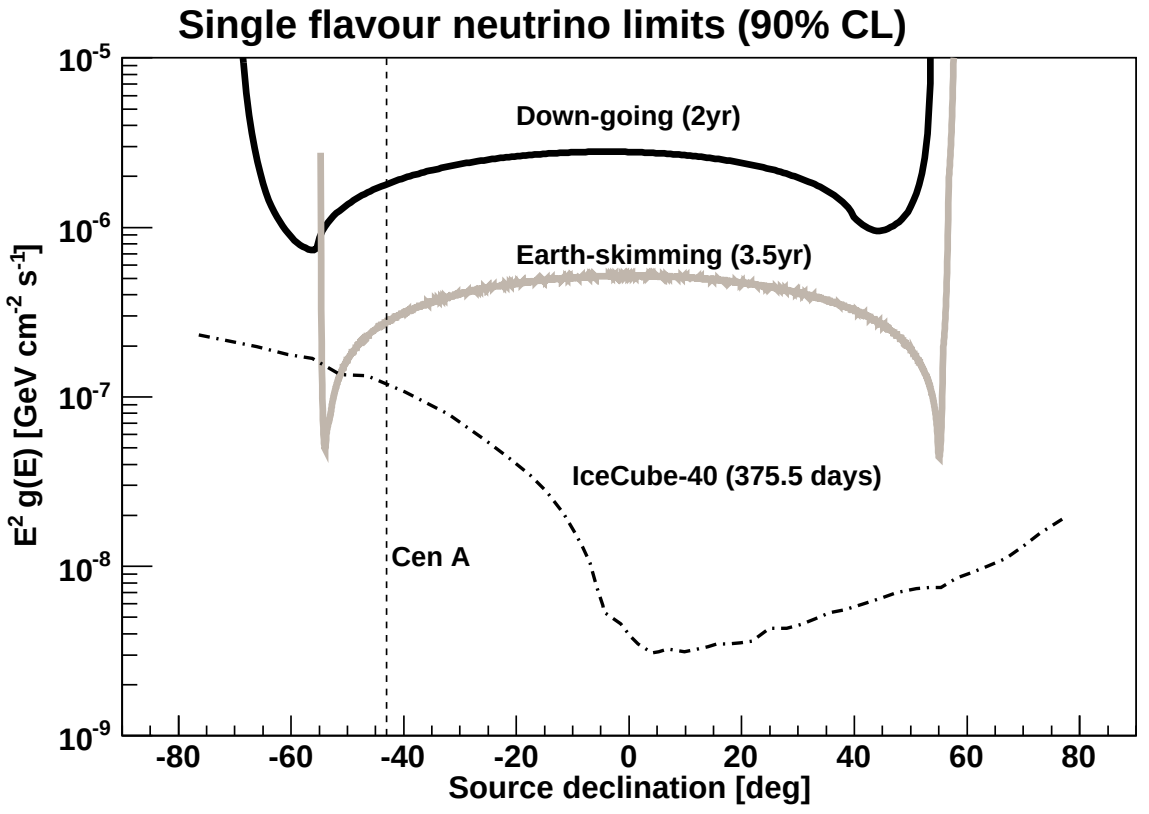

Figure 4: Neutrino flux limits to a $E^{-2}$ differential neutrino flux from a point source as a function of the declination of the source, as obtained with the SD of the Pierre Auger Observatory for $1.6 \times 10^{17}$ eV $- 2.0 \times 10^{19}$ eV (Earth-skimming) and $1 \times 10^{17}$ eV $- 1 \times 10^{20}$ eV (down-going). Also shown is the limit obtained by IceCube [19] that applies below $10^{17}$ eV (or lower depending on declination).

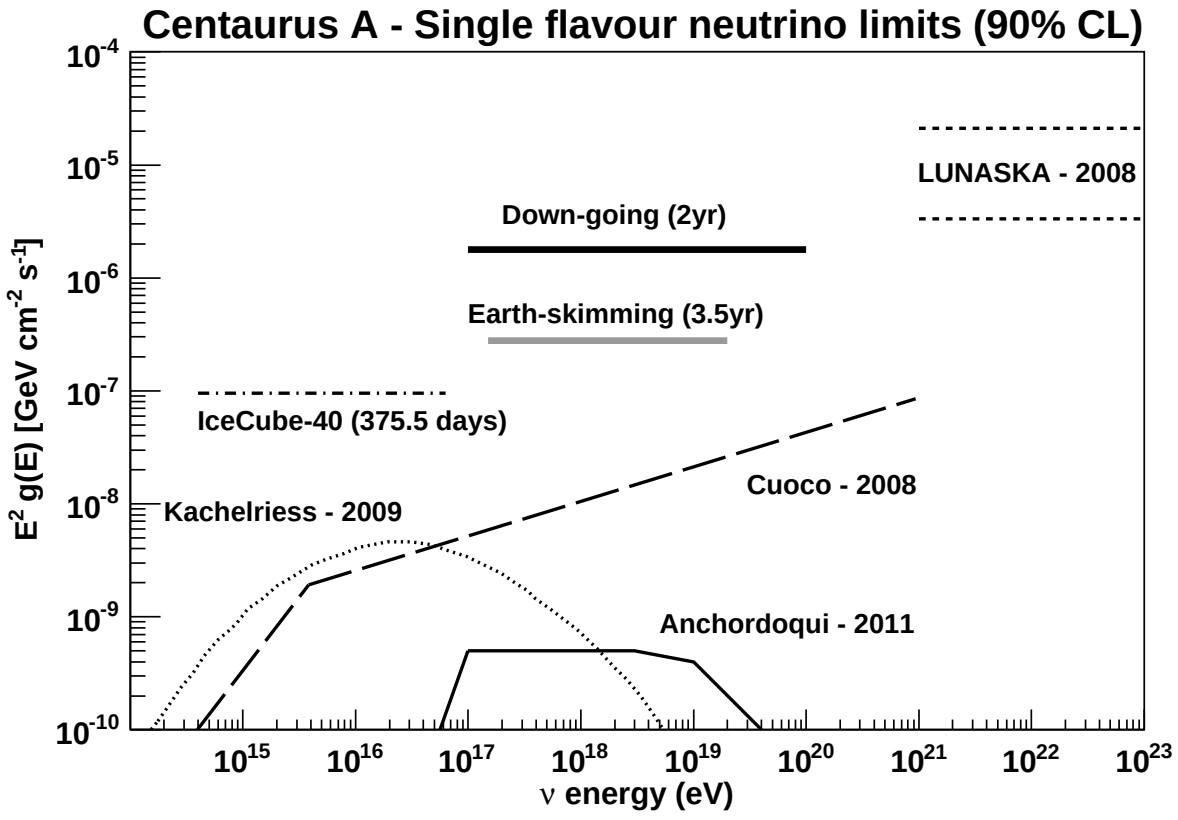

Figure 5: Limits on Cen A coming from the Earth-skimming and down-going analyses. Also shown are limits from IceCube [19] and LUNASKA [20] in different energy ranges and three theoretical predictions [21].

We perform the integration over time and we obtain the point source exposure which depends not only on $E_\nu$ but also on $\delta$. Assuming now a point source flux which decreases in energy as $g(E_\nu) = k^{\rm PS} \cdot E_\nu^{-2}$ and a 1:1:1 flavour ratio, we can obtain a point source upper limit $k^{\rm PS}(\delta)$.

In Fig. 4 we show the value of $k^{\rm PS}$ as a function of the declination of the source. In both Earth-skimming and down-going analyses the sensitivity has a broad "plateau" spanning $\Delta\delta \sim 100°$ in declination. We also show the sensitivity of IceCube which is at a lower neutrino energy.

In Fig. 5 we show the constraints on $k$ for the case of the active galaxy Centaurus A (CenA) at a declination $\delta \sim -43°$. We also show three models of UHE$\nu$ production in the jets and the core of CenA [21]. The expected number of events from each of these models with the current exposure is given in Table 2.

Table 2: Expected number of events for two diffuse neutrino flux models and two CenA neutrino flux models.

| Diffuse flux model | Earth-skimming | Down-going |
|---|---|---|
| Cosmogenic | 0.71 | 0.14 |
| Exotic | 3.5 | 0.97 |

| CenA flux model | Earth-skimming | Down-going |
|---|---|---|
| Cuoco *et al.* | 0.10 | 0.02 |
| Kachelriess *et al.* | 0.006 | 0.001 |

# Analysis of the modulation in the first harmonic of the right ascension distribution of cosmic rays detected at the Pierre Auger Observatory

HARIS LYBERIS[1,2], FOR THE PIERRE AUGER COLLABORATION[3]
[1] *IPN Orsay, CNRS/IN2P3 & Université Paris Sud, Orsay, France*
[2] *Università degli Studi di Torino, Torino, Italy*
[3] *Observatorio Pierre Auger, Av. San Martín Norte 304, 5613 Malargüe, Argentina*
*(Full author list: http://www.auger.org/archive/authors_2011_05.html)*
*auger_spokespersons@fnal.gov*

**Abstract:** We present an update of the results of searches for first harmonic modulations in the right ascension distribution of cosmic rays detected with the surface detector of the Pierre Auger Observatory over a range of energies. The upper limits obtained provide the most stringent bounds at present above $2.5 \times 10^{17}$ eV. The infill surface detector array which is now operating at the Pierre Auger Observatory will allow us to extend this search for large scale anisotropies to lower energy thresholds.

## 1 Introduction

The large scale distribution of arrival directions of cosmic rays represents one of the main tools for understanding their origin, in particular in the EeV energy range - where 1 EeV $\equiv$ $10^{18}$ eV. Using the large statistics provided by the surface detector (SD) array of the Pierre Auger Observatory, upper limits below 2% at 99% $C.L.$ have been recently reported [1] for EeV energies on the dipole component in the equatorial plane. Such upper limits are sensible, because cosmic rays of galactic origin, while escaping from the galaxy in this energy range, might generate a dipolar large-scale anisotropy with an amplitude at the % level as seen from the Earth [2, 3]. Even for isotropic extragalactic cosmic rays, a large scale anisotropy may be left due to the motion of our galaxy with respect to the frame of extragalactic isotropy. This anisotropy would be dipolar in a similar way to the *Compton- Getting effect* [4] in the absence of the galactic magnetic field, but this field could transform it into a complicated pattern as seen from the Earth, described by higher order multipoles [5].

Continued scrutiny of the large scale distribution of arrival directions of cosmic rays as a function of the energy is thus important to constrain different models for the cosmic rays origin. To do so, we present an update of the results of searches for anisotropies by applying first harmonic analyses to events recorded by the SD array data from 1 January 2004 to 31 December 2010, with the same criteria for event selection as in [1].

## 2 First harmonic analyses

### 2.1 Analysis methods

A dipolar modulation of *experimental origin* in the distribution of arrival times of the events with a period equal to one solar day may induce a spurious anisotropy in the right ascension distribution. Such spurious variations can be accounted for thanks to the monitoring of the number of unitary cells $n_{\rm cell}(t)$ recorded every second by the trigger system of the Observatory, reflecting the array growth as well as the dead periods of each surface detector. Here, accordingly to the fiducial cut applied to select events [6], a unitary cell is defined as an active detector surrounded by six neighbouring active detectors. For any periodicity $T$, the total number of unitary cells $N_{\rm cell}(t)$ as a function of time $t$ within a period and summed over all periods, and its associated relative variations are obtained from :

$$N_{\rm cell}(t) = \sum_j n_{\rm cell}(t + jT), \quad \Delta N_{\rm cell}(t) = \frac{N_{\rm cell}(t)}{\langle N_{\rm cell}(t)\rangle}. \tag{1}$$

with $\langle N_{\rm cell}(t)\rangle = 1/T \int_0^T dt N_{\rm cell}(t)$. Hence, to perform a first harmonic analysis accounting for the slighlty non-uniform exposure in different parts of the sky, we weight each event with right ascension $\alpha_i$ by the inverse of the integrated number of unitary cells for computing the Fourier coefficients $a$ and $b$ as :

$$a = \frac{2}{\mathcal{N}} \sum_{i=1}^{N} \frac{\cos(\alpha_i)}{\Delta N_{\rm cell}(\alpha_i^0)}, \; b = \frac{2}{\mathcal{N}} \sum_{i=1}^{N} \frac{\sin(\alpha_i)}{\Delta N_{\rm cell}(\alpha_i^0)}, \tag{2}$$

where $\mathcal{N} = \sum_{i=1}^{N} [\Delta N_{\text{cell}}(\alpha_i^0)]^{-1}$ and $\alpha_i^0$ is the local sidereal time expressed here in radians and chosen so that it is always equal to the right ascension of the zenith at the center of the array. The amplitude $r$ and phase $\varphi$ are then given by $r = \sqrt{a^2 + b^2}$ and $\varphi = \arctan(b/a)$, and follow respectively a Rayleigh and uniform distributions in the case of an underlying isotropy.

Changes in the air density and pressure have been shown to affect the development of extensive air showers and consequently to induce a temporal variation of the observed shower size at a fixed energy [7]. Such an effect is important to control, because any seasonal variation of the modulation of the daily counting rate induces sidebands at both the sidereal and the anti-sidereal frequencies, which may lead to misleading measures of anisotropy in case the amplitude of the sidebands significantly stands out from the background noise [8]. To eliminate these variations, the conversion of the shower size into energy is performed by relating the observed shower size to the one that would have been measured at reference atmospheric conditions. Above 1 EeV, this procedure is sufficient to control the size of the sideband amplitude to well below $\simeq 10^{-3}$ [1].

Below 1 EeV, as weather effects affect the detection efficiency to a larger extent, spurious variations of the counting rate are amplified. Hence, we adopt the differential *East-West method* [9]. Since the instantaneous exposure for Eastward and Westward events is the same, the difference between the event counting rate measured from the East sector, $I_E(\alpha^0)$, and the West sector, $I_W(\alpha^0)$, allows us to remove at first order the direction independent effects of experimental origin without applying any correction, though at the cost of a reduced sensitivity. This counting difference is directly related to the right ascension modulation $r$ by [9] :

$$I_E(\alpha^0) - I_W(\alpha^0) = -\frac{N}{2\pi}\frac{2}{\pi}\frac{\langle\sin(\theta)\rangle}{\langle\cos(\delta)\rangle} r \sin(\alpha^0 - \varphi). \quad (3)$$

where $\delta$ is the declination and $\theta$ the zenith angle of the detected events. The amplitude $r$ and phase $\varphi$ can thus be calculated from the arrival times of $N$ events using the standard first harmonic analysis slightly modified to account for the subtraction of the Western sector to the Eastern one. The Fourier coefficients $a_{EW}$ and $b_{EW}$ are thus defined by :

$$a_{EW} = \frac{2}{N}\sum_{i=1}^{N}\cos(\alpha_i^0 + \zeta_i),$$
$$b_{EW} = \frac{2}{N}\sum_{i=1}^{N}\sin(\alpha_i^0 + \zeta_i), \quad (4)$$

where $\zeta_i$ equals 0 if the event is coming from the East or $\pi$ if coming from the West (so as to effectively subtract the events from the West direction). This allows us to recover the right ascension amplitude $r$ and the phase $\varphi_{EW}$ from $r = \frac{\pi\langle\cos(\delta)\rangle}{2\langle\sin(\theta)\rangle}\sqrt{a_{EW}^2 + b_{EW}^2}$ and $\varphi_{EW} = \arctan(b_{EW}/a_{EW})$. Note however that $\varphi_{EW}$, being the phase corresponding to the maximum in the differential of the East and West fluxes, is related to $\varphi$ through $\varphi = \varphi_{EW} + \pi/2$.

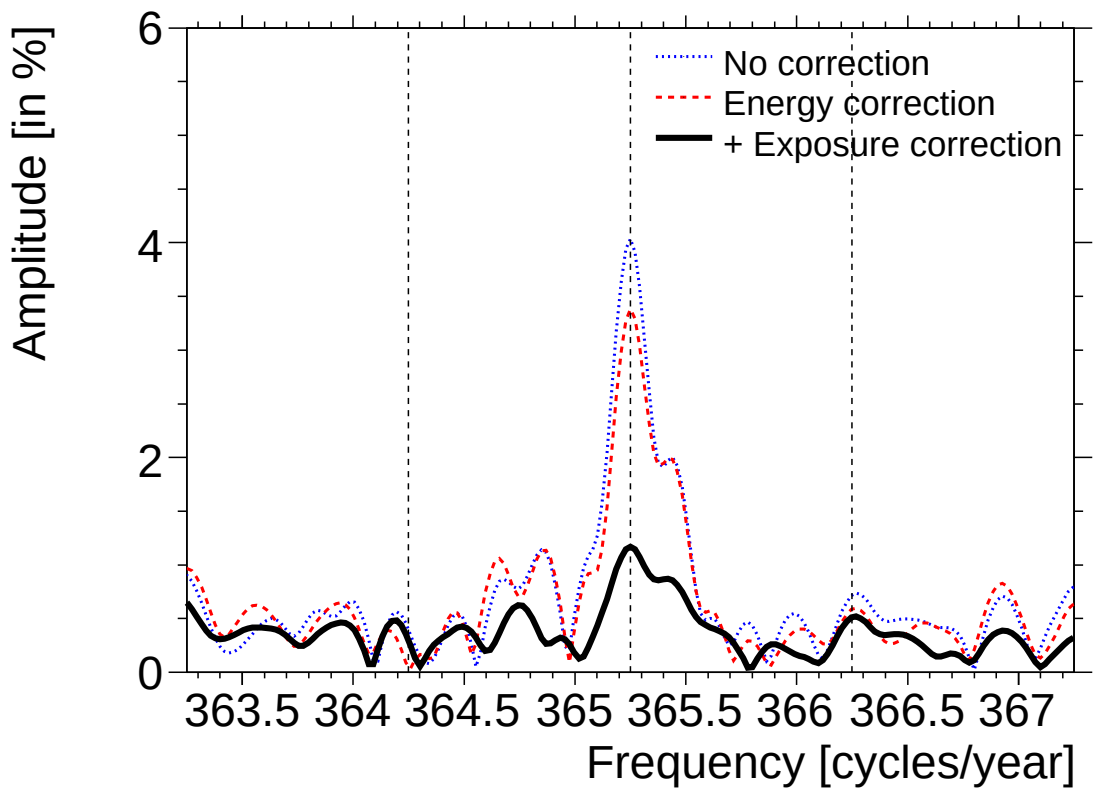

Figure 1: Amplitude of the Fourier modes as a function of the frequency above 1 EeV (see text).

## 2.2 Analysis of solar frequency above 1 EeV

Over a 7-years period, spurious modulations are partially compensated in sidereal time. Though, since the amplitude of an eventual sideband effect is *proportional* to the solar amplitude, it is interesting to look at the impact of the corrections at and around the solar frequency by performing the Fourier transform of the modified time distribution [10] :

$$\tilde{\alpha}_i^0 = \frac{2\pi}{T_{sid}} t_i + \alpha_i - \alpha_i^0. \quad (5)$$

The amplitude of the Fourier modes when considering all events above 1 EeV are shown in Fig. 1 as a function of frequencies close to the solar one (dashed line at 365.25 cycles/year). The thin dotted curve is obtained without accounting for the variations of the exposure and without accounting for the weather effects. There is a net solar amplitude of $\sim 4\%$, highly significant. The impact of the correction of the energies is evidenced by the dashed curve within the resolved solar peak (reduction of $\simeq 20\%$ of the spurious modulations). In addition, when accounting also for the exposure variation at each frequency, the solar peak is then reduced at a level close to the statistical noise, as evidenced by the thick curve. This provides support that the variations in the exposure and weather effects are under control.

## 2.3 Analysis of the sidereal frequency

The amplitude $r$ at the sidereal frequency as a function of the energy is shown in Fig. 2. The size of the energy intervals was chosen to be $\Delta\log_{10}(E) = 0.3$ below 8 EeV, so that it was larger than the energy resolution (about 15% [11]) even at low energies. Above 8 EeV, to guarantee the determination of the amplitude measurement within an uncertainty $\sigma \simeq 2\%$, all events ($\simeq 5,000$) where gathered in a single energy interval. The dashed line indicates the 99% $C.L.$ upper bound on the amplitudes that could result from fluctuations of an isotropic distribution. There is no evidence of any significant signal in any energy range. The probability with which the 6 observed amplitudes could have arisen from an underlying isotropic distribution can be made by combining the amplitudes in all bins. It is found to be 45%.

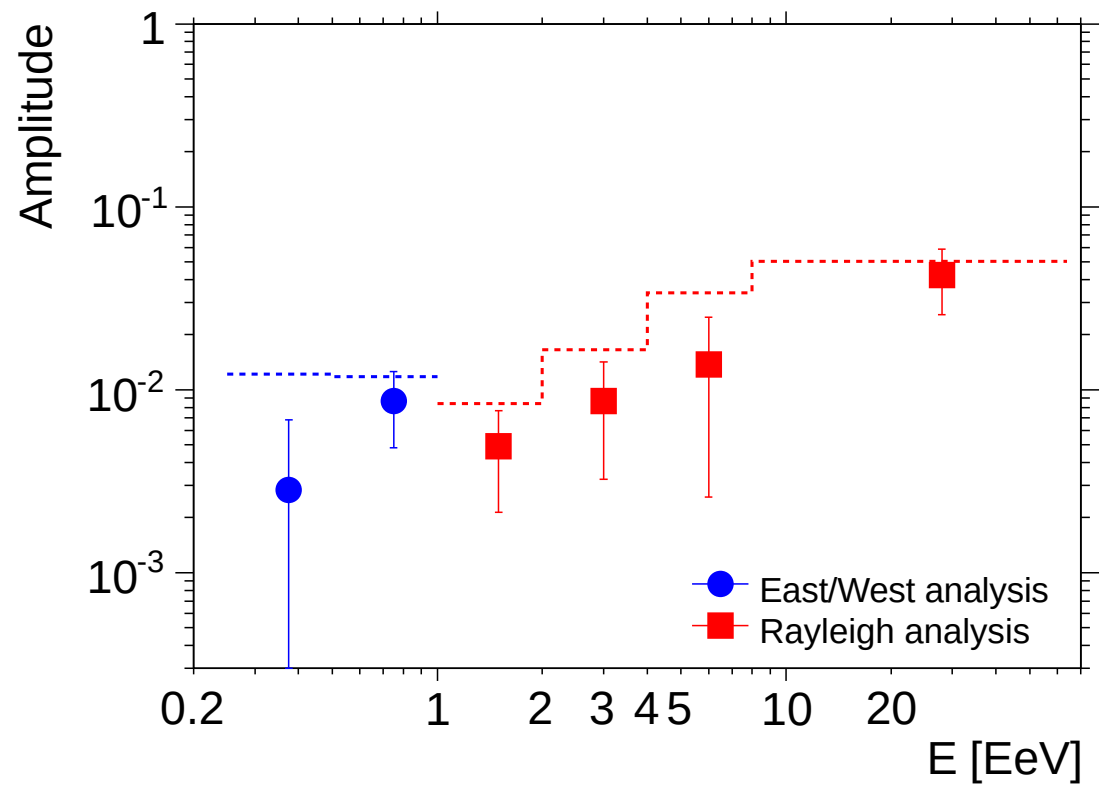

Figure 2: Amplitude of the first harmonic as a function of energy. The dashed line indicates the 99% $C.L.$ upper bound on the amplitudes that could result from fluctuations of an isotropic distribution.

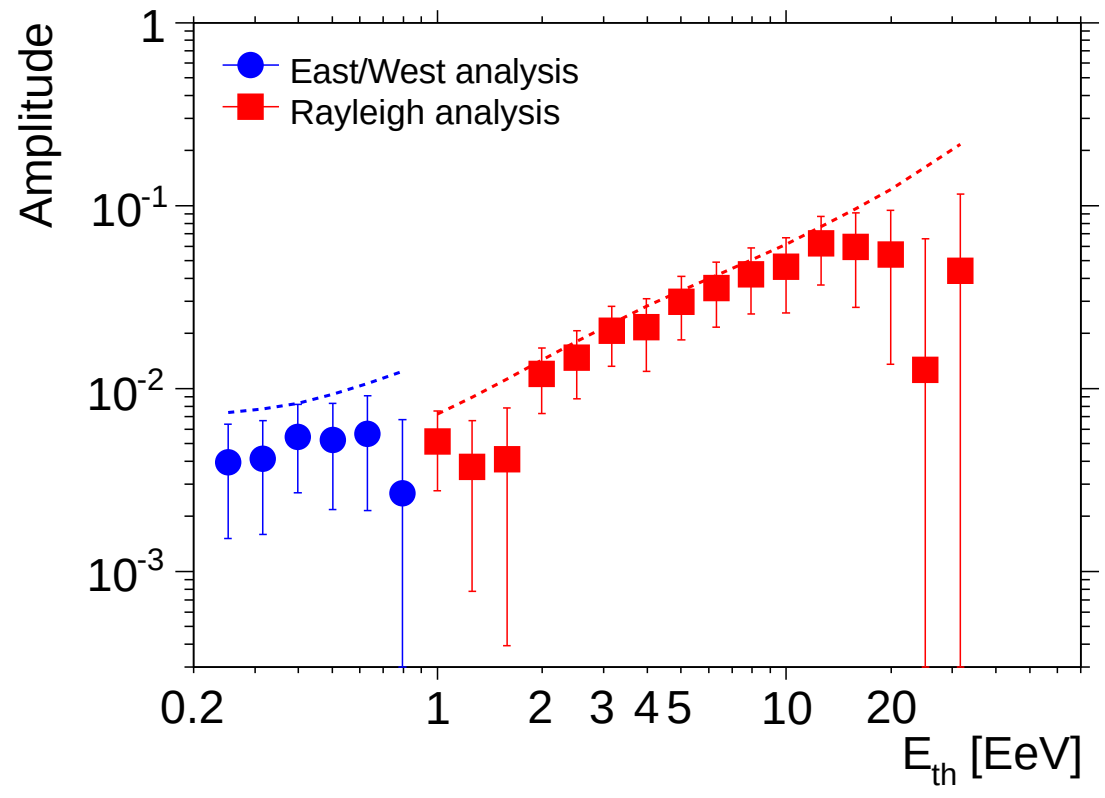

Figure 3: Same as Fig. 2, but as a function of energy thresholds.

Results of the analysis performed in terms of energy thresholds (strongly correlated bins) are shown in Fig. 3. They provide no further evidence in favor of a significant amplitude.

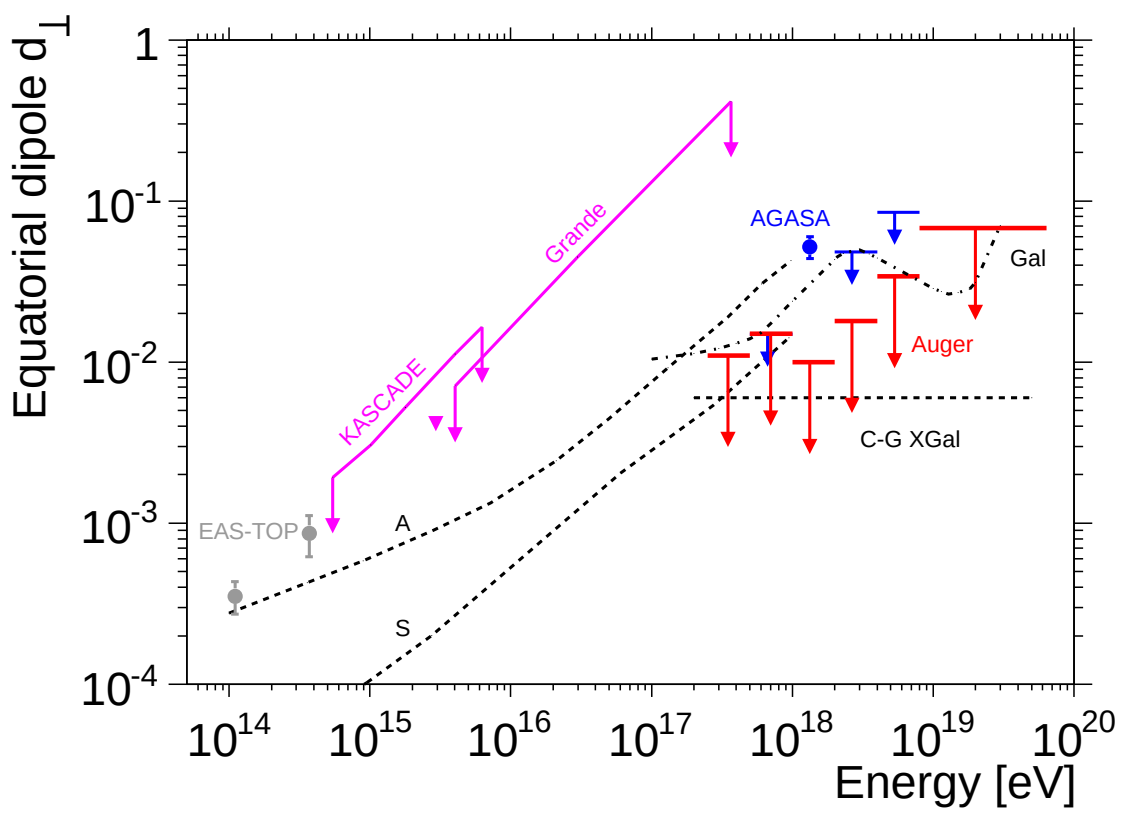

Figure 4: Upper limits on the anisotropy : equatorial dipole component $d_\perp$ as a function of energy from this analysis. Results from EAS-TOP, AGASA, KASCADE and KASCADE-Grande experiments are also displayed, in addition to several predictions (see text).

## 3 Upper limits

From the analyses reported in the previous Section, upper limits on amplitudes at 99% $C.L.$ can be derived according to the distribution drawn from a population characterised by an anisotropy of unknown amplitude and phase as derived by Linsley [12]. The Rayleigh amplitude measured by an observatory depends on its latitude and on the range of zenith angles considered. The measured amplitude can be related to a real equatorial dipole component $d_\perp$ by $d_\perp \simeq r/\langle\cos\delta\rangle$, where $\delta$ is the declination of the detected events, allowing a direct comparison of results from different experiments and from model predictions [1]. The upper limits on $d_\perp$ are shown in Fig. 4, together with previous results from EAS-TOP [13], KASCADE [14], KASCADE-Grande [15] and AGASA [16], and with some predictions for the anisotropies arising from models of both galactic and extragalactic cosmic ray origin. In models $A$ and $S$ ($A$ and $S$ standing for 2 different galactic magnetic field symmetries) [3], the anisotropy is caused by drift motions due to the regular component of the galactic magnetic field, while in model $Gal$ [17], the anisotropy is caused by purely diffusive motions due to the turbulent component of the field. Some of these amplitudes are challenged by our current sensitivity. For extragalactic cosmic rays considered in model $C$-$G\,Xgal$ [18], the motion of our galaxy with respect to the CMB (supposed to be the frame of extragalactic isotropy) induces the small dipolar anisotropy (neglecting the effect of the galactic magnetic field).

## 4 Phase of first harmonic analyses

The phase of the first harmonic is shown in Fig. 5 as a function of the energy. While the measurements of the amplitudes do not provide any evidence for anisotropy, it does not escape our notice that these measurements suggest a smooth transition between a common phase of $\simeq 270°$ below 1 EeV and another phase (right ascension $\simeq 100°$) above 5 EeV. This is potentially interesting, because with a real underlying anisotropy, a consistency of the phase measurements in ordered energy intervals is indeed expected with lower statistics than that required for the amplitudes to significantly stand out of the background noise [19]. To quantify whether or not a parent random distribution of arrival directions reproduces the phase measurements in adjacent energy intervals better than an alternative dipolar parent distribution, we introduced a likelihood ratio test in our previous report [1]. When applied to data points of Fig. 5, this test leads to a probability of $\sim 10^{-3}$ to accept the random distribution compared to the alternative one. Since we did not perform an *a priori* search for such a smooth transition in the phase measurements, no confidence level can be derived from this result. With an independent data set of comparable size, we will be able to confirm whether this effect is real or not.

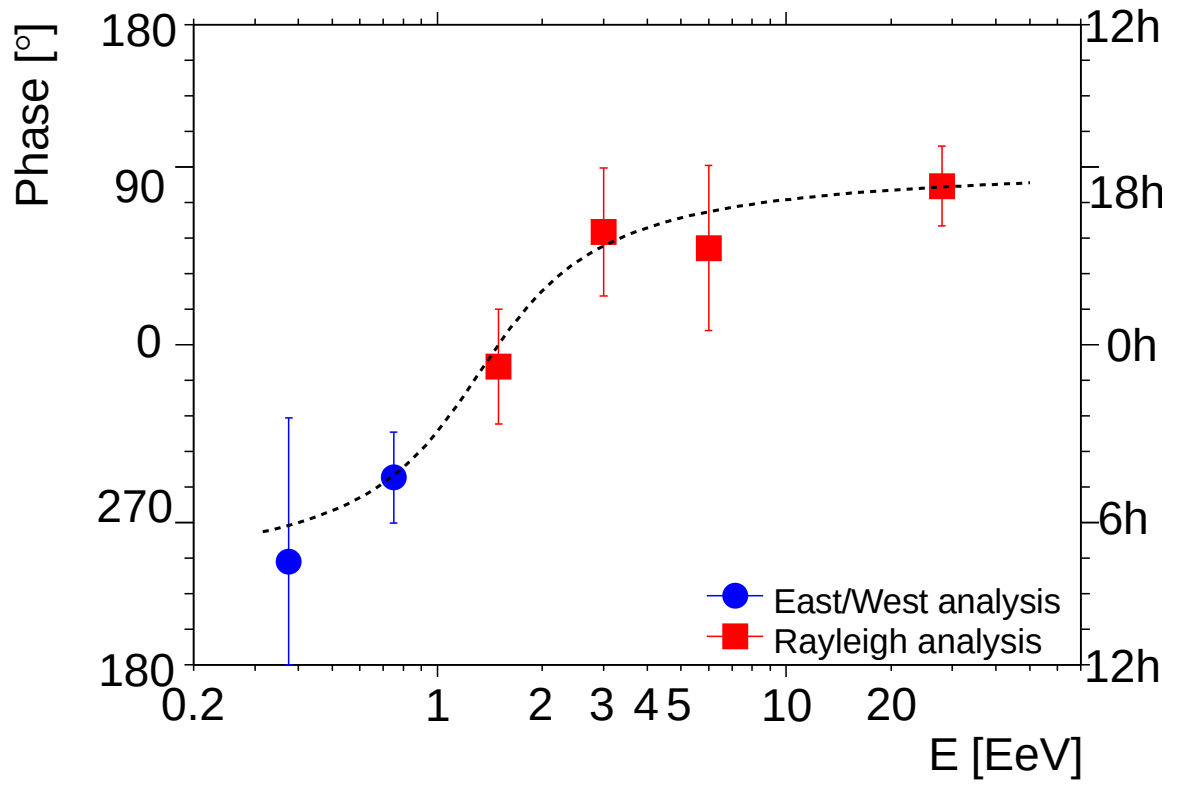

Figure 5: Phase of the first harmonic as a function of energy. The dashed line, resulting from an empirical fit, is used in the likelihood ratio test (see text).

It is important to note that an apparent constancy of phase, even though the significances of the amplitudes are relatively small, has been pointed out previously in surveys of measurements made in the range $10^{14} < E < 10^{17}$ eV [20]. A clear tendency for maxima to occur around 20 hours l.s.t. was stressed, not far from our own measurements in the energy range $2.5 \times 10^{17} < E < 10^{18}$ eV. Greisen *et al.* pointed out that most of these experiments were conducted at northern latitudes, and therefore regarded the reality of such sidereal waves as not yet established due to possible atmospheric effects leading to spurious waves. It is important that the Auger measurements are made with events coming largely from the southern hemisphere. In future analyses, we will benefit from the lower energy threshold now available at the Pierre Auger Observatory thanks to the infill array [21], allowing a better overlap with the energy ranges presented in Ref. [20]. Preliminary analyses of this data with the East-West method show also an apparent constancy of the phase.

# Influence of geomagnetic effects on large scale anisotropy searches

MORITZ MÜNCHMEYER[1] FOR THE PIERRE AUGER COLLABORATION[2]
[1]*Laboratoire de Physique Nucléaire et de Hautes Energies, Universités Paris 6 et Paris 7, CNRS-IN2P3, Paris, France*
[2]*Observatorio Pierre Auger, Av. San Martin Norte 304, 5613 Malargüe, Argentina*
*(Full author list: http://www.auger.org/archive/authors_2011_05.html)*
*auger_spokespersons@fnal.gov*

**Abstract:** We discuss the influence of the geomagnetic field on the energy estimate of extensive air showers with zenith angles smaller than $60^\circ$, detected with the Surface Detector array of the Pierre Auger Observatory. The geomagnetic field induces a modulation of the energy estimator, depending on the shower direction, at the $\sim 2\%$ level at large zenith angles. We present a method to account for this modulation in the reconstruction of the energy of the cosmic rays. We analyse the effect of the energy shift on large scale anisotropy searches in the arrival direction distributions of cosmic rays above the energy threshold at which the detection efficiency of the surface detector array is saturated (3 EeV). At a given energy, the geomagnetic effect is shown to induce a pseudo-dipolar pattern at the percent level in the declination distribution that needs to be accounted for before performing large scale anisotropy searches.

## 1 Introduction

The development of extensive air showers in the Earth's atmosphere is influenced by the geomagnetic field, which acts on the charged particles in the shower. This results in broadening of the spatial distribution of the particles in the direction of the Lorentz force. Current empirical models, used in the reconstruction of the primary energy and other parameters for showers with zenith angle $\theta < 60^\circ$ detected with the Surface Detector array of the Pierre Auger Observatory, assume a radial symmetry of the particle distribution in the plane perpendicular to the shower axis. The geomagnetic field induces a systematic effect on the energy estimate, depending on the angle between geomagnetic field and the shower direction. This effect is currently neglected in the measurement of the energy spectrum with the Pierre Auger Observatory based on showers with zenith angles smaller than $60^\circ$. This is reasonable since the magnitude of the effect is well below the statistical uncertainty of the energy reconstruction, which is of order $15\%$ [1]. However, in the search for large scale anisotropies at the percent level it induces a modulation of the measured cosmic ray event rate [2], resembling a true dipolar asymmetry in the North-South direction. The local magnetic field vector is approximately time independent, so this effect has no influence on a large scale anisotropy search in the right ascension distribution of cosmic rays [3, 4]. An analysis of the geomagnetic effect in the framework of horizontal air showers can be found in [5, 6].

## 2 Influence of the geomagnetic field on extensive air showers

The primary interaction of a cosmic ray in the atmosphere is followed by a hadronic cascade generating the muonic and electromagnetic shower components. The shower muons are produced by the decay of charged pions and have a typical energy $E_\mu$ of a few GeV. The production point of these muons is within tens of metres of the shower axis and their energy loss, mainly due to ionisation, is relatively small (about 2 MeV g$^{-1}$ cm$^2$). Unlike the electrons in the electromagnetic cascade, muons are weakly scattered and a large fraction of them reaches the ground. The geomagnetic effect will be therefore dominated by the action of the Lorentz force on the shower muons. In this analysis we treat the geomagnetic field **B** at the Pierre Auger Observatory site as a constant field

$$B = 24.6\,\mu\mathrm{T}, \qquad D_B = 2.6^\circ, \qquad I_B = -35.2^\circ, \quad (1)$$

$D_B$ and $I_B$ being the field's declination and inclination.

### 2.1 Distortion of the shower symmetry

Using a simple toy model we aim at understanding the main features of the muon density distortion induced by the geomagnetic field. In the absence of this field and neglecting scattering processes, a relativistic muon of energy $E_\mu$ and transverse momentum $p_\mathrm{T}$ that travels a distance $d$ will have

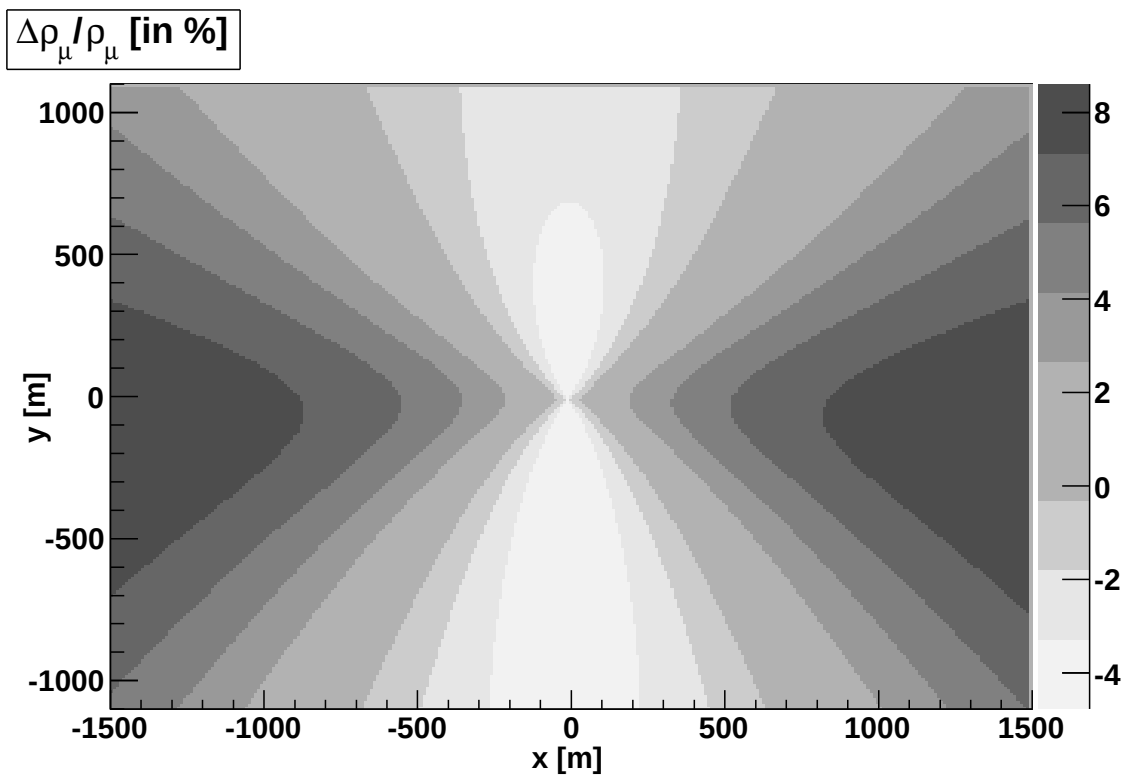

Figure 1: Relative changes of $\Delta\rho_\mu/\rho_\mu$ in the transverse shower front plane due to the presence of the geomagnetic field, for a zenith angle $\theta = 60^\circ$ and with the azimuth angle aligned along $D_B + 180^\circ$.

a radial deviation $r$ from the shower axis given by

$$r \simeq \frac{p_\mathrm{T}}{p_\mu}\, d \simeq \frac{c p_\mathrm{T}}{E_\mu}\, d. \tag{2}$$

The deflection of a relativistic muon in the presence of a magnetic field with transverse component $B_\mathrm{T}$ can be approximated with

$$\delta x_\pm \simeq \pm \frac{ecB_\mathrm{T} d^2}{2E_\mu}, \tag{3}$$

where the $x$-axis is oriented along the direction of the deflection. Given a muon density $\rho_\mu(x, y)$ in the shower plane in the absence of the geomagnetic field, the corresponding density $\overline{\rho}_\mu(\overline{x}, \overline{y})$ in the presence of the geomagnetic field is given by

$$\overline{\rho}_\mu(\overline{x}, \overline{y}) = \rho_{\mu_+}(\overline{x} - \delta x_+, \overline{y}) + \rho_{\mu_-}(\overline{x} - \delta x_-, \overline{y}), \tag{4}$$

neglecting a dependence of $\delta x_\pm$ on $x$ and $y$, which is only valid for a restricted range in $r$. Here we are interested in the shower size at $r \simeq 1000$ m, which is used to estimate the primary energy [8]. Assuming a symmetry in the distribution of positive and negative muons, we can further simplify this equation to

$$\overline{\rho}_\mu(\overline{x}, \overline{y}) \simeq \rho_\mu(\overline{x}, \overline{y}) + \frac{(\delta x)^2}{2} \frac{\partial^2 \rho_\mu}{\partial \overline{x}^2}(\overline{x}, \overline{y}), \tag{5}$$

where we used $\delta x = \delta x_+ = -\delta x_-$ and $\rho_{\mu_-} = \rho_{\mu_+} = \rho_\mu/2$. The geomagnetic field thus changes the muon density by a factor proportional to $B_\mathrm{T}^2(\theta, \varphi)$. This term describes the azimuthal behavior of the effect, as verified in the next section by Monte Carlo shower simulations.

### 2.2 Observation of the distortion

We illustrate the relative change $\overline{\rho}_\mu/\rho_\mu$ by shower simulations in the presence and in the absence of the geomagnetic field. A predominantly quadrupolar asymmetry is visible, corresponding to the separation of positive and negative charges in the direction of the Lorentz force (Fig. 1).

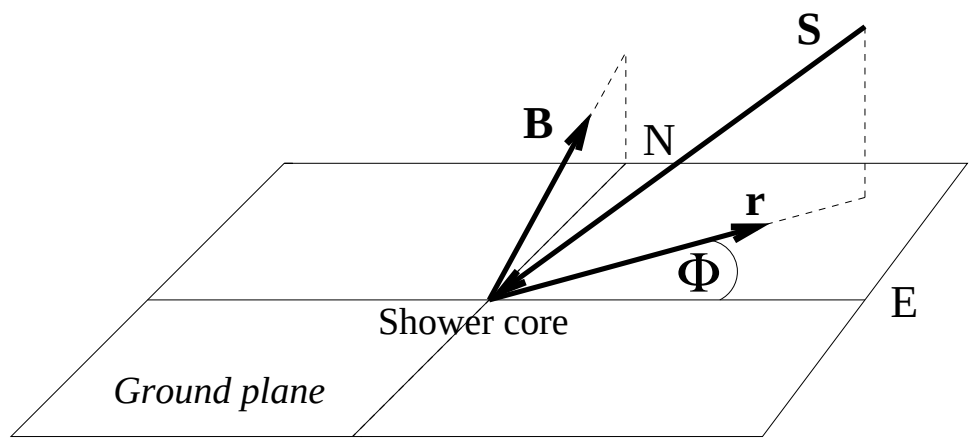

Figure 2: Definition of the polar angle $\Phi$, with respect to the shower core of a shower **S** and the magnetic East E.

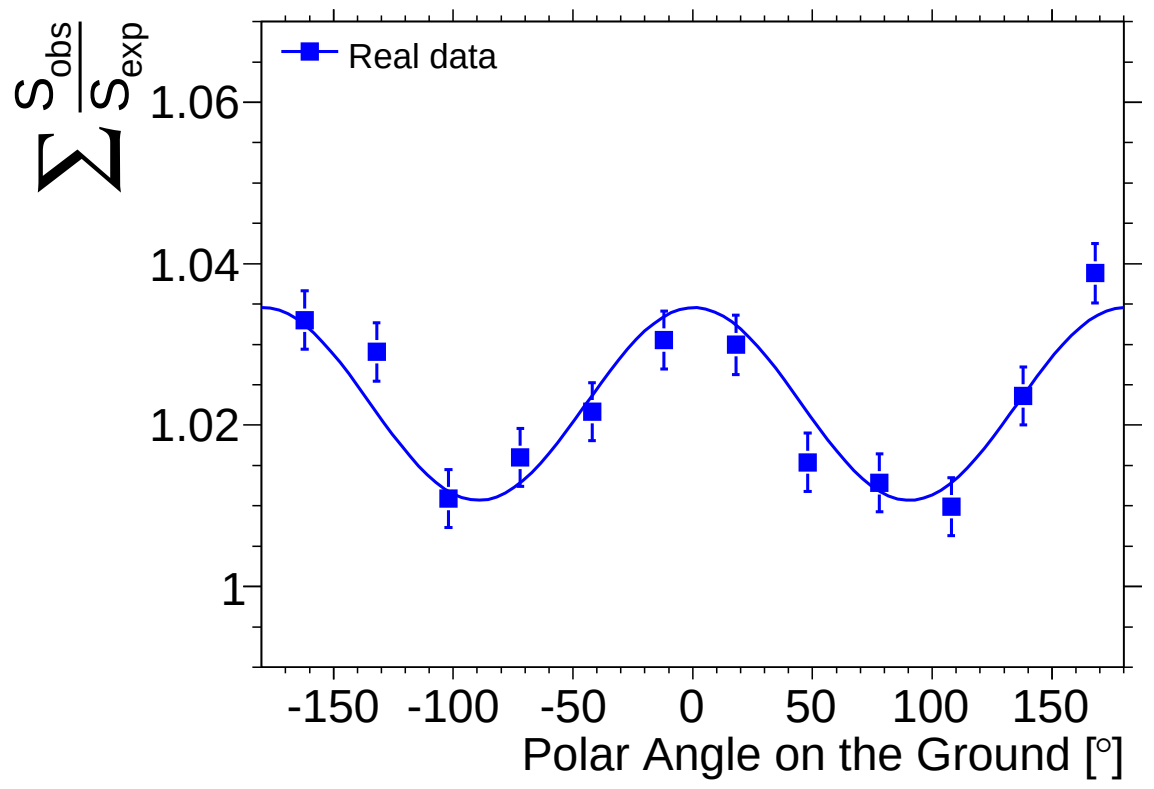

Figure 3: Ratio between observed and expected signal in the surface detectors (with radial distance $r$ to the shower core larger than 1000 m) as a function of the polar angle $\Phi$. The solid line is a fit of a quadrupolar modulation.

This effect is expected to manifest itself in a quadrupolar modulation of the surface detector signals as a function of the polar angle $\Phi$, defined with respect to the magnetic East as shown in Fig. 2. The ratio between observed and expected signal (which is calculated assuming radial shower symmetry) as a function of $\Phi$ is shown in Fig. 3, drawn from approximately 30 000 showers with energies larger than 4 EeV, observed by the Pierre Auger Observatory until December 2010 and passing standard fiducial cuts [7]. A significant quadrupolar modulation of $(1.2 \pm 0.2)\%$ is observed in the data. Its origin can be ascribed to the geomagnetic field, as was verified by an end-to-end Monte Carlo simulation, that was constructed to be similar to the real data in terms of shower energies and arrival directions. The quadrupolar amplitude in the case of simulations is $(1.1 \pm 0.2)\%$ in the presence of the geomagnetic field (phase consistent with the real data case) and $(0.1 \pm 0.2)\%$ in its absence. Details of this analysis can be found in [9].

## 3 Geomagnetic distortions of the energy estimator

The energy estimates of showers detected with the Surface Detector array are done in a three step procedure [8]. First,

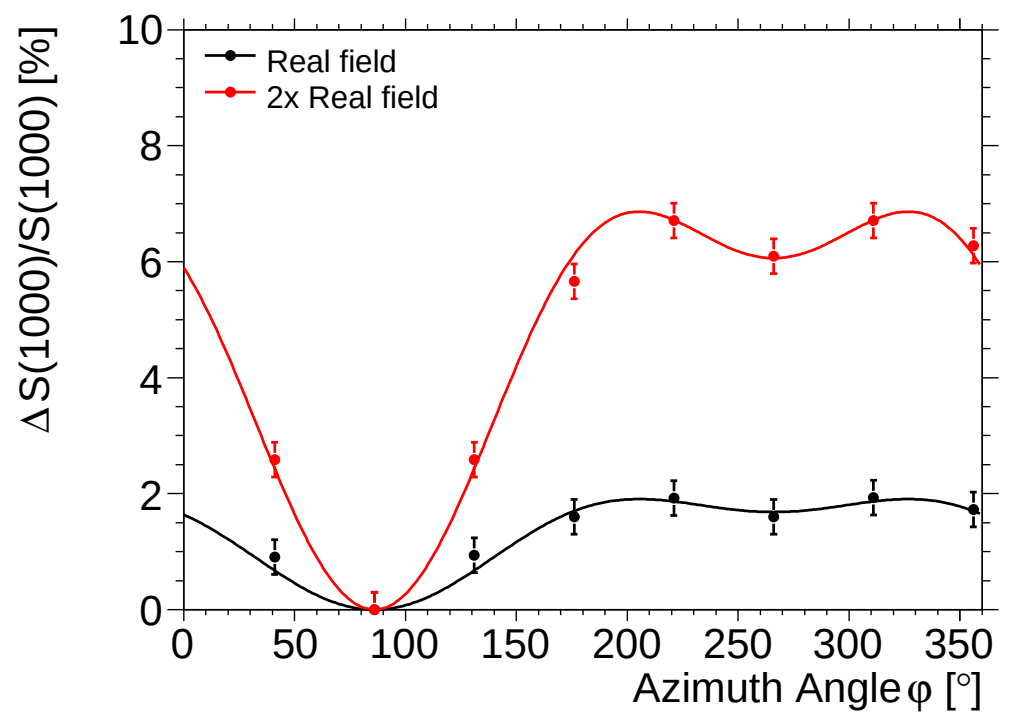

Figure 4: $\Delta S(1000)/S(1000)$ (in %) as a function of the azimuth angle $\varphi$, at zenith angle $\theta = 55°$.

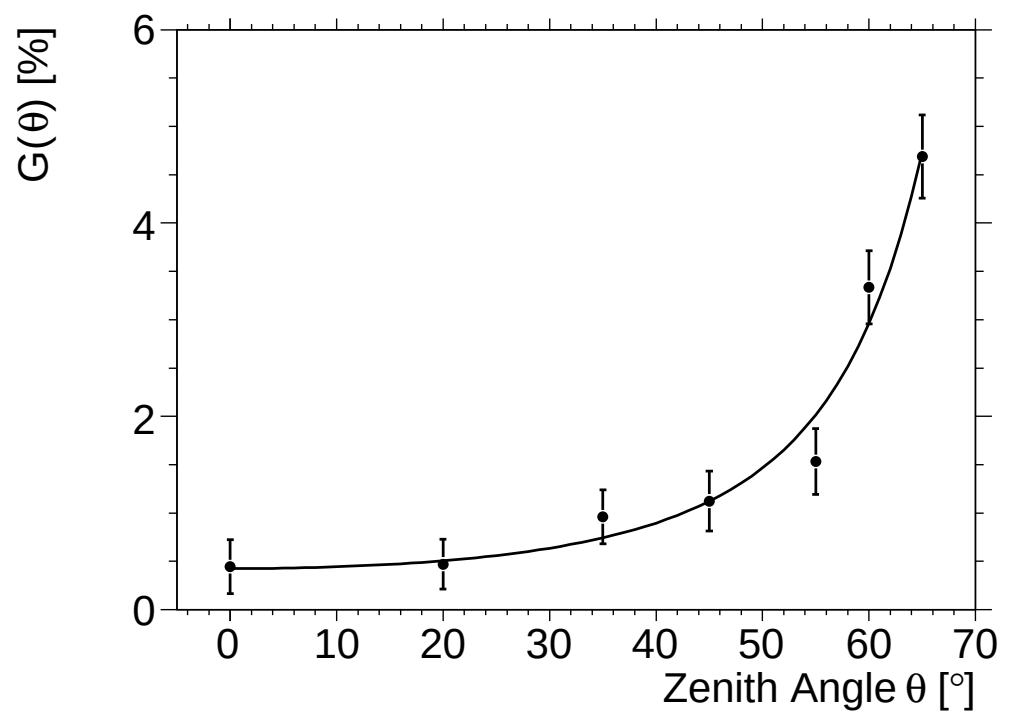

Figure 5: $G(\theta) = \Delta S(1000)/S(1000)/\sin^2(\widehat{\mathbf{u},\mathbf{b}})$ as a function of zenith angle $\theta$.

the shower size $S(1000)$ at 1000 m from the shower core is calculated, by fitting the lateral distribution function to the detector signals. Then the dependence of $S(1000)$ on the zenith angle, arising from the attenuation of the shower in the atmosphere and from the surface detector geometry is quantified by applying the *Constant Intensity Cut* (CIC) method, resulting in the shower size $S_{38}$ at reference zenith angle $\theta = 38°$. The conversion of $S_{38}$ to energy $E$ is then achieved using a relation of the form $E = AS_{38}^B$ which is calibrated using hybrid events that have an independent energy measurement from the Fluorescence Detector [1].

As predicted by the toy model above, the shower size $S(1000)$ shifts proportionally to $B_\mathrm{T}^2 \propto \sin^2(\widehat{\mathbf{u},\mathbf{b}})$, the hat notation denoting the angle between shower direction **u** and magnetic field direction **b**. We verified this prediction by simulating sets of showers with fixed directions, each set containing 1000 showers simulated in the presence and in the absence of the geomagnetic field. The showers were generated with the AIRES program [10] and the hadronic interaction model QGSJET, simulating protons of 5 EeV. We find a systematic shift of the reconstructed $S(1000)$ values, that follows the predicted azimuthal behavior (Fig. 4).

The zenithal behavior of the $S(1000)$ shift depends on the muon distribution properties and cannot be predicted by the toy model. To obtain it by a Monte Carlo calculation, we created further sets of 1000 showers, for different zenith angles. The result is shown in Fig. 5, where the superimposed curve $G(\theta)$ is an empirical fit of the data points. To obtain the pure zenithal dependency, the $S(1000)$ shift was divided by $\sin^2(\widehat{\mathbf{u},\mathbf{b}})$.

Placing the azimuthal and the zenithal dependence together, we arrive at a parametrisation of the geomagnetic shower size shift given by

$$\frac{\Delta S(1000)}{S(1000)}(\theta,\varphi) = 4.2 \times 10^{-3} \cos^{-2.8}(\theta) \sin^2(\widehat{\mathbf{u},\mathbf{b}}). \tag{6}$$

Note that these results were obtained by simulating protons of 5 EeV. It is shown in [9] that the above parametrisation depends only weakly on energy, composition and the hadronic interaction model used in the simulations.

Part of the zenithal shift of $S(1000)$ induced by the geomagnetic field is already corrected for by the CIC procedure, which assumes a uniform flux. By construction the CIC averages over the azimuthal variation. We therefore obtain the following correction formula that gives the reconstructed energy $E$ in terms of the value $E_0$ that is reconstructed if the effect of the geomagnetic field is not accounted for

$$E = \frac{E_0}{(1+\Delta(\theta,\varphi))^B}, \tag{7}$$

with

$$\Delta(\theta,\varphi) = G(\theta)\left[\sin^2(\widehat{\mathbf{u},\mathbf{b}}) - \left\langle \sin^2(\widehat{\mathbf{u},\mathbf{b}})\right\rangle_\varphi\right] \tag{8}$$

where $\langle\cdot\rangle_\varphi$ denotes the average over $\varphi$, taking the influence of the CIC procedure into account and $B$ is one of the parameters used in the $S_{38}$ to $E$ conversion described above.

## 4 Consequences for large scale anisotropy searches

The influence of the geomagnetic effect on large scale anisotropy analysis is caused by the angular dependence of the energy estimate, that translates into a shift in the measured event rate at a fixed estimated energy.

### 4.1 Impact on the event rate

Above 3 EeV, the surface array has full acceptance, so the exposure is geometrical [7] and given by

$$\omega(\theta) \propto \cos(\theta)\, H(\theta - \theta_\mathrm{max}) \tag{9}$$

where $H$ is the Heaviside step function that imposes a maximum observed zenith angle $\theta_\mathrm{max}$. The event rate at a given declination $\delta$ and above an energy threshold $E_\mathrm{th}$ is obtained by integrating in energy and right ascension $\alpha$

$$N(\delta) \propto \int_{E_\mathrm{th}}^{\infty} \mathrm{d}E \int_0^{2\pi} \mathrm{d}\alpha\, \omega(\theta)\, \frac{\mathrm{d}N(\theta,\varphi,E)}{\mathrm{d}E} \tag{10}$$

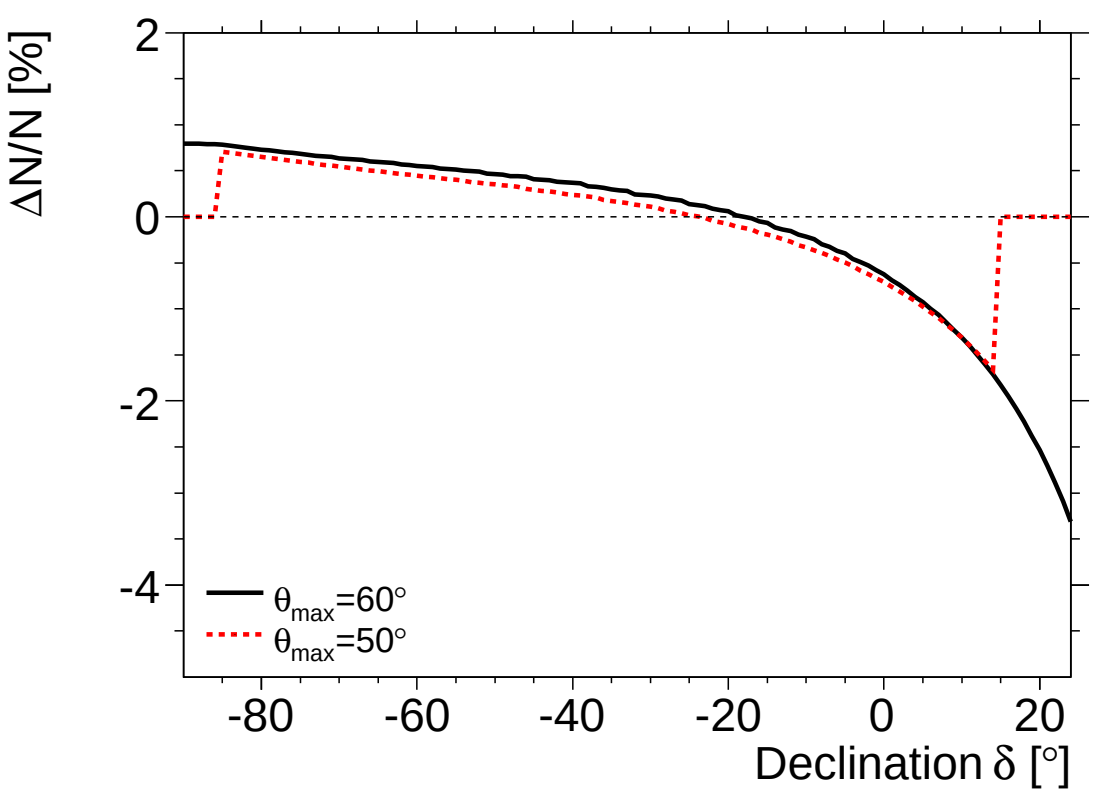

Figure 6: Relative differences $\Delta N/N_{\rm corr}$ as a function of the declination.

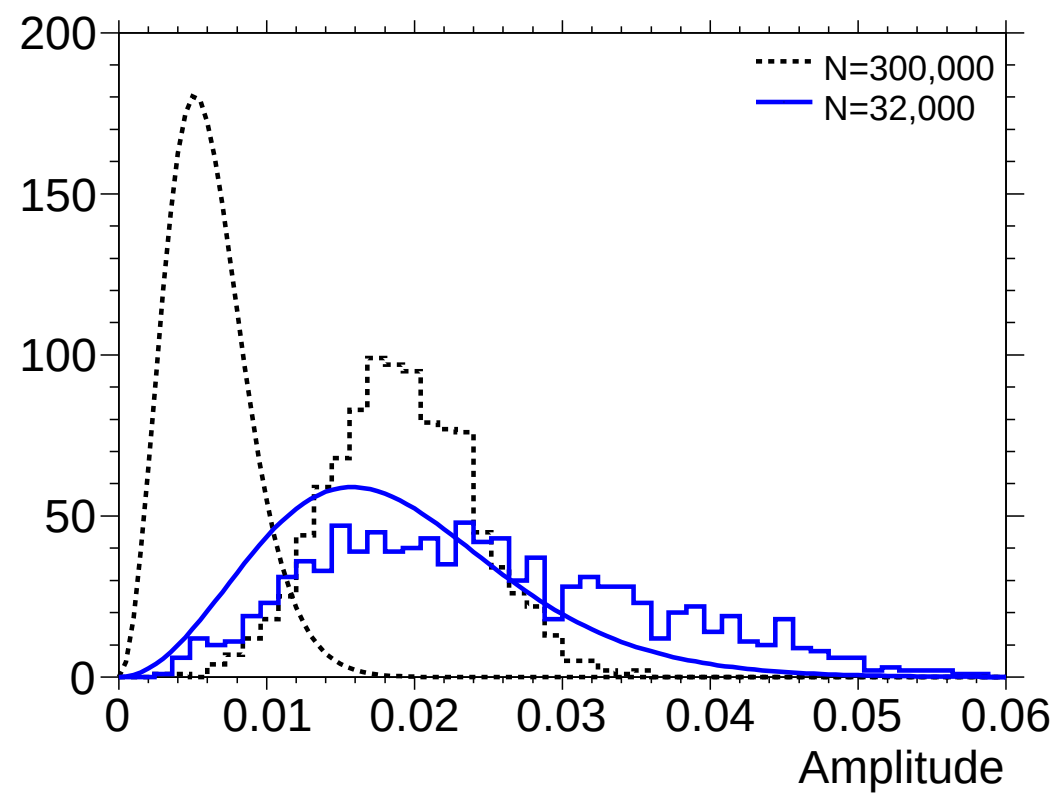

Figure 7: Two distributions of dipolar amplitudes reconstructed from arrival directions of mock data sets whose event rate is distorted by the geomagnetic effect, and their expected isotropic distribution.

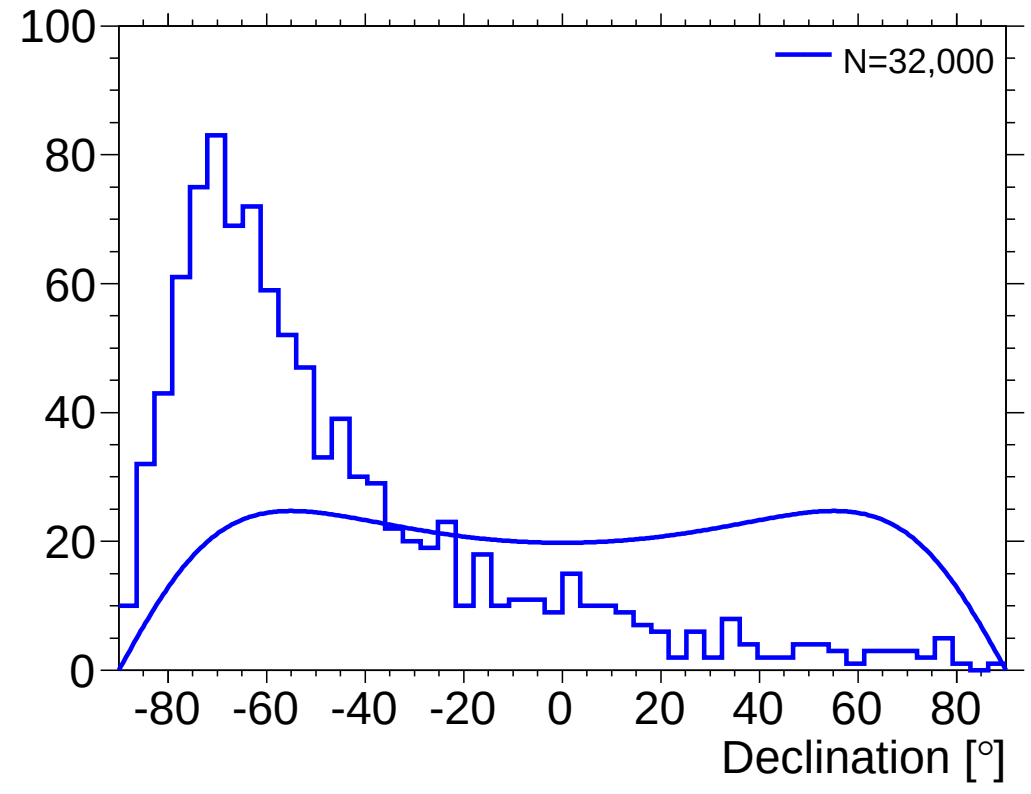

Figure 8: Distribution of reconstructed dipolar declinations and expected isotropic distribution.

We assume that the cosmic ray spectrum is a power law, i.e. $dN/dE \propto E^{-\gamma}$. From Eqn. (7) it follows that if the effect of the geomagnetic field were not accounted for, the spectral distribution would have a directional modulation given by

$$\left(\frac{\mathrm{d}N}{\mathrm{d}E}\right)_0 \propto [1+\Delta(\theta,\varphi)]^{B(\gamma-1)} \times E_0^{-\gamma}, \qquad (11)$$

The event rate $N_0(\delta)$ as a function of declination is then calculated using Eqn. (11) in Eqn. (10). The relative difference $\Delta N/N$ is shown in Fig. 6 as a function of the declination, with spectral index $\gamma = 2.7$. It corresponds to the deviation from isotropy that would be observed above a fixed energy threshold if the geomagnetic effect were not accounted for in the reconstruction of the energy. The pattern displayed in Fig. 6 is similar to that produced by a dipole anisotropy in North-South direction with an amplitude at the percent level.

### 4.2 Impact on dipolar anisotropy searches

To study the effect of the modulation in the energy estimator on dipolar anisotropy searches we drew samples of simulated data from the ”uncorrected” event rate $N_0(\delta)$. For the dipolar anisotropy search we used the method described in [11], which is adapted to a partial sky coverage. The results of the reconstructed dipolar amplitudes for 1000 mock data sets are shown in Fig. 7, for two different sample sizes. The expected isotropic distribution is plotted in the curves with solid lines. Its analytical expression is derived in [9]. For $N = 300\,000$ events, we find a strong deviation from the expected distribution. The condition $N = 32\,000$ is the number of events, for which the mean of the histogram is of the same order as the mean noise amplitude from the isotropic distribution.

In addition to having a non-isotropic amplitude distribution, the reconstructed dipole is preferentially oriented towards the South. For $N = 32\,000$ events, the declination distribution is shown in Fig. 8.

# Measurement of Energy-Energy-Correlations with the Pierre Auger Observatory

PETER SCHIFFER[1, 2] FOR THE PIERRE AUGER COLLABORATION[3]
[1]*III. Physikalisches Institut A, RWTH Aachen University, Otto-Blumenthal-Str., 52074 Aachen, Germany*
[2]*II. Institut für Theoretische Physik, Universität Hamburg, Luruper Chausee 149, 22761 Hamburg, Germany*
[3]*Observatorio Pierre Auger, Av. San Martín Norte 304, 5613 Malargüe, Argentina*
*(Full author list: http://www.auger.org/archive/authors_2011_05.html)*
*auger_spokespersons@fnal.gov*

**Abstract:** To investigate energy ordering effects close to the most energetic cosmic rays, a measurement of Energy-Energy-Correlations (EEC) has been performed with the data of the surface detector of the Pierre Auger observatory. The measurement includes the ultra high energy cosmic rays (UHECRs) with energies above E=5 EeV arriving within a small solid angular region around UHECRs with $E > 60$ EeV. The measured EEC distribution is compared to the expectation for isotropic arrival directions of UHECRs.

## 1 Introduction

The origin of ultra-high energy cosmic rays (UHECRs) is unknown. While there are strong hints that UHECRs are accelerated in discrete sources [1] and that the source distribution follows the large scale structure of the universe [2, 3], it has not been possible so far to identify individual sources. The main obstacle has been that galactic and extragalactic magnetic fields (GMF & EGMF) deflect the UHECRs as they propagate. On the other hand, the deflection offers a chance to constrain these magnetic fields. In many source and magnetic field scenarios, a characteristic energy-ordering of the arrival directions relative to the source position is expected.

The Energy-Energy-Correlation (EEC) distribution is an observable that is sensitive to such energy-ordering effects and was recently proposed as an observable sensitive to turbulent cosmic magnetic fields [4]. The EEC is a quantity originally developed in high energy physics analyses for testing strong interaction phenomena (e.g. [5, 6, 7]).

## 2 Energy-Energy-Correlations

In this section the EEC is introduced and its properties are studied for a scenario in which the arrival directions are isotropically distributed.

### 2.1 Definition

The EEC distribution is calculated as described in [4], except from one adjustment accounting for the field of view of the Pierre Auger observatory.

Firstly, regions of interest (ROI) are defined as cones with an opening angle of $0.2$ rad, centered around each UHECR with an energy above 60 EeV. A cone jet-algorithm is then applied to these ROIs:

1. The "center of mass" of the UHECRs within each ROI is calculated using the arrival directions weighted by the UHECR energies and the inverse exposure at each arrival direction.

2. Each ROI is moved to the corresponding center of mass.

3. The algorithm concludes after three iterations.

The Energy-Energy-Correlation $\Omega$ is calculated for every pair of UHECRs within each ROI using

$$\Omega_{ij} = \frac{(E_i - \langle E_i(\alpha_i)\rangle) \cdot (E_j - \langle E_j(\alpha_j)\rangle)}{E_i \cdot E_j}. \qquad (1)$$

Here $E_i$ is the energy of an UHECR $i$ with an angular distance $\alpha_i$ to the center of the ROI. $\langle E_i(\alpha_i)\rangle$ is the average energy of all UHECRs at the angular distance $\alpha_i$ from the centers of the ROIs.

The angular distribution of the EEC is determined by averaging over all $\Omega_{ij}$ calculated in all ROIs. Each $\Omega_{ij}$ is taken into account twice, once at the angular distance $\alpha_i$ and once at $\alpha_j$.

## 2.2 Isotropic Expectation

Here we evaluate the EEC distribution for a scenario in which the arrival directions are isotropic. Later, we will compare this to the distribution obtained from data (see section 3.3 below).

An isotropic set is realized using 18744 directions and the same energy spectrum as the Auger data above 5 EeV. To achieve this, the original energies of the actual events (see section 3.1) are reassigned to new directions according to the acceptance [8] of the Pierre Auger Observatory.

The average EEC distribution of 100 isotropic data sets is shown in Figure 1 as the black triangular symbols. The error bar denotes the RMS of these realizations. In this distribution two distinct features can be seen.

Firstly, there is a plateau at larger angles which is characteristic of the energy range and spectrum used. The level can be estimated by removing the angular dependence from (1) and calculating the expectation value

$$\langle \Omega_{ij} \rangle = \left( 1 - \langle E \rangle \left\langle \frac{1}{E} \right\rangle \right)^2 . \qquad (2)$$

Secondly, an increase towards smaller angles is observed. This is caused by the jet algorithm used for the determination of the ROI (see section 2.1), which leads to a systematic overdensity of the most energetic events near the center of the ROIs. This effect increases the average value of the EEC at smaller angles.

A signal of energy-ordering of events would be a broader increase of the distribution, beyond that seen for isotropic sets, near the center of the ROI. The shape will depend on the scale of the ordering.

# 3 Data Analysis

The Pierre Auger Observatory is a hybrid air shower detector located in Malargüe, Argentina. The Surface Detector (SD) consists of a $3000\,\mathrm{km}^2$ array of 1660 surface detectors overlooked by the 27 fluorescence telescopes of the Fluorescence Detector (FD) grouped at 4 sites on the array boundary. This allows for complementary measurements of the lateral distribution of air shower particles at ground level by the SD and the longitudinal development of the air shower by the FD.

## 3.1 Event Selection

For the measurement of the EEC, all events with energies above 5 EeV measured between 1 January 2004 and 31 December 2010 SD are used. These event energies are above the so-called spectral ankle [9], and can thus be reasonably hypothesized to be of extragalactic origin [10].

The following additional cuts are applied to the SD data set:

- A reconstructed zenith angle of less than $60°$.
- The SD tank with the highest signal has to be surrounded by 6 operating tanks during the time the UHECR is measured.
- Time periods in which the data acquisition was unstable are excluded. These are associated to unavoidable problems in the construction phase, or more generally to hardware instabilities [11].

This results in a set of $N_{events} = 18744$ UHECRs.

## 3.2 Experimental Uncertainties

In this section the relevant experimental uncertainties of the Pierre Auger Observatory are discussed and propagated to the EEC analysis. The error propagation is performed either by variation of the data itself, if possible, or with Monte Carlo (MC) data of isotropic arrival directions. Since the EEC distribution depends on the total number of events in the data set, this number is fixed for all the studies performed below.

### 3.2.1 Energy Resolution

Events with energies larger than 3 EeV are measured by the SD with an energy resolution of 14.8% [13]. To model the effect on the EEC distribution the energies of the events are varied or "smeared" by a Gaussian with this width. By varying all Auger events, including those below 5 EeV, events may cross the imposed threshold from above or from below. Due to the steep spectrum, this variation will slightly increase the number of events exceeding the threshold. To keep the number of events fixed, events are randomly removed.

This variation of the energies is performed 100 times and each time the EEC distribution is calculated. The RMS of these distributions then is considered to be the statistical uncertainty resulting from the energy resolution.

### 3.2.2 Angular Resolution

The angular resolution of the SD is better than $1°$ for energies above 5 EeV [12], where the angular resolution is given in terms of the 68% quantile of a two-dimensional Gaussian distribution. The angular resolution is propagated in the same way as in section 3.2.1. The RMS of 100 data sets gives the statistical uncertainty resulting from the angular resolution.

### 3.2.3 Absolute Energy Scale

The energy measurement of the SD is calibrated using the fluorescence detector of the Pierre Auger Observatory [9].

There is a systematic uncertainty of 22% on the overall energy scale. In order to keep the number of UHECRs constant the corresponding uncertainty has been studied using 100 isotropic MC data sets of 70000 UHECRs above 3 EeV. The MC set has a spectrum according to [9] and the geometrical coverage of the Pierre Auger observatory given in reference [8], using a latitude of $-35°$ S for the observatory site. In a second step, the following three types of data sets are produced:

- For the nominal value of energy, the first $N_{events}$ above 5 EeV are taken from each data set.
- All energies are shifted up by 22%, then the first $N_{events}$ above 5 EeV are taken from each data set.
- All energies are shifted down by 22%, then the first $N_{events}$ above 5 EeV are taken from each data set.

The average of the EEC distributions of the unshifted data sets is taken as the mean value and the average of the shifted EEC distributions as the uncertainty of the mean. The relative uncertainties of this MC study are used to quantify the effect of a systematic shift of the energy scale.

#### 3.2.4 Detector Acceptance

The detector acceptance has no direct influence on the measurement of the EEC distribution, but the effects are important for a comparison with models of UHECR propagation like the one performed in section 2.2. At energies larger than 5 EeV the Pierre Auger observatory has reached full trigger efficiency [11], so a geometrical acceptance model [8] can be assumed. Historically, data was taken while the Observatory was being constructed. The effect from the growing SD before 2008 and from bad periods of operation is much smaller than the uncertainty from the angular and energy resolutions. Therefore it is sufficient to use a geometrical acceptance model for MC comparisons.

### 3.3 Measurement of the Energy-Energy-Correlations

The measurement of Energy-Energy-Correlation in the data set as defined in section 3.1 is shown in Figure 1 by the red circular symbols.

The statistical uncertainty has been determined as described above (sections 3.2.1 and 3.2.2). The arrival directions and the energies of the UHECRs have been varied simultaneously by their respective uncertainty. The RMS of the average distribution is the statistical uncertainty denoted by the error bars. The systematic uncertainty is calculated as described in section 3.2.3, by varying the energy scale of isotropic MC data sets. It is denoted by the blue error band.

For comparison the expectation from isotropically arriving UHECRs with the same energy spectrum as the data set is shown as the black triangular symbols. The error bars indicate the RMS of 100 realizations.

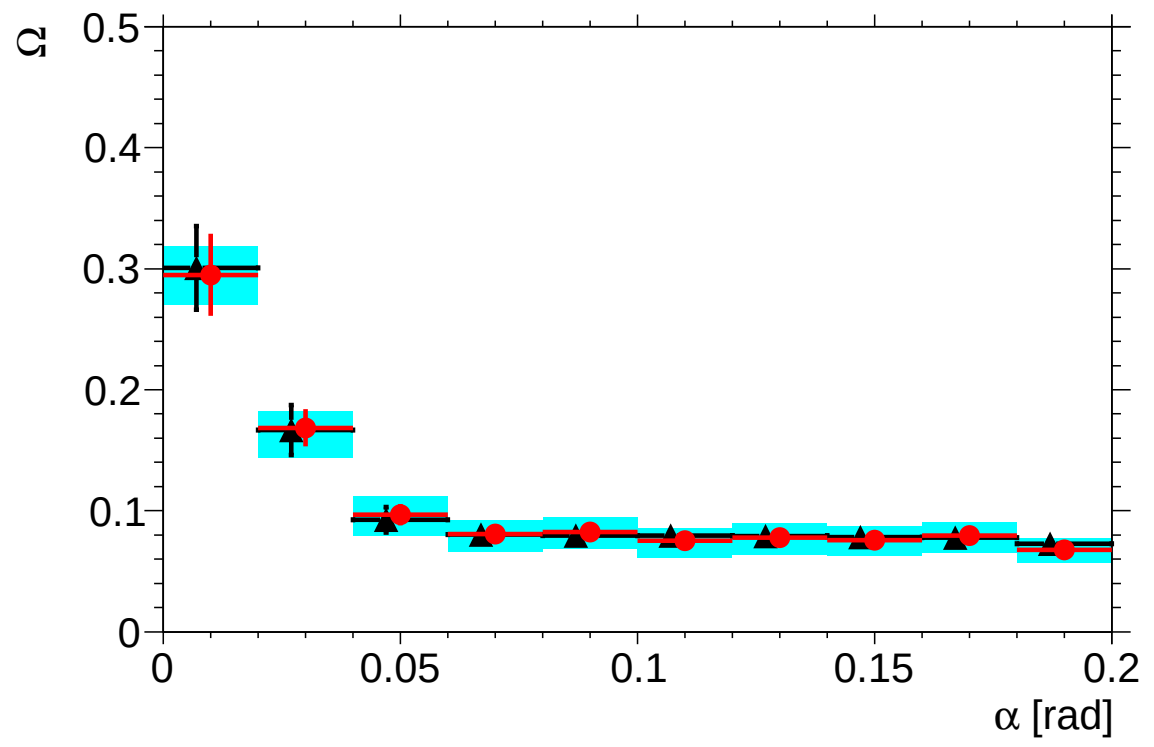

Figure 1: Measurement of the EEC distribution by the Pierre Auger Observatory (red circular symbols). The error bars denote the statistical uncertainty from angular and energy resolution effects, the band denotes the systematic uncertainty from the overall energy scale. For comparison an EEC distribution from a simulated data set with isotropic arrival directions is shown (black triangular symbols). The error bars denote the RMS of 100 realizations.

### 3.4 Discussion

As can be seen in figure 1 the measured EEC distribution is compatible with the expectation from isotropic arrival directions. This means, in particular, that in this analysis no energy-ordered deflections are observed near the most energetic UHECRs. Such a distribution can be caused either by a high source density for an isotropic source distribution or by large deflections of the UHECRs in cosmic magnetic fields.

## 4 Conclusions

The observable $\Omega$ of the Energy-Energy-Correlations has been used to investigate the strength of energy-ordering effects close to the most energetic UHECRs above $E = 60$ EeV. The measurement presented in this contribution includes UHECRs above 5 EeV arriving within regions of interest (ROI), each of size 0.2 rad, near the most energetic events. The average value of $\Omega$ has been measured as a function of the angular distance in each ROI. In this measurement no energy ordering has been found.

# Back-tracking studies of the arrival directions of UHECR detected by the Pierre Auger Observatory

MICHAEL S. SUTHERLAND[1], FOR THE PIERRE AUGER COLLABORATION[2]
[1]*Louisiana State University, Baton Rouge, LA 70803-4001, United States*
[2]*Observatorio Pierre Auger, Av. San Martín Norte 304, 5613 Malargüe, Argentina*
*(Full author list: http://www.auger.org/archive/authors_2011_05.html)*
*auger_spokespersons@fnal.gov*

**Abstract:** The Pierre Auger Observatory has performed precise measurements of ultra-high energy cosmic rays (UHECRs) which carry information on their source and propagation environments. We present an analysis that explores various features of the geometry and intensity of magnetic fields that influence the trajectories of ultra-high energy cosmic rays. Under the hypothesis of pure primary protons at the energy of interest (i.e., above 50 EeV), using different catalog-based assumptions on sources and a variety of simplistic Galactic magnetic field (GMF) models, we backtrack the arrival directions of UHECRs observed with the Pierre Auger Observatory. We quantify self-consistency by comparing to results for a simulated set of isotropic directions. The obtained results represent an illustrative example of the potential of UHECR data for obtaining information on their origin, their propagation, and on the properties of the GMF.

## 1 Introduction

The observation of cosmic rays at the highest energies (UHECRs) and the precise measurement of their kinematic properties with the Pierre Auger Observatory enables approaches towards the fundamental questions on the UHECR origin and propagation. As the physics of particle propagation is well established through laboratory studies, comparison of UHECR measurements with astrophysical models constitutes a promising method for obtaining information on their sources and characterization of the magnetic fields traversed from those sources.

In this contribution, we present a first comparison of the Pierre Auger dataset with specific astrophysical models using an analysis method previously shown in [1] which has been extended in [2]. For a variety of hypothesis sets each comprised of Galactic magnetic field (GMF) models and UHECR sources, we perform a search for hypothesis set self-consistency when folded with the Pierre Auger measurements. The primary focus of this analysis is the GMF, whose properties have been measured previously using Faraday rotation and starlight polarization techniques [3, 4, 5, 6, 7]. Observed UHECRs are backtracked as single-charged particles through simple GMF geometries. In order to quantify a self-consistent hypothesis set we compare the resulting particle trajectories with source candidates using different catalogues and test this source correlation against the expectation of backtracked isotropic simulations.

The paper is structured as follows. We first introduce the method for determining hypothesis set self-consistency. We then specify the data set recorded by the Pierre Auger Observatory. In the third section we explain the different astrophysical model components of catalog-based sources and GMF parameterizations. The fourth section contains results obtained by scanning the different models in comparison to the hypothesis of isotropic expectations. Finally, we discuss the effects of the measurement uncertainties and the influence of additional magnetic turbulent components on a representative hypothesis set.

## 2 Method Description

The Field Scan Method (FSM) [2] assesses the self-consistency of a set of UHECR and GMF hypotheses. It compares the correlation between source locations and event trajectories after backtracking both a dataset of interest (DOI) and isotropic simulations. The results of the method are explicitly dependent on the choice of the individual components. The comparison with isotropy accounts for chance correlation with sources as well as for lensing due to the GMF configuration, building on the procedure used in [1]. The test statistic (TS)

$$\Psi_i = \Theta_i/(1+\Delta_i) \tag{1}$$

is computed for each event $i$, where $\Delta_i$ is the deflection magnitude and $\Theta_i$ the angular distance to the nearest source object. A Kolmogorov-Smirnov (KS) test is performed between the DOI and isotropic TS distributions, using the largest signed difference $D_{\max}$ between the cumulative TS distributions. $|D_{\max}|$ maps to the probability $P_{\mathrm{KS}}$ that the TS distributions are drawn from the same parent distribution.

The hypothesis set is deemed *self-consistent* to the extent that the $P_{\mathrm{KS}}$ value indicates *inconsistency* with isotropy and that the DOI correlates well with the source hypothesis. A large positive $D_{\max}$, resulting in small $P_{\mathrm{KS}}$, located at a small TS value indicates that the DOI better correlates with the source hypothesis than the isotropic expectation and is inconsistent with the isotropic expectation. Conversely, if the dataset differs little from the isotropic expectation (small $D_{\max}$ and large $P_{\mathrm{KS}}$), then one or more of the hypothesis components may be incorrect, or perhaps the method is probing a regime where self-consistency cannot be identified (e.g., strong lensing that hinders identification of significant source correlation beyond the isotropic expectation). Positive $D_{\max}$ at large TS values and any negative $D_{\max}$ are also indicators of these scenarios.

## 3 Dataset

The Pierre Auger Observatory is a hybrid air shower detector located in Malargüe, Argentina. The Surface Detector (SD) consists of a 3000 km$^2$ array of 1660 surface detectors overlooked by the 27 fluorescence telescopes of the Fluorescence Detector (FD) grouped at 4 sites on the array boundary. This allows for complementary measurements of the lateral distribution of air shower particles at ground level by the SD and the longitudinal development of the air shower by the FD.

The data used here consists of 126 events recorded between 1 January 2004 and 31 December 2010 with reconstructed energies greater than 50 EeV and zenith angles smaller than $60°$. These events are required to have at least five active detectors surrounding the detector reporting the highest signal and that the reconstructed core location lie within an equilateral triangle of active detectors.

## 4 Hypothesis Sets

### 4.1 Composition

The data and simulations are hypothesized here to be entirely protons ($Z_p = 1$). This approach appears limited with respect to measurements of the air shower characteristics which are consistent with a heavy composition at the highest energies [8]. Another measurement, namely the correlation of the observed arrival direction at the highest energies ($E \geq 55$ EeV) with extragalactic objects, is suggestive of a light composition and deflection magnitudes of order a few degrees [9].

| Parameter | Min. Value | Max. Value | Step Size |
|---|---|---|---|
| $B_{\odot}$ | -2.0 $\mu$G | 10.0 $\mu$G | 0.5 $\mu$G |
| $p$ | $-20°$ | $-1°$ | $1°$ |
| $Z_1$ | 0.2 kpc | 4.0 kpc | 0.2 kpc |

Table 1: GMF Parameter Space Grid

### 4.2 Source Distributions

Choices for source distributions are drawn from different expectations. Four (4) distinct source distributions are assumed. We use Active Galactic Nuclei (AGN) from the Veron Catalog of Quasars & AGN, 12th Edition [10] (VCV). The 39 month SWIFT-BAT catalog [11] provides a comprehensive all-sky AGN survey in hard X-rays (Swift39). We also use the 2MRS compilation [12] of redshifts of the $K_{mag} < 11.25$ brightest galaxies from the 2MASS catalog [13]. This catalog provides an excellent tracer of the nearby matter distribution in the universe. We apply a variety of redshift cuts $z \leq z_{cut}$ to these three catalogs using the values $z_{cut} = (0.010, 0.011, ..., 0.024)$. Additionally for the 2MRS catalog, we apply an absolute brightness cut that scales with the redshift cut to prevent a bias towards faint galaxies at small distances (2MRS-VS); this cut would be equivalent to $M_K < -25.25$ at redshift $z = 0.048$ ($d$ =200 Mpc). Finally, the radio galaxy Centaurus A is treated as a sole source (CenA).

### 4.3 Galactic Magnetic Field Models

In this work, we implement logarithmic symmetric spiral field models as models for the large-scale regular GMF. Logarithmic symmetric spirals represent *a priori* reasonable models for the functional form of the regular component of the Galactic magnetic field [14]. Such spiral models have been explored in previous studies of UHECR deflection in GMF models [15, 16, 17, 18]. Turbulent and halo components are not considered here, nor are extra-galactic magnetic fields [19, 20, 21].

We investigate the axisymmetric (ASS_A) and bisymmetric models (BSS_S) as described in [16]. These are smoothed versions of models given by [15]. The ASS_A (BSS_S) model exhibits (anti-) symmetry under rotations of $\pi$ around the Galactic pole and is antisymmetric (symmetric) under reflection across the Galactic mid-plane. Three model parameters are scanned using the volume defined by the range shown in Table 1: the field strength in the local solar vicinity $B_{\odot}$, the pitch angle $p$ giving the orientation of the local field vector in the mid-plane, and the scale height $Z_1$ giving the exponential attenuation of the field strength with distance from the mid-plane. All other parameters are the same as in [16].

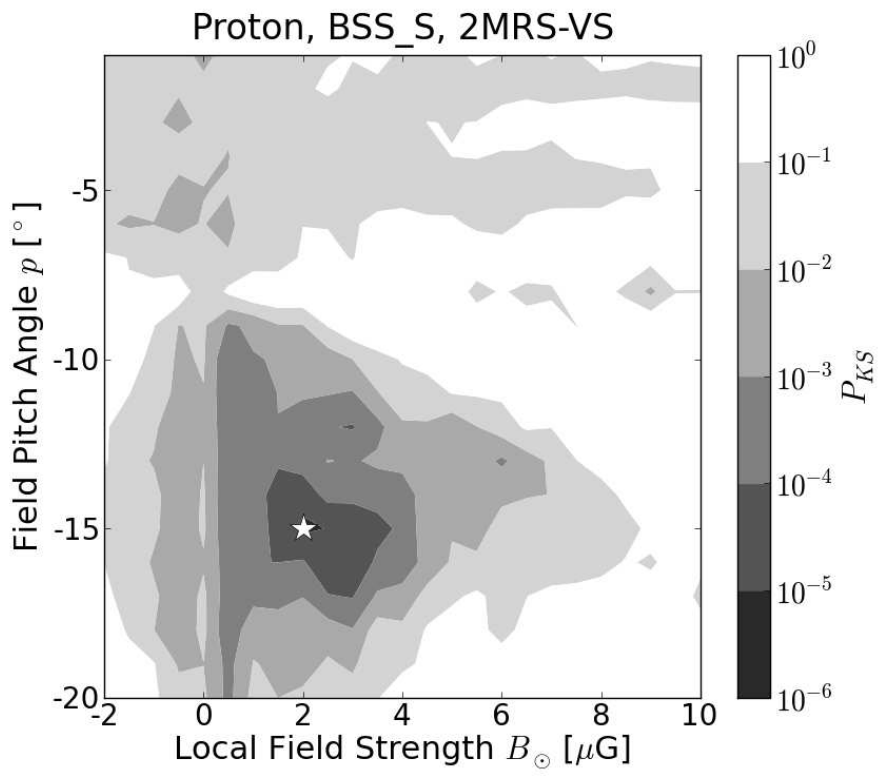

Figure 1: $P_{\rm KS}$ contours for (BSS_S, Proton, 2MRS-VS) hypothesis set with $z_{cut} = 0.017$ at $Z_1 = 2.0$ kpc.

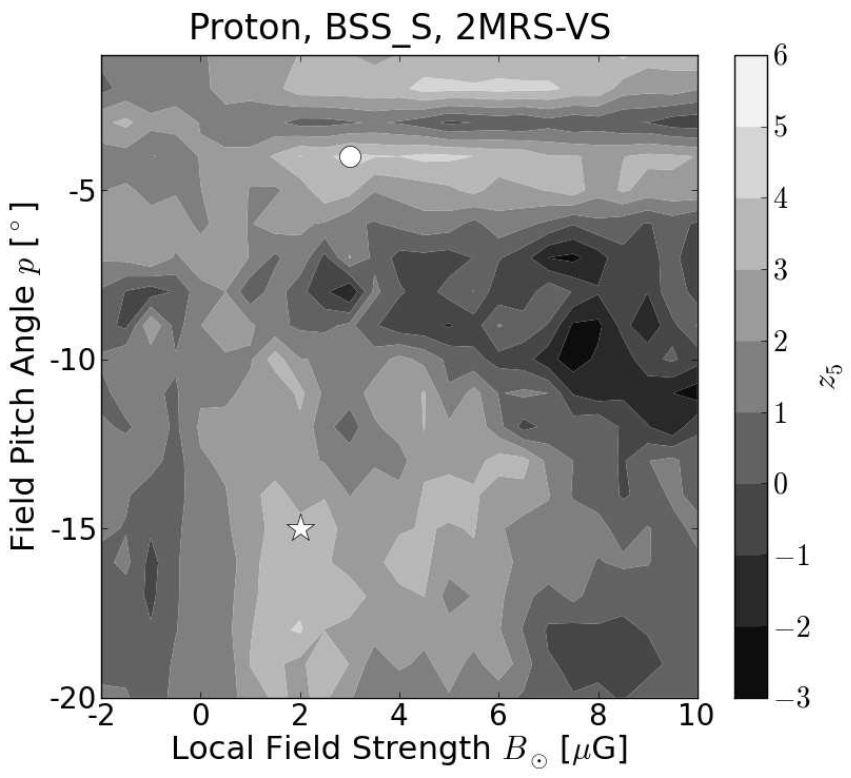

Figure 2: Deviation from isotropic source correlation expectation for (BSS_S, Proton, 2MRS-VS) hypothesis set with $z_{cut} = 0.017$ at $Z_1 = 2.0$ kpc.

## 5 Results

The isotropic TS expectation is comprised of 100 simulations[1] generated by randomly reassigning event directions while respecting the detector exposure. Data and simulations are backtracked using the *CRT* propagation code [22].

Regions of hypothesis set self-consistency will appear as sets of neighboring points in the GMF parameter space indicating comparable small values of $P_{\rm KS}$, positive $D_{\max}$ and excess source correlation at small angular scales. Such regions are observed for sets comprised of both field models and the VCV, Swift39, and 2MRS-VS catalogs implementing redshift cuts between roughly 0.014 and 0.020. For the BSS_S model, these regions typically encompass positive field strengths less than about +4 $\mu$G and $p \approx -15° \pm$ few degrees. Regions in the ASS_A parameter space are found for slightly smaller field strengths of the opposite sign, possibly resulting from a nearby field reversal inside the solar Galactic orbit induced by the model change. Using smaller or larger redshift cuts, these regions are smaller in extent and shallower in $P_{\rm KS}$ as well as shifted within the parameter space. Values of the scale height range from 1 - 3 kpc depending on the specific hypothesis set. These parameter values are compatible with those determined from radio astronomical measurements [3, 4, 6].

Figure 1 depicts an example self-consistent region within the (Proton, BSS_S, 2MRS-VS) hypothesis set. The smallest value of $P_{\rm KS}$ in this set is $6.2 \times 10^{-6}$ at (2.0 $\mu$G, -15.0°), marked by the white star. Low values of $P_{\rm KS}$ indicate that the dataset is behaving differently than isotropy for an extensive region of parameter space, although themselves do not indicate actual source correlation.

Figure 2 depicts the dimensionless deviation from the isotropic expectation of source correlation within 5°, defined as $z_5 = (n_5^d - n_5^i)/s_5^i$, for the same hypothesis set. At each point $n_5^d$ is the number of dataset events. The number of isotropic events is calculated for each of the 100 simulations; $n_5^i$ and $s_5^i$ are the mean and standard deviation of this distribution. At (2.0 $\mu$G, -15.0°) marked by the white star, $z_5 = 3.25$, resulting from 36 events correlating within 5° compared to 21.47 expected. The combination of this small-scale correlation and small $P_{\rm KS}$ indicates that this hypothesis set is self-consistent. Points in the surrounding parameter space possess similar $n_5^d$ and $P_{KS}$ values. Larger $z_5$ are observed outside this region although marginal TS inconsistency with isotropy is found using $P_{\rm KS}$. One such point is (3.0 $\mu$G, -4.0°) marked by the white circle where 35 events comprise $n_5^d$ (18.76 expected) giving $z_5 = 4.35$ but $P_{\rm KS} = 0.051$. Here the sky distribution of the data closely matches that of the isotropic simulations.

Hypothesis set combinations with Cen-A indicate no regions of self-consistency for either field model. The method returns a minimum $P_{\rm KS}$ value of 0.3% (0.05%) for the BSS_S (ASS_A) model. No parameter point in either hypothesis set returns $n_5^d \geq 12$ and large $z_5$ is not observed in conjunction with small $P_{\rm KS}$.

We note that the average $P_{\rm KS}$ for $B_\odot = 0$ $\mu$G in Figure 1 is 0.0123 indicating marginal inconsistency with isotropy. This is in accordance with a previous study of correlations of UHECR arrival directions with extragalactic objects reported in [9] using similar catalogs.

### 5.1 Energy and Angular Resolutions

We also investigate the effects of the energy and angular resolutions to determine the robustness of $P_{\rm KS}$. The Auger Observatory energy ($\sigma_E$) and angular ($\sigma_\psi$) resolutions are 15% and 0.9° [23, 24]. $10^3$ "mock" datasets are generated where the event energies and directions are simultaneously resampled from gaussian distributions with widths $\sigma_E$ and $\sigma_\psi$ centered on the reconstructed energies and arrival directions, respectively. Then, 100 isotropic simulations are constructed for each "mock" dataset by keeping the resampled energies and reassigning an exposure-modulated isotropic direction for each event. A $P_{\rm KS}$ distri-

1. Simulations are unique for each parameter point. For $B_\odot = 0$ $\mu$G this will naturally induce variation in $P_{\rm KS}$.

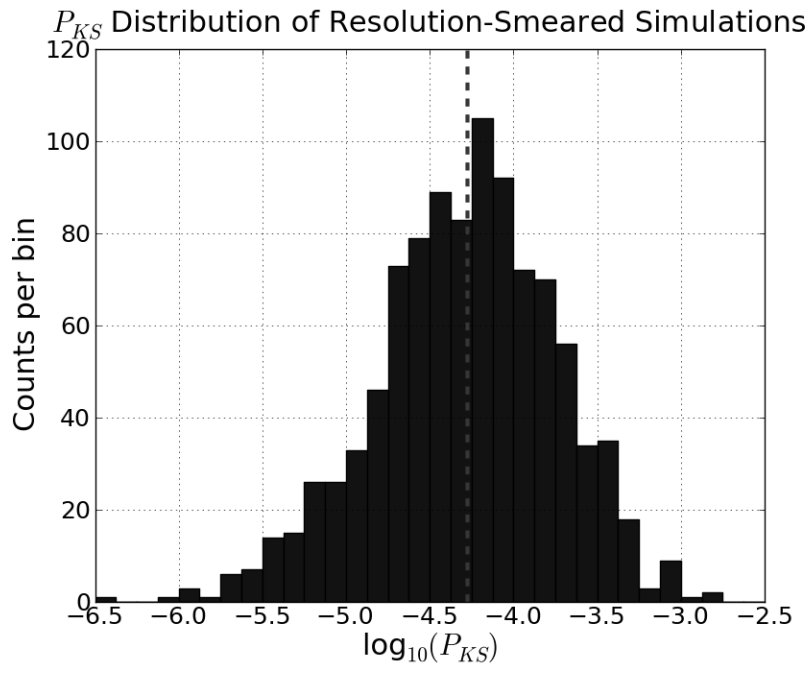

Figure 3: $P_{\rm KS}$ distribution of resolution-smeared "mock" datasets. The $P_{\rm KS}$ value for the unsmeared dataset is depicted by the vertical dotted line.

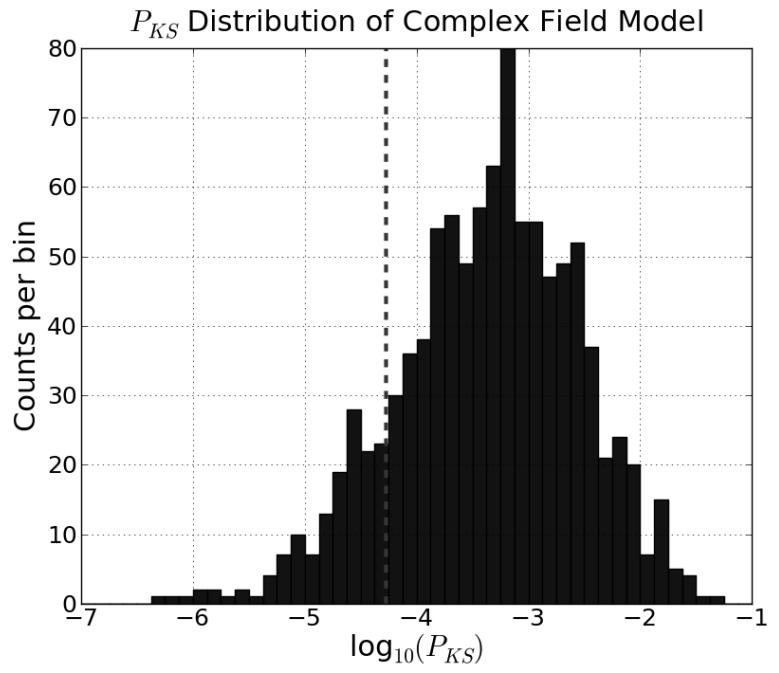

Figure 4: $P_{\rm KS}$ distribution for the regular plus turbulent GMF model. The $P_{\rm KS}$ value for the regular component-only model is depicted by the vertical dotted line.

bution sharply peaked about the value calculated using the reconstructed energy and direction would indicate strong robustness against experimental uncertainties. Figure 3 shows the $P_{\rm KS}$ distribution of the (Proton, BSS_S, 2MRS-VS) hypothesis set under $z_{cut} = 0.017$ for $(B_\odot,\ p,\ Z_1) = (2.0\ \mu{\rm G}, -15^\circ,\ 3.0\ {\rm kpc})$. The determination of self-consistency using the reconstructed energies and directions appears robust against the experimental resolutions.

### 5.2 Turbulent GMF Component

The addition of a turbulent field component is expected to induce isotropization with respect to the hypothesized source distribution. We compare the $P_{\rm KS}$ distribution of $10^3$ realizations of a GMF with regular and turbulent components to the $P_{\rm KS}$ value of the sole regular field. The regular component is the same as in Section 5.1. The turbulent component consists of independent spherical cells of varying sizes drawn from a gaussian distribution with mean 0.1 kpc and rms 0.06 kpc. The field within individual cells has constant magnitude and direction, which is drawn from a gaussian distribution centered at $2.5\ \mu$G with rms $1\ \mu$G. Figure 4 shows the $P_{\rm KS}$ distribution for a particular BSS_S and 2MRS-VS hypothesis set. The addition of a turbulent component tends to strongly isotropize the data.

## 6 Conclusion

In this contribution we have presented a comparison of the Pierre Auger data with specific astrophysical models of the origin, composition, and propagation of UHECRs putting special emphasis on the GMF. In the comparisons we folded the measurements with the astrophysical hypothesis sets and performed a quantitative search for self-consistency. Interesting self-consistent descriptions are found for GMF parameter values that are consistent with contemporary radio astronomical measurements. By scanning a broad phase space of conventional GMF model parameters and several cosmic ray source hypotheses, we have shown an illustrative example of the potential of UHECR precision measurements to obtain new and important information on the fundamental characteristics of the high energy universe.

# Measurement of Low Energy Cosmic Radiation with the Water Cherenkov Detector Array of the Pierre Auger Observatory

HERNÁN ASOREY[1] FOR THE PIERRE AUGER COLLABORATION[2]
[1] *Centro Atómico Bariloche (CNEA), U. N. de Cuyo and U. N. de Río Negro, Bariloche, Río Negro, Argentina*
[2] *Observatorio Pierre Auger, Av. San Martín Norte 304, 5613 Malargüe, Argentina*
*(Full author list: http://www.auger.org/archive/authors_2011_05.html)*
*auger_spokespersons@fnal.gov*

**Abstract:** The flux of secondary cosmic ray particles recorded in the 1660 water Cherenkov detectors of the Pierre Auger Observatory is continuously monitored and analysed for calibration purposes. It is possible to study the flux of primary cosmic rays within the GeV to TeV energy range using calibration histograms of the total energy deposited in the detectors, or data from low threshold scaler counters, which are publicly available. The observed flux is affected by several factors, such as solar activity, the directional properties of Galactic cosmic rays, and local meteorological conditions. The measured secondary flux (of the order of $10^8$ counts per minute) can provide important data on these factors by virtue of the high counting rates which are possible due to the large collecting area of the whole array. Current results of the analysis of each of these factors are presented.

## 1 Introduction

The flux of low energy Galactic cosmic rays (GCRs) is modulated by transient solar eruptions and by changes of the global structure and polarity of the magnetic field in the heliosphere. Variations in the intensity of secondary cosmic rays observed at the surface of the Earth also provide valuable information about the transport of particles in the inner and outer heliosphere, as well as about particles coming into the solar system from the local interstellar medium [1].

The flux of GCRs shows a long-term modulation associated with the solar cycle and short-term variations produced by the passage near the Earth of solar ejecta (i.e., the interplanetary manifestation of a coronal mass ejection, ICME), known as Forbush decreases (Fds) [1]. When observed with neutron monitors and with muon detectors, these Fds exhibit an asymmetrical structure: a characteristic fast decrease of the cosmic ray flux with a time-scale of some hours, and a smooth recovery with a timescale of several days. In some cases, Fds can show complex structures [2] due to the interaction of ICMEs with fast streams of plasma or with other ICMEs during their propagation through the interplanetary space [3].

In this article we present measurements of low energy cosmic radiation, in the range from GeV to several TeV, performed with the Pierre Auger Observatory. In section 2 we describe the water Cherenkov detectors of the Pierre Auger Observatory and their calibration with background muons. In sections 3 and 4 we describe the scaler mode for measuring the flux of low energy radiation and how the calibration histograms can be used to study the dependence of the intensity on energy, while in section 5 the detector response is evaluated. We present conclusions in section 6.

## 2 The Pierre Auger Observatory

The Pierre Auger Observatory [4], located at Malargüe, Argentina (69.3° W, 35.3° S, 1400 m a.s.l.), was designed for the study of cosmic rays of the highest energies. It combines two complementary techniques for the detection of the secondary particles in extensive atmospheric showers (EAS), produced by the interaction of cosmic rays with the atmosphere. In this hybrid design, two types of detectors register EAS: the fluorescence detector (FD) consists of twenty-seven telescopes located at four sites for the observation of ultraviolet fluorescence radiation produced by the shower, and the surface detector array (SD) measures the lateral distribution of secondary particles at ground level.

The SD consists of an arrangement of 1660 water Cherenkov detectors. The detectors of the main array are placed in a triangular grid with a spacing of 1500 m, distributed over an area of 3000 km$^2$. It has an operation duty cycle of nearly 100%. Each water Cherenkov detector consists of a tank containing 12 m$^3$ of high-purity water with an area of 10 m$^2$, providing the full array with a total detec-tor area of about $16\,600\,\mathrm{m}^2$. Cherenkov radiation is generated by the passage of charged, ultra-relativistic EAS particles through the water in the detector. While each detector works as a calorimeter for $e^{\pm}$ and photons (which create $e^{\pm}$ pairs in water), typical muons possess enough energy to go through the full detector, and their Cherenkov emission is proportional to their track length within the water volume.

Three 9" Photonis photomultiplier tubes (PMTs) collect the Cherenkov light in each detector and their signals are processed with a sampling rate of 40 MHz by six 10-bit flash analog-to-digital converters (FADC). Each detector is an autonomous station linked to the central data acquisition system (CDAS) in Malargüe through a dedicated radio network with a bandwidth of 1200 bps per station.

As detailed in [5], the detector is self-calibrated by measuring the pulse signals produced by the particles interacting in the water volume and by building one-minute histograms of their total charge. Since the total signal from a muon depends mainly on its track length, muons produce a characteristic peak. The position of the peak corresponds to $(1.03 \pm 0.02)$ times the total signal deposited by a vertical and central through-going muon [6]. Since the energy loss for energetic muons is $dE/dX \simeq 2\,\mathrm{MeV\,g^{-1}\,cm^2}$, it is possible to calibrate the charge histograms, originally in arbitrary units of FADC counts, in units of energy deposited within the water volume, $E_d$. Figure 1 shows a typical charge histogram, with deposited energy measured in MeV.

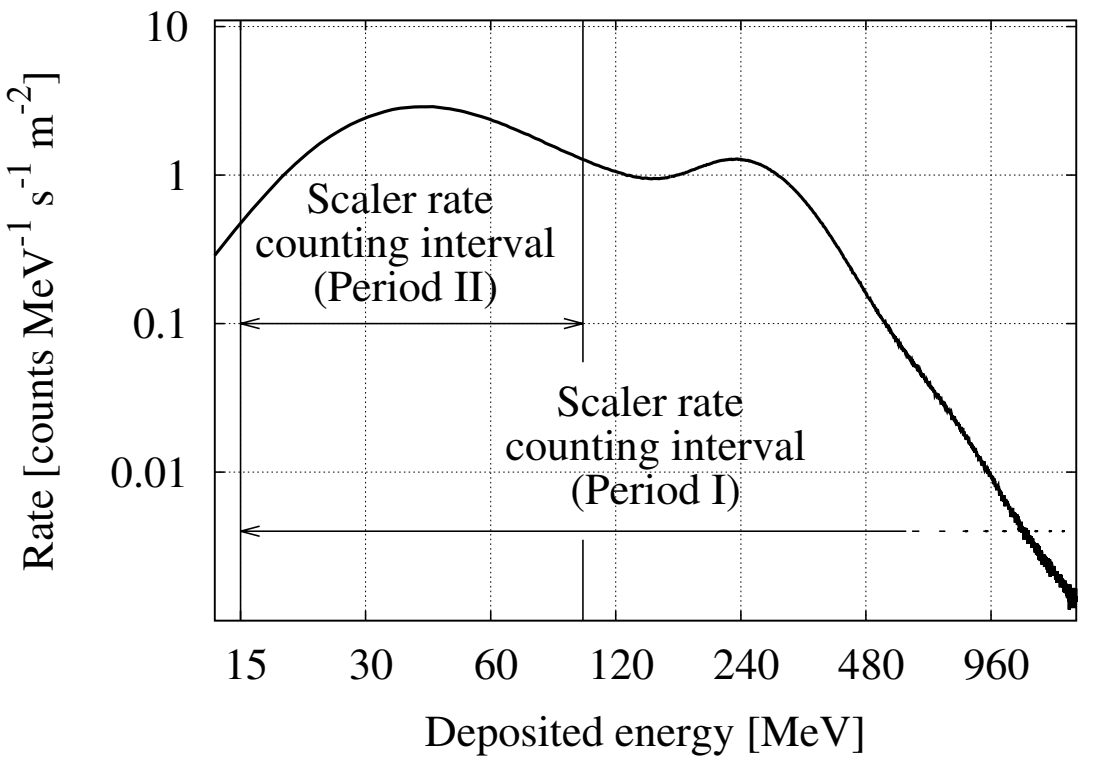

Figure 1: Charge histogram of the signals recorded by one PMT of a water Cherenkov detector of the SD, in bins of energy $E_d$. Both indicated energy regions ($15 \leq (E_d/\mathrm{MeV}) < \infty$ for Period I, and $15 \leq (E_d/\mathrm{MeV}) \leq 100$ for period II) correspond to the counting interval of the scaler mode of the SD, as described in section 3.

Each time the SD detects an EAS, the current one-minute histograms of all nearby detectors with significant signal are sent to CDAS for their storage in order to be used for an off-line calibration of the stations. In this way, on average 10 one-minute histograms of the flux of secondary particles at ground level are registered every minute.

## 3 The Scaler Mode

In March 2005, a new detection mode known as "single particle technique" was implemented in all the detectors of the SD at the Pierre Auger Observatory. This mode consists in the recording of low threshold rates (scalers) for the surface detectors of the array. It is intended for measurements of low energy radiation, long term stability and monitoring studies, and the search of transient events such as gamma ray bursts or Forbush decreases [7].

Two different configurations of the scaler mode were implemented at Auger, in different time periods. In Period I, from 01 Mar 2005 to 20 Sep 2005, the scaler mode counted the total number of signals per second in each detector above a threshold of 3 FADC counts above the baseline (corresponding to $E_d \sim 15\,\mathrm{MeV}$), with a typical rate of about $380\,\mathrm{counts\,s^{-1}\,m^{-2}}$. In Period II, starting at the end of Sep 2005, an upper bound of 20 FADC counts ($E_d \sim 100\,\mathrm{MeV}$) was introduced in order to diminish the sensitivity of the scalers to muon signals. This produced a reduction of the counting rate to about $200\,\mathrm{counts\,s^{-1}\,m^{-2}}$. The main characteristics of the scaler rates for both periods are summarised in table 1. As both periods include the construction phase of the Observatory, the total detector collecting area ranged from $6\,660\,\mathrm{m}^2$ at the beginning of 2005 to $16\,600\,\mathrm{m}^2$ after its completion in 2008, with counting rates of $\sim 2 \times 10^8\,\mathrm{counts\,min^{-1}}$ for the full SD.

The flux of low energy particles at ground level, produced by the interaction of primary cosmic rays at the top of the atmosphere, is intrinsically non-constant. It is furthermore modulated by several atmospheric factors, such as atmospheric pressure. As expected, a strong anti-correlation is observed between the scaler rate and atmospheric pressure, corresponding to $(-2.7 \pm 0.2)$ ‰ per hPa for Period I, and $(-3.6 \pm 0.2)$ ‰ per hPa for Period II [7].

A comparison of the pressure-corrected Auger scaler rate with data from the close-by Los Cerrillos Observatory 6NM64 neutron monitor [8] (Chile, 33.3° S, 70.4° W, 10.8 GV cut-off rigidity) is shown in figure 2. Peaking at 15 May 2005 08:05 UTC, Auger scalers show a decrease of $2.9\%$ with respect to the reference rate for May 2005. The fit of an exponential function for the recovery phase gives a time constant of $2.21 \pm 0.18$(stat) days. A decrease of $4.8\%$ is found in the Los Cerrillos neutron rate, with a time constant of $3.52 \pm 0.12$(stat) days. The observatories possess a similar cut-off rigidity. The differences in the observed time constants result from the higher energy threshold of the Auger detectors compared to neutron monitors.

Instead of using averaged scaler rates for the whole array, it is also possible to study the scaler rate of individual stations, in order to study the propagation of some phenomena across the Auger SD, like the crossing of a storm over the $3000\,\mathrm{km}^2$ of the SD (see [9]). The flux of secondary particles changes as the pressure front moves from the SW towards the NE border of the SD. Additional analyses to study the influence of the variation of electric fields on the flux of EAS particles are currently being carried out.

| Period | Energy range [MeV] | Average scaler rate [counts $s^{-1}$ $m^{-2}$] | Total collection area [$m^2$] |
|---|---|---|---|
| **I**: 01 Mar 2005 - 20 Sep 2005 | $E \gtrsim 15$ | $\sim 380$ | 6 660 - 8 420 |
| **II**: After 20 Sep 2005 | $15 \lesssim E \lesssim 100$ | $\sim 200$ | 8 420 - 16 600 |

Table 1: Count rates for Auger scalers in both periods as defined in section 3. The collecting area range is due to the installation of new detectors in the SD, up to its completion in 2008.

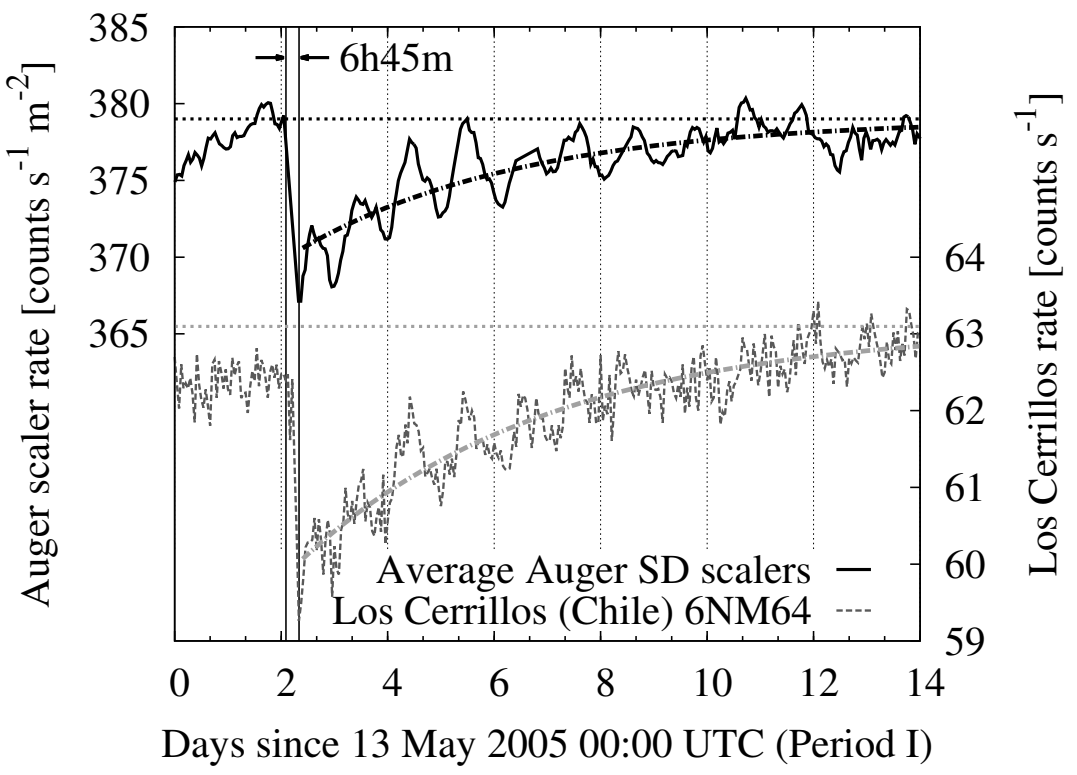

Figure 2: Auger scaler rate (solid line) for the 15 May 2005 Forbush event, compared with the Los Cerrillos (Chile) neutron monitor rate (dashed line). A 2.9% decrease is observed in the Auger data, peaked at 15 May 2005 08:05 UTC, taking 6h45 (solid vertical lines) to reach the peak value. Similar daily variations in the flux are seen at both observatories.

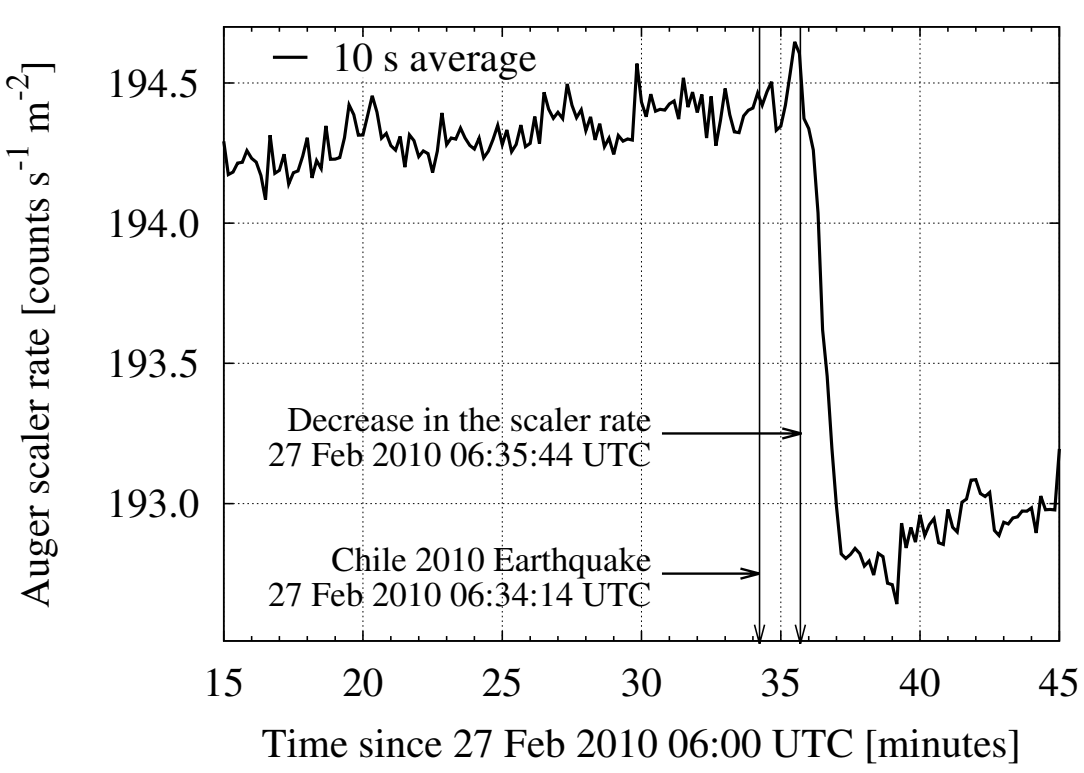

Figure 3: Ten seconds average of the Auger scaler rate for the 27 Feb 2010 Chile major 8.8 magnitude earthquake. A strong $24\,\sigma$ decrease is found $90 \pm 2$ (stat) seconds afterwards, compatible with the time delay expected for seismic S-waves traversing the distance from the epicentre to the Auger Observatory.

It has been suggested that an increment in the flux of low energy cosmic rays could be expected as a precursor to the occurrence of a major earthquake, and some low significance correlations have been found with low altitude spacecraft measurements [10]. At 27 Feb 2010 06h34 UTC an 8.8 magnitude earthquake occurred in Chile, with the epicentre located at 35.9° S, 72.7° W in the Bio-Bio Region, 300 km SW from the Auger Observatory. The averaged scaler rate for the whole array and also for individual stations showed a $24\,\sigma$ decrease beginning $(90 \pm 2)$ seconds after the earthquake. This delay is compatible with the propagation of seismic S-waves over that distance. The scaler rate from 6h15 to 6h45 UTC is shown in figure 3. Although other minor quakes have been recorded by seismographs near the SD, no other similar effects have been found in 6 years of data. Detailed analyses to identify the causes of the observed drop in the scaler rate are underway. These include simulations and shaking tests of selected detectors in the array. After 6 hours, the scaler rate recovered to the mean value for February 2010.

## 4 The Histogram Mode

Except for a strong Forbush decrease observed on 13 Dec 2006, no other significant activity in the heliosphere was recorded in the period 2006–2009. This period was therefore selected to study the influence of atmospheric conditions on the charge histograms. A subset of detectors of the complete array was selected and, for each PMT of each of those stations, a fit of the correlation of the observed rate of particles in each 20 MeV bin of deposited energy with atmospheric pressure was performed, and the fitted parameters were averaged over the selected sub array.

The scaler rate in the "single particle technique" mode is related to the integral of the calibration histogram between two limits defined by the lower and upper scaler trigger bounds (see figure 1). By integrating the calibration histograms with other bounds, it is possible to obtain a rate related to the flux of secondary particles in a specific range of deposited energy. As shown by simulations (see section 5), this flux is related to the number of incident primary cosmic rays of different energies.

The pressure corrected histogram data of the sub array detectors for the 15 May 2005 Fd are shown in figure 4. Six-hour averages in five 20 MeV deposition energy bands of the charge histogram are shown, centred at 140 MeV, 240 MeV, 480 MeV, 840 MeV and 1 GeV. Since a vertical and central through-going muon deposits $\sim$ 240 MeV in the water volume, the integrated counting rate of this band is strongly related to the counting of muons at ground level. The Fd is clearly visible in all the energy bands.

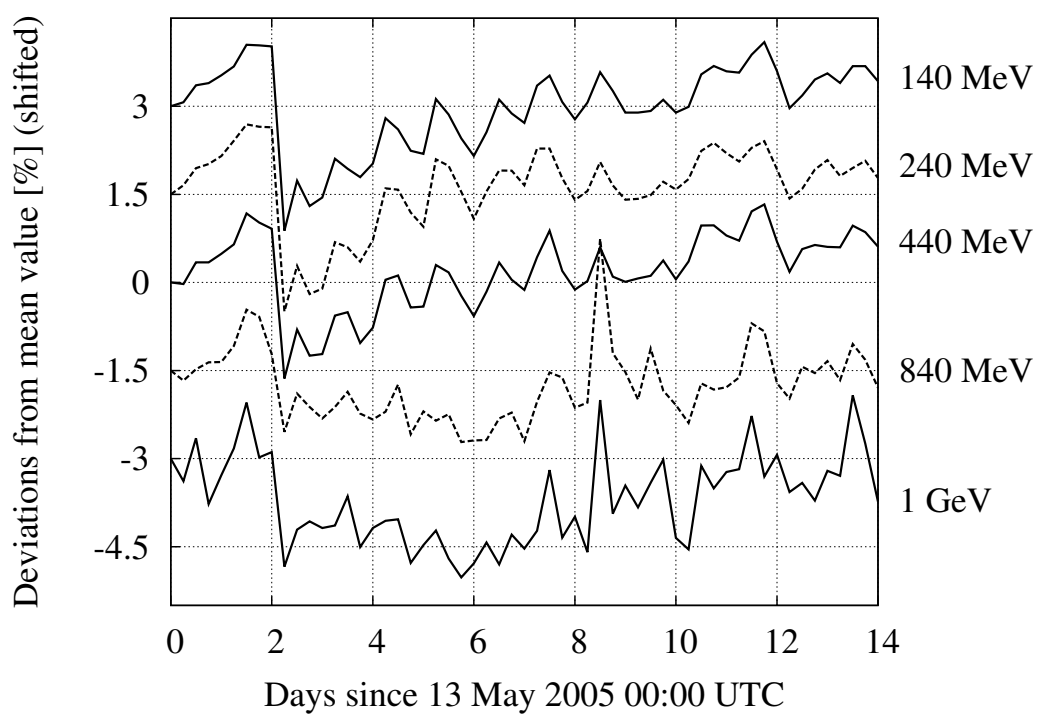

Figure 4: May 2005 Forbush decrease observed by the Pierre Auger Observatory. Each curve shows, as a function of time, the integral of the pressure corrected charge histogram over a 20 MeV bin of deposited energy $E_d$, centred at 140 MeV, 240 MeV, 480 MeV, 840 MeV and 1 GeV. Each energy band was offset by a value of 3%, 1.5%, 0%, $-1.5\%$, and $-3\%$ resp.

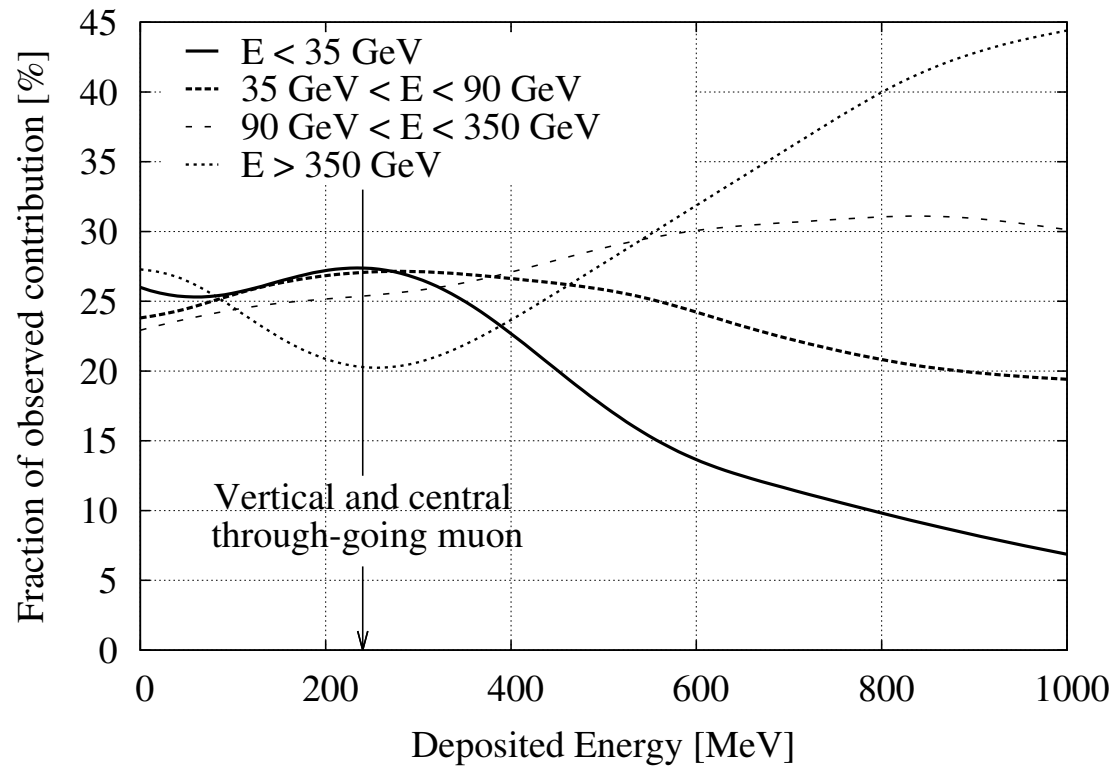

Figure 5: Simulated response of the water Cherenkov detector for four primary GCRs energy bands as a function of the deposited energy of EAS particles within the detector volume. At higher values of $E_d$, the contribution is dominated by GCRs of $E_p > 350$ GeV.

## 5 Detector response

To determine the energies of primary GCRs to which the Auger Observatory low energy modes are sensitive, a set of low energy EAS simulations was performed using CORSIKA 6.980 [11] with QGSJET-II model for the high energy hadronic interactions and GHEISHA low energy interaction routines. The flux of primaries at the top of the atmosphere for all nuclei in the range $1 \leq Z_p \leq 26$ $(1 \leq A_p \leq 56)$ was assumed to be a power law of the form $j(E_p) = j_0(E_p/\mathrm{TeV})^{-\gamma}$. The values for $j_0$ and the spectral index $\gamma$ were obtained from [12], from the measured spectra in the range $(10 \times Z_p) < (E_p/\mathrm{GeV}) < 10^6$, and for $0^\circ \leq \theta_p \leq 88^\circ$ in zenith angle. The detector response was simulated using a simple simulator developed within the Auger data analysis framework. Results are shown in figure 5, where the fraction of the observed contribution for four primary GCR energy bands is plotted as a function of the energy deposition within the detector volume. Different regimes are visible in the figure: while for $E_d \sim 240$ MeV, the typical deposited energies for single muons, the contribution is dominated by primaries of $E_p < 350$ GeV, at $E_d \gtrsim 600$ MeV the contribution becomes dominated by GCRs of higher energies.

## 6 Conclusions

The study of variations in the galactic cosmic ray flux is important because it carries information about the local interstellar and interplanetary media, and about the physical mechanisms involved in the interaction between charged particles and plasma in the heliosphere.

In this work, measurements of low energy cosmic radiation in the GeV–TeV range using the surface detector array of the Pierre Auger Observatory are described. The capabilities of water Cherenkov detectors for the study of transient solar events at the Earth surface has been demonstrated using the scaler data.

The scaler mode is now complemented by the analysis of the calibration charge histograms, which enable the study of the time evolution of transient solar events at the same rigidity cut-off for different bands of deposited energy.

The full scaler data set, averaged every 15 minutes for the whole surface detector array, is publicly available and can be downloaded from the Pierre Auger Observatory Public Event Display web site [13]. A user-friendly web interface has been set up to handle, visualise and download the data.

# Acknowledgments

The successful installation, commissioning and operation of the Pierre Auger Observatory would not have been possible without the strong commitment and effort from the technical and administrative staff in Malargüe.

We are very grateful to the following agencies and organizations for financial support:

Comisión Nacional de Energía Atómica, Fundación Antorchas, Gobierno De La Provincia de Mendoza, Municipalidad de Malargüe, NDM Holdings and Valle Las Leñas, in gratitude for their continuing cooperation over land access, Argentina; the Australian Research Council; Conselho Nacional de Desenvolvimento Científico e Tecnológico (CNPq), Financiadora de Estudos e Projetos (FINEP), Fundação de Amparo à Pesquisa do Estado de Rio de Janeiro (FAPERJ), Fundação de Amparo à Pesquisa do Estado de São Paulo (FAPESP), Ministério de Ciência e Tecnologia (MCT), Brazil; AVCR AV0Z10100502 and AV0Z10100522, GAAV KJB100100904, MSMT-CR LA08016, LC527, 1M06002, and MSM0021620859, Czech Republic; Centre de Calcul IN2P3/CNRS, Centre National de la Recherche Scientifique (CNRS), Conseil Régional Ile-de-France, Département Physique Nucléaire et Corpusculaire (PNC-IN2P3/CNRS), Département Sciences de l'Univers (SDU-INSU/CNRS), France; Bundesministerium für Bildung und Forschung (BMBF), Deutsche Forschungsgemeinschaft (DFG), Finanzministerium Baden-Württemberg, Helmholtz-Gemeinschaft Deutscher Forschungszentren (HGF), Ministerium für Wissenschaft und Forschung, Nordrhein-Westfalen, Ministerium für Wissenschaft, Forschung und Kunst, Baden-Württemberg, Germany; Istituto Nazionale di Fisica Nucleare (INFN), Ministero dell'Istruzione, dell'Università e della Ricerca (MIUR), Italy; Consejo Nacional de Ciencia y Tecnología (CONACYT), Mexico; Ministerie van Onderwijs, Cultuur en Wetenschap, Nederlandse Organisatie voor Wetenschappelijk Onderzoek (NWO), Stichting voor Fundamenteel Onderzoek der Materie (FOM), Netherlands; Ministry of Science and Higher Education, Grant Nos. 1 P03 D 014 30, N202 090 31/0623, and PAP/218/2006, Poland; Fundação para a Ciência e a Tecnologia, Portugal; Ministry for Higher Education, Science, and Technology, Slovenian Research Agency, Slovenia; Comunidad de Madrid, Consejería de Educación de la Comunidad de Castilla La Mancha, FEDER funds, Ministerio de Ciencia e Innovación and Consolider-Ingenio 2010 (CPAN), Xunta de Galicia, Spain; Science and Technology Facilities Council, United Kingdom; Department of Energy, Contract Nos. DE-AC02-07CH11359, DE-FR02-04ER41300, National Science Foundation, Grant No. 0450696, The Grainger Foundation USA; ALFA-EC / HELEN, European Union 6th Framework Program, Grant No. MEIF-CT-2005-025057, European Union 7th Framework Program, Grant No. PIEF-GA-2008-220240, and UNESCO.